\documentclass[12pt,twoside]{mitthesis}
\usepackage{lgrind}
\usepackage{cmap}
\usepackage[T1]{fontenc}
\usepackage{amssymb}
\usepackage{graphicx}
\usepackage{booktabs}
\usepackage{dcolumn}
\usepackage{amsmath}
\usepackage{subcaption}
\usepackage{booktabs}
\usepackage{graphicx}
\usepackage[position=t,singlelinecheck=off]{subfig}
\usepackage{caption}
\usepackage{float}
\usepackage{multirow}
\usepackage{amsmath}
\usepackage{relsize}

\usepackage{color}
\definecolor{Blue}{rgb}{0,0,1}

\definecolor{Orange}{rgb}{1,0.5,0}

\definecolor{Green}{rgb}{0,1,0}

\begin{document}

%
%
%
%
%
%
%
%
%
%
%

\title{Complex Systems and A Computational Social Science Perspective on the Labor Market}

\author{Abdullah Almaatouq}
       \prevdegrees{BSc., University of Southampton (2012)}
\department{Center for Computational Engineering \& Program in Media Arts and Sciences}
\degree{Master of Science in Computation for Design and Optimization \& Master of Science in Media Arts and Sciences}

\degreemonth{June}
\degreeyear{2016}
\thesisdate{May 6, 2016}


\supervisor{Alex (Sandy) Pentland}{Toshiba Professor of Media Arts and Sciences}
\supervisor{Olivier de Weck }{Professor of Aeronautics and Astronautics and Engineering Systems}

\chairman{Nicolas Hadjiconstantinou}{Co-Director, Computation for Design and Optimization}

\chairmand{Pattie Maes}{Academic head, Program in Media Arts and Sciences}

\maketitle



\begin{abstractpage}
	\vspace{-1.5mm}
Labor market institutions are central for modern economies, and their polices can directly affect unemployment rates and economic growth. At the individual level, unemployment often has a detrimental impact on people's well-being and health. At the national level, high employment is one of the central goals of any economic policy, due to its close association with national prosperity. The main goal of this thesis is to highlight the need for frameworks that take into account the complex structure of labor market interactions. In particular, we explore the benefits of leveraging tools from computational social science, network science, and data-driven theories to measure the flow of opportunities and information in the context of the labor market. First, we investigate our key hypothesis, which is that opportunity/information flow through weak ties, and this is a key determinant of the length of unemployment. We then extend the idea of opportunity/information flow to clusters of other economic activities, where we expect the flow within clusters of related activities to be higher than within isolated activities. This captures the intuition that within related activities there are more ``capitals'' involved and that such activities require similar ``capabilities.'' Therefore, more extensive clusters of economic activities should generate greater growth through exploiting the greater flow of opportunities and information. We quantify the opportunity/information flow using a complexity measure of two economic activities (i.e. jobs and exports). 
\end{abstractpage}

\pagestyle{plain}
\smaller{\tableofcontents}
\chapter{Introduction}
Unemployment is a central issue in all countries. At the individual level, unemployment often has a detrimental impact on people's well-being and health~\cite{chang2013impact}. At the national level, high employment is one of the central goals of any economic policy, due to its close association with the economic, social and political prosperity of countries~\cite{lopez2015network}. 

In the United States, during normal economic times, 3 million people change jobs each month, while 1.8 million people separate (either voluntarily or involuntarily) from their employers without new jobs to go to (thus becoming unemployed), and another 1.8 million people move out of unemployment into new jobs~\cite{axtell2013endogenous}. That is an overall monthly turnover of 4\% of the total workforce of 150 million. 

The 2008 economic crisis has had far-reaching consequences for labor markets around the world~\cite{chang2013impact}. According to the International Labour Organization, the number of unemployed individuals worldwide reached around 212 million in 2009, and youth unemployment increased more rapidly than overall unemployment~\cite{bell2011young}. Conventional labor models attempt to explain unemployment flows by aggregating job hiring and job separations across companies~\cite{davis1998job,rogerson2004search}, which ignores the varieties of job seeking behavior, firm strategies, and their consequences on national economic growth.

\section{The Need for a New View of the Labor Market}
Employment dynamics are the product of complex interactions taking place within and between people, firms, cities, and countries. Traditionally, labor economies are studied through search and matching models, which aggregate job hirings and job separations (i.e., employees being fired, quitting their jobs, etc.) across companies to get pools of job seekers (i.e., job changers or the unemployed), and then matching them to vacancies~\cite{davis1998job,rogerson2004search}. These models typically consist of a stochastic process~\cite{mortensen1994job} that matches unemployed individuals and vacancies using aggregate matching functions~\cite{petrongolo2001looking,stevens2007new}. However, such search and matching models have inherent limitations concealed by the way they are constructed.

One of the biggest limitations is the \textit{averaging} of the behavior of independent actors (often assumed to have a normal distribution), which determines where the market eventually settles (reaches equilibrium)~\cite{pentland2015economics}. Such aggregation can lead to a loss of critical information about variations in the behaviors of job-seekers and firms, and also ignores the contribution of actor-specific characteristics in generating aggregate fluctuations (especially from shocks~\cite{di2014firms}). In addition, policies are usually designed and implemented at the level of individuals, demographic classes, and organizations, so ignoring realistic micro-mechanisms could lead to unfavorable consequences in the real world~\cite{guerrero2016understanding}.

Another limitation is the \textit{assumption of independence}. The idea that actors (people or firms) act independently can be problematic. Evidently, people and companies can synchronize behaviors by learning from each other (i.e., copying the successful actions of others), and by reacting to external conditions such as shocks~\cite{di2014firms}. This synchronization in market-like situations can amplify fluctuations and result in financial bubbles, political upheavals and health fads~\cite{pentland2014social}.

For instance, traditional policies rely on Pigouvian taxation or subsidies as economic incentives that change individual behaviors by internalizing the externalities they produce (e.g., reducing energy consumption, improving the work environment, motivating the unemployed to find jobs, cleaning up the neighborhood, etc). Ideally, economic incentives would lead to long-lasting behavioral change and allow us to eradicate some of the most important problems in modern society, such as pollution, global warming, and rising health care and insurance costs ~\cite{mani2013inducing}. However, changing the behavior of self-interested individuals is extremely difficult ~\cite{dietz2003struggle}. A fundamental assumption of Pigouvian taxation or subsidies is that the individual is the correct target for incentives. In this thesis, we will present evidence that this assumption may have limited utility, that social networks containing the individuals are an important additional target for incentives, and that these ``network incentives'' can amplify the effect of the economic incentives.

A major challenge for studying economic systems is the \textit{lack of granular data} at the individual level and describing the connections or relationships between them (i.e., social networks). Sherwin Rosen and Robert Willis, two pioneers in labor economics, explored this challenge in two independent articles in the mid-1980s, focusing on high resolution datasets that capture the interactions between individuals and firms in order to advance the study of the labor market~\cite{rosen1987theory,willis1985wage}. Yet, much of our understanding of the factors that affect employability has been traditionally obtained through complex and costly surveys, with an update rate ranging from months to decades, which limits the scope of the studies and potentially biases the data~\cite{henrich2001search}. In addition, as participation rates in unemployment surveys drop, serious questions regarding the declining accuracy and increased bias in unemployment numbers have been raised~\cite{krueger2014evolution}. Another challenge for studying the labor market from a social network framework is that social ties are not usually susceptible to the traditional tools of economic policy, and are often difficult to characterize empirically.

Therefore, unlike classical labor-market models and despite the measurement challenges, in reality, job search behavior and employment outcomes are closely related to the personalities, expectancies, social networks, motives, and individual differences~\cite{kanfer2001job} of \textit{the job seekers}. In addition, hiring and separation occur at \textit{individual companies}~\cite{guerrero2013employment}, which adapt to the urban developments and economic changes in \textit{cities}, and translate into human, infrastructural, physical, and institutional capitals that could determine the economic performance of entire systems such as \textit{countries}~\cite{hidalgo2007product}. 

Generally, economic science is perceived as a theoretically driven discipline, where data is mainly used for hypothesis testing instead of serving as an instrument for the construction of theories and models~\cite{guerrero2016understanding}. This has restricted the discipline to developing easily tractable models, rather than taking full advantage of the big data that is continuously pushing the development of other social sciences.

\section{Networks and the Era of Data-Driven Theory}
Recent widespread adoption of electronic and pervasive technologies, the development of e-government, and open data movements have enabled the study of human behavior at an unprecedented level and helped uncover universal patterns underlying human activity. Lazer, Pentland et al.~\cite{lazer2009life} formally introduced computational social science (CSS) as a new field of research that studies individuals and groups in order to understand populations, organizations, and societies using big data, i.e. phone call records~\cite{aleissawired,alhasouncity}, GPS traces~\cite{zheng2011learning}, credit card transactions~\cite{sobolevsky2014money,almaatouq2014malicious}, web page visits~\cite{deshpande2004selective}, emails~\cite{karsai2011small}, and data from social media~\cite{nouh2014social}. Driven by the ubiquitous availability of data and inexpensive data storage capabilities, the concept of ``big data'' has permeated the public discourse and led to surprising insights across the sciences and humanities. Such understanding can answer epistemological questions on human behavior in a data-driven manner, and provide prescriptive guidelines for persuading people to undertake certain actions in real-world social scenarios. In particular, this availability of data over the past fifteen years has shed on the important role  networks play in human society.

Networks are everywhere. The cell is a network of molecules connected by biochemical reactions~\cite{barabasi2004network}. The brain is a network of nerve cells (neurons) connected by axons~\cite{hawrylycz2015canonical,van2010exploring}. Societies are networks of people connected according to their social relationships~\cite{pentland2014social}, shared acquaintances~\cite{wood2014agricultural}, classmates~\cite{eagle2006reality}, friendships~\cite{almaatouq2016b}, collaborations~\cite{newman2001structure}, social media followers~\cite{almaatouq2014twitter,almaatouq2016if}, and sexual relationships~\cite{kretzschmar2000sexual}. Networks also pervade technology: the internet~\cite{albert1999internet}, power grids~\cite{pagani2013power}, and transportation systems~\cite{gonzalez2008understanding,simini2012universal}, to mention just a few notable examples. Even the language we use to convey our thoughts is a network of words connected by significant co-occurrence in text~\cite{koelsch2002bach}. 

Some economic research has analyzed the job search as a specialized social network, where information about vacancies travels along social ties~\cite{boorman1975combinatiorial,calvo2004effects}. This approach was inspired by social scientist Mark Granovetter's work on the `strength of weak ties' ~\cite{granovetter1973strength}. The resulting economic models are consistent with Granovetter's hypothesis, which states that the strength of a tie between A and B increases with the overlap of their friendship circles, making weak ties an important connector between communities. Another approach focuses on the dynamics of hiring firms, looking at the problem through a purely theoretical model where worker transitions between firms are viewed as a Markov process~\cite{karunanithi1994neural}. However, this modeling approach mixes aggregate and disaggregate features, and lacks empirical verification~\cite{lopez2015network}. More recently, researchers have proposed the use of labor flow networks (i.e., workers' mobility) and studied their empirical properties~\cite{guerrero2013employment} alongside the construction of a basic model that predict job mobility at high resolution (down to the firm level). These results were consistent with the collected empirical evidence~\cite{lopez2015network}.

\section{Main Contributions}
The main goal of this thesis is to highlight the need for frameworks that take into account the complex structure of labor market interactions. As Alex (Sandy) Pentland demonstrated in his book \textit{Social Physics: How Good Ideas Spread---The Lesson From a New Science}, the best decision-making environments are the ones with high levels of ``idea flow.'' This can be captured by quantitatively measuring: i) how often people in a group communicate with each other (i.e., sharing social knowledge); and ii) how often they seek out new ideas and new people ~\cite{pentland2014social}. According to the theory of idea flow, the emergence of innovative ideas or the coordination of collective action depends on network structure. The ability to measure communication and transactions on an unprecedented scale and use this knowledge about the flow of ideas (or information and opportunities) is the central element of this thesis. In particular, we explore the benefits of leveraging advancements and tools from computational social science, network science, and data-driven theories to  measure the opportunity/information flow in the context of the labor market. 

First, we investigate our key hypothesis, which is that opportunity/information flow through weak ties (at the individual level), and this flow is a key determinant of the length of unemployment. This implies that behaviors that promote social exploration and/or lower the cost of this exploration are key strategies for job-seekers to find employment. In Chapter~\ref{ch1:indiv}, we provide evidence relating unemployment to individual behavior by examining the signatures of unemployment through mobile phone usage. We find that mobile phone behavioral indicators are consistently associated with unemployment rates and this relationship persists even when we include detailed controls. We construct a simple model to produce accurate, easily interpretable reconstructions of district-level unemployment from individuals' mobile phones through usage patterns alone. We also investigate the role of social cohesion in increasing the effectiveness of incentives. Incentives are of utmost importance in promoting behaviors that are likely to gain job-seekers employment and succeed in their chosen occupations. This benefits not only the job seekers themselves, but also the workforce, the community and the economy. In particular, we present evidence that social cohesion---defined as the density of social ties---can amplify the effect of incentives through the mechanism of social pressure. Therefore, we expect that the campaigns that target lowering unemployment to be most effective in areas with high social capital.

Finally, we extend the idea of opportunity/information flow beyond job-seekers to clusters of other economic activities. Similarly, we expect the flow within clusters of related activities to be higher than within isolated activities. This captures the intuition that more ``capitals'' are involved and that such activities require similar ``capabilities.'' Therefore, more extensive clusters of economic activities should generate greater growth, by exploiting the greater flow of opportunities/information. We quantify the opportunity/information flow using a ``complexity measure'' of two economic activities (i.e. jobs and exports). We find that this measure indeed predicts economic growth. More precisely, we investigate how jobs are related to each other and introduce a novel measure that captures the opportunity/information flow by quantifying the complexity of the job market. Understanding the ecosystem of job markets and possible changes within them is key when planning for the future. We introduce and investigate a measure associated with \textit{job complexity}. A job with high complexity means that for a city to have significant prominence in this job, it also needs to have significant prominence in many other associated jobs. This measure captures the idea that there are strong dependencies between jobs. We find it to be a strong predictor of a city's economic performance (measured in GDP), and a good predictor of a particular job's susceptibility to automation. We also investigate the concept of the opportunity/information flow by examining the main features of the Economic Complexity Index (ECI) and highlight the ability of this single measure to capture many factors regarding the capabilities of a country---particularly the basis of the labor economy, its human capital. 

\chapter{The Individual}
\label{ch1:indiv}
In this chapter, we investigate our key hypothesis, which is that opportunity/information at the individual level flow through weak ties, and this is a key determinant of the length of unemployment. This implies that behaviors that promote social exploration and/or lower the cost of this exploration are key strategies for job-seekers to find employment. In particular, in Section~\ref{mobile} we provide evidence relating unemployment to individual behavior by examining the signatures of unemployment through mobile phone usage. Furthermore, in Section~\ref{reciprocity} we present evidence on the role of social cohesion---defined as the density of social ties---in increasing the effectiveness of incentives through the mechanism of social pressure. Therefore, we expect that campaigns that focus on lowering unemployment will be most effective in areas with high social cohesion.
\section{The Behavioral Signature of Unemployment}
\label{mobile}
The mapping of a population's socioeconomic well-being is highly constrained by the logistics of censuses and surveys. Consequently, spatially detailed changes across scales of days, weeks, or months, or even year to year, are difficult to assess; thus, the speed at which policies can be designed and evaluated is limited.
However, recent studies have shown the value of mobile phone data as an enabling methodology for demographic modeling and measurement. In this work, we investigated whether indicators extracted from mobile phone usage can reveal information about the socioeconomic status of microregions such as districts (average spatial resolution $< 2.7km$). We examined anonymized mobile phone metadata combined with beneficiaries' records from an unemployment benefit program, and found that aggregated activity, social, and mobility patterns strongly correlate with unemployment. Furthermore, we constructed a simple model to produce accurate reconstructions of district-level unemployment from individuals' mobile communication patterns alone. 
Our results suggest that reliable and cost-effective economic indicators could be based on passively collected and anonymized mobile phone data. With similar data collected every day by telecommunication services across the world, survey-based methods of measuring community socioeconomic status could potentially be augmented or replaced by such passive sensing methods.

In this chapter, we focus on Saudi Arabia as a case study. In particular, the Saudi Arabian labor market is characterized by an emergence of high youth participation, and high levels of expatriate workers (mostly in the private sector), leading to a depression in the levels of average general wages, lower productivity, and potentially high unemployment levels.
Rapid economic and population growth occurring throughout Saudi Arabia are posing new opportunities and changes for the Kingdom. The labor economy is one of the foremost challenges facing the country, especially with the national population increasing by close to half a million people per year and the unemployment rate among Saudis hovering around 10\%--and above 30\% for females and those aged 20 to 24.

\subsection{Computational Social Science Approach}
A major challenge in the development space is the lack of access to reliable and timely socioeconomic data. Much of our understanding of the factors that affect the economic development of cities has been traditionally obtained through complex and costly surveys, with an update rate ranging from months to decades, which limits the scope of the studies and potentially biases the data~\cite{henrich2001search}. In addition, as participation rates in unemployment surveys drop, serious questions regarding the declining accuracy and increased bias in unemployment numbers have been raised~\cite{krueger2014evolution}. However, recent widespread adoption of electronic and pervasive technologies (e.g., mobile penetration rate of 100\% in most countries), and the development of Computational Social Science~\cite{lazer2009life}, has enabled digital `breadcrumbs' (e.g., phone records, GPS traces, credit card transactions, web page visits, and online social networks) to act as in situ sensors for human behavior, allowing researchers to quantify social actions and study of human behavior on an unprecedented scale~\cite{eagle2006reality,gonzalez2008understanding,pentland2014social}.

Scientists have long suspected that human behavior is closely linked with socioeconomic status, as many of our daily routines are driven by activities afforded by such status, or related to maintaining or improving it~\cite{becker1976economic,granovetter1973strength,Granovetter85economicaction}. Recent studies have provided empirical support and investigated these theories in vast and rich datasets (e.g., social media~\cite{llorente2014social} and phone records~\cite{toole2015tracking,eagle2010network}) with varying scales and granularities~\cite{blumenstock2015predicting}.

In this work, we provide empirical results that support the use of individual communication patterns drawn from Call Detail Records (CDRs) to infer district-level behavioral indicators and examine their ability to explain unemployment as a socioeconomic output. In order to achieve this result, we combine a large dataset of CDRs with records from the unemployment benefit program. We quantify individual behavioral indicators from over 1.8 billion logged mobile phone activities, which were generated by 2.8 million unique phone numbers and distributed among 148 different districts in Riyadh, the capital of Saudi Arabia. We extract aggregated mobile phone usage indicators (e.g., activity patterns, social interactions, and spatial markers), as detailed in Section~\ref{sec:indicators}, and examine the relationships between the district-level behaviors and unemployment rates. Then, we address whether the identified variables with strong correlations suffice to explain the observed unemployment (explanatory power) and determine which of those variables are the most important (i.e., which has more explanatory power than the others). We then explore the performance of several predictive models in reconstructing unemployment at the district level. Our approach is different from prior work that has examined the relation between regional wealth and regional phone use (at the city scale), as we focus on microregions composed of just a few households with unprecedentedly high-quality ground truth labels. This type of work can provide critical input to social and economic research and policy as well as the allocation of resources.

In summary, we frame our contributions as follows:
\begin{itemize}
	\item We find that CDR indicators are consistently associated with unemployment rates and this relationship persists even when we include detailed controls for a district's area, population, and mobile penetration rate.
	\item We compare several categories of indicators with respect to their performance in predicting unemployment rates at the district level.
\end{itemize}

\begin{figure}[h!]
	\centering
	\includegraphics[width=1\columnwidth]{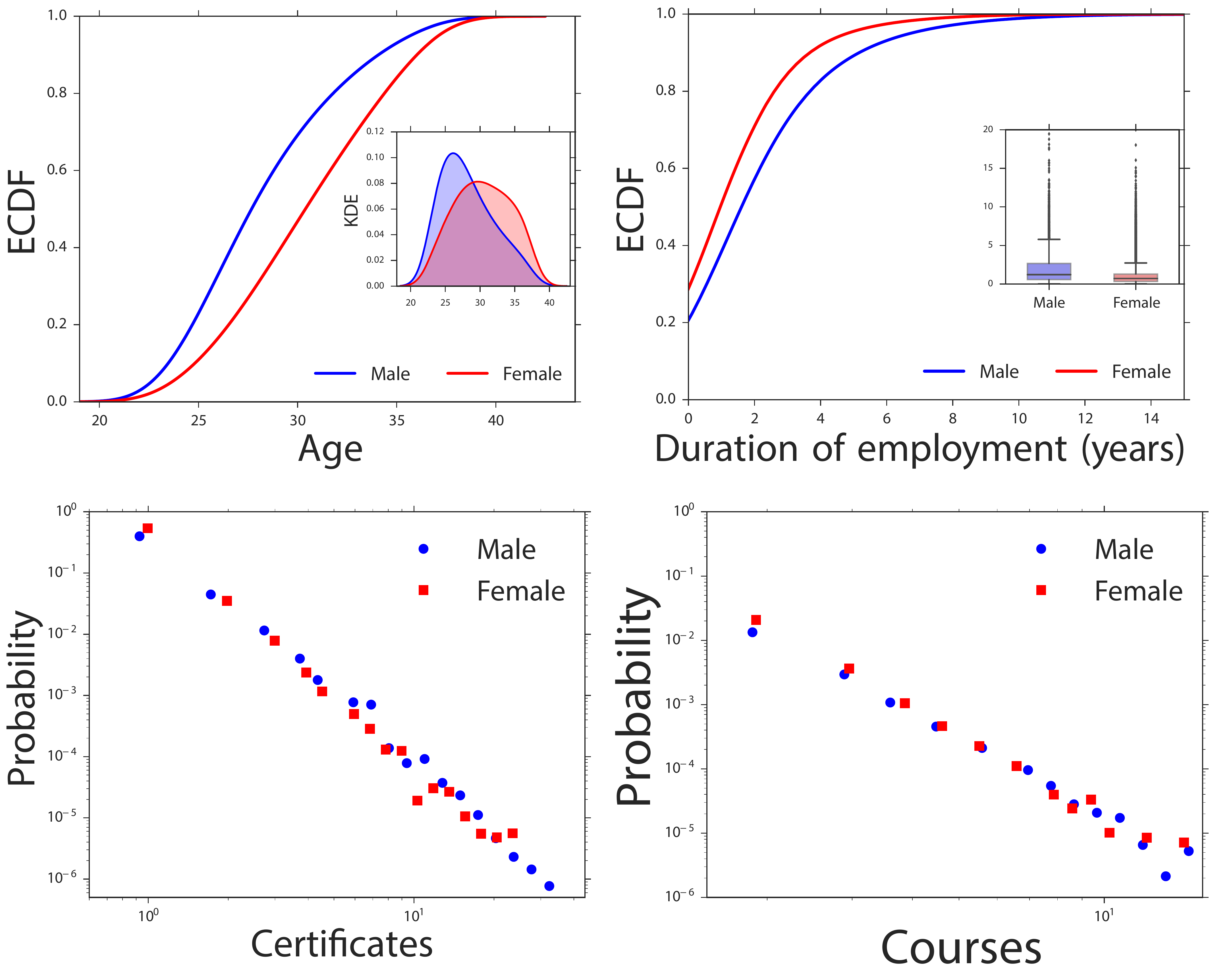}
	\caption{The distribution of age, previous years of experience and qualification of the Hafiz eligible applicants. We find that unemployed female population is older and with less years of work experience. However, there are no obvious differences in terms of the number qualifications.}
	\label{fig:basic_hafiz_statistics}
\end{figure}

\begin{figure}[h!]
	\centering
	\includegraphics[width=0.9\columnwidth]{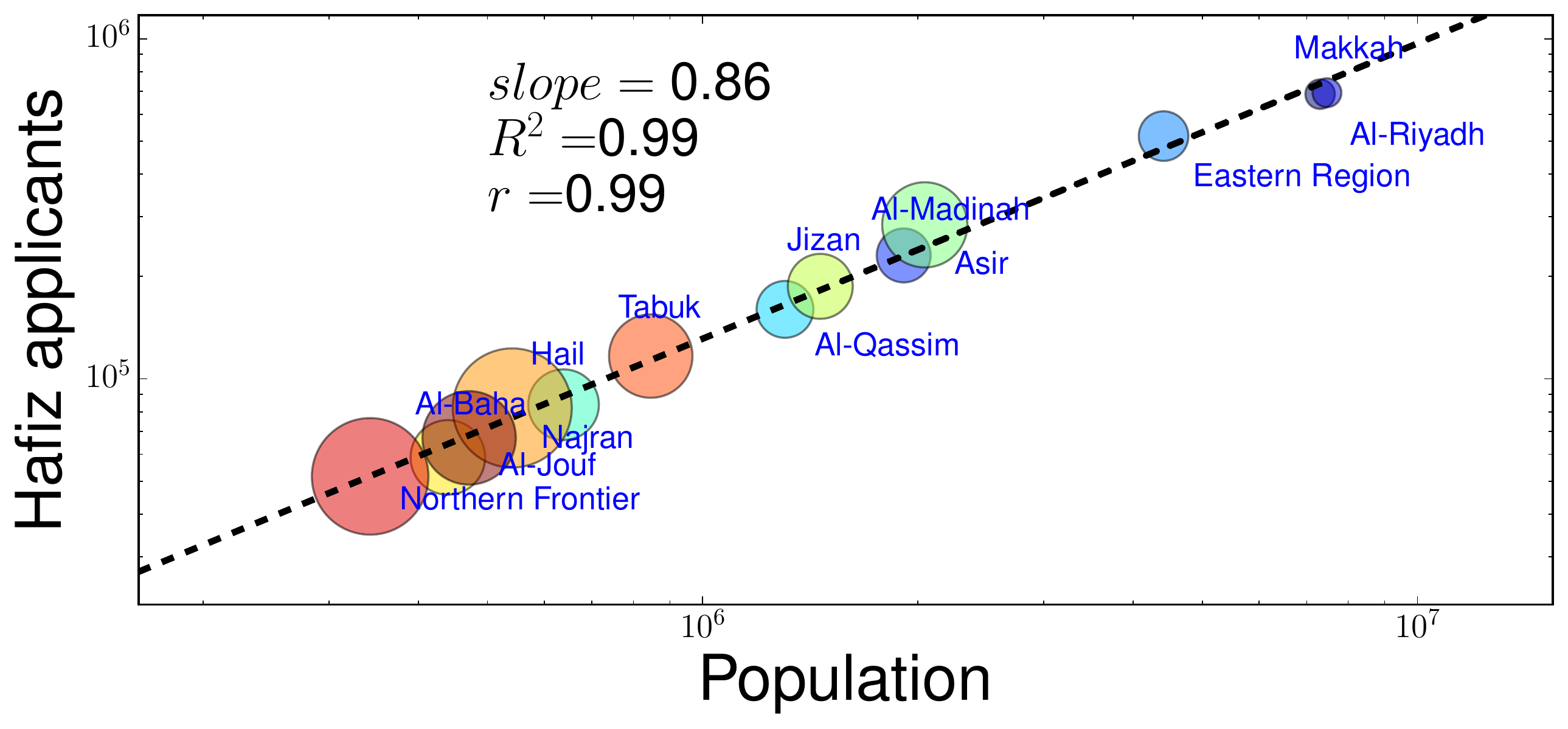}
	\caption{scaling relationships between population at the regional level versus number of Hafiz applications. Best-fit scaling relations are shown as dashed line.}
	\label{fig:hafiz_scaling}
\end{figure}

\begin{figure}[h!]
	\centering
	\includegraphics[width=0.8\columnwidth]{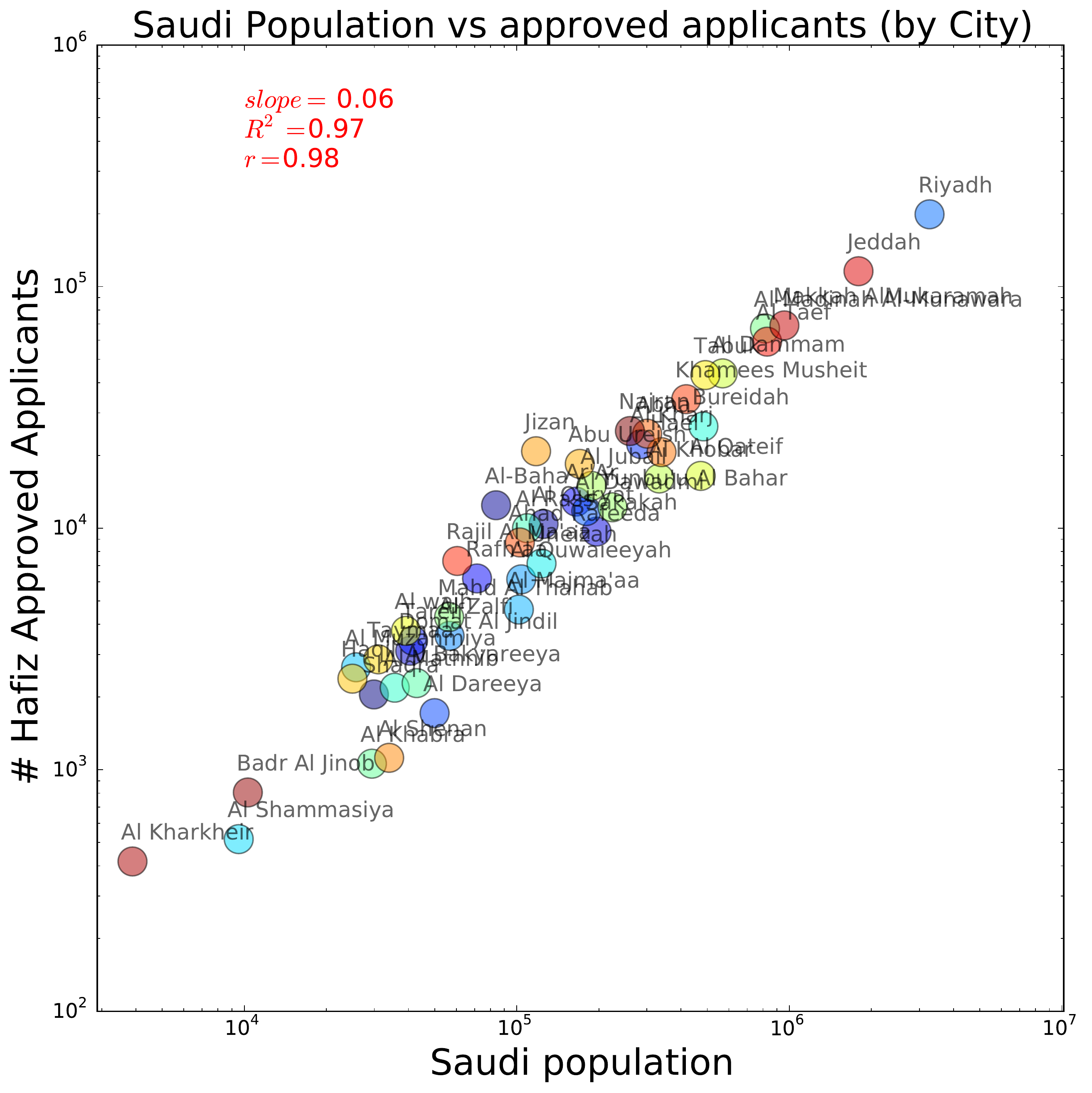}
	\caption{scaling relationships between cities' population versus number of Hafiz eligible applicants. Best-fit scaling relations are shown as dashed line.}
	\label{fig:hafiz_scaling_city}
\end{figure}

\begin{figure}[h!]
	\centering
	\includegraphics[width=0.8\columnwidth]{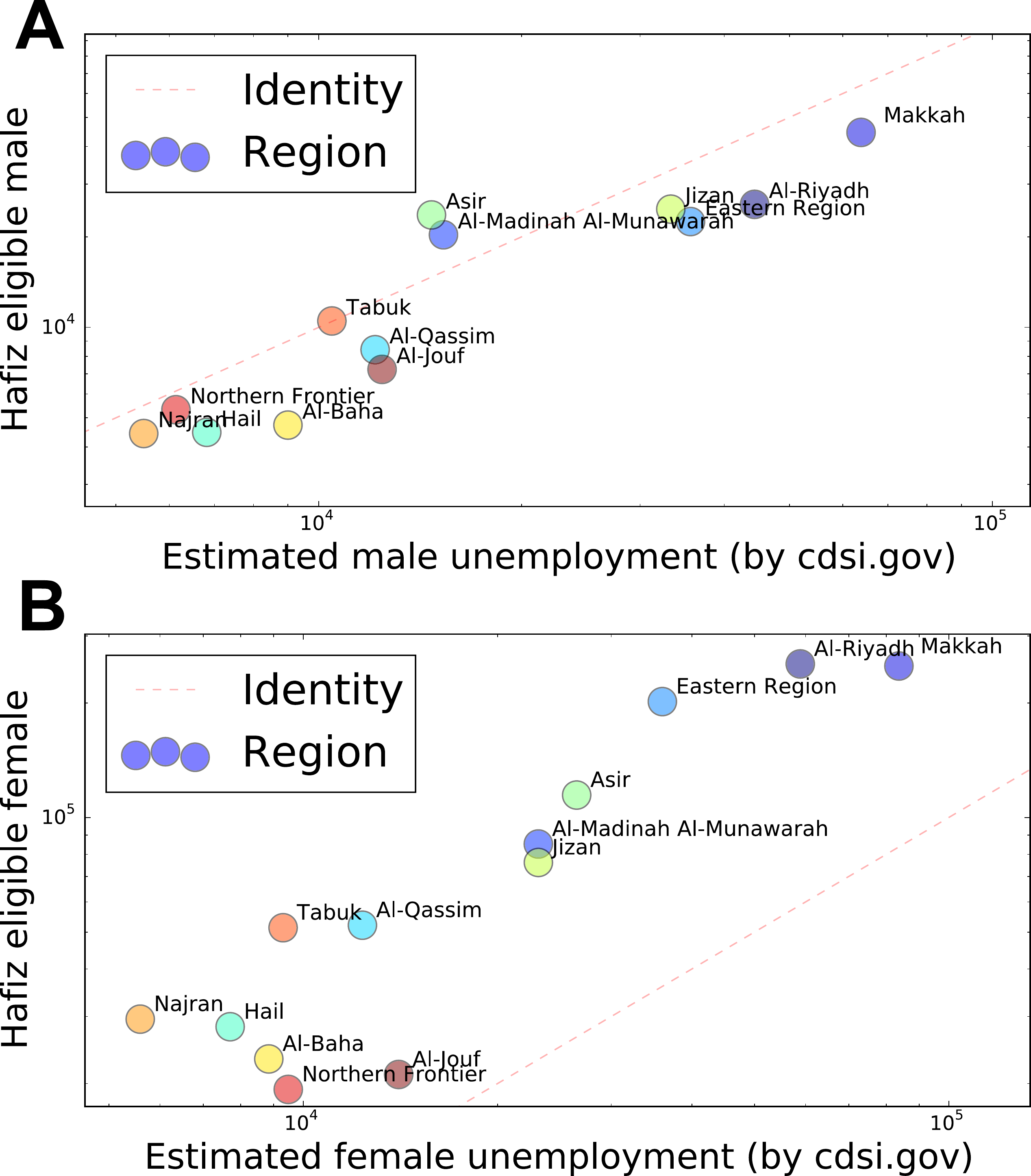}
	\caption{Panel~\textbf{(A)} shows the relationship between the estimated male unemployment as reported by the Central Department of Statistics \& Information (CDSI) versus the number of male unemployment benefit precipitants at the regional level. On the other hand, Panel~\textbf{B} shows the same relationship for the female population. We observe a significant bias in estimating the female unemployment by the CDSI (i.e., underestimating unemployment)}
	\label{fig:hafiz_bias}
\end{figure}

For this study, we used anonymized mobile phone meta data known as Call Detail Records (CDRs), combined with records from an unemployment benefit program (Hafiz) and census information (see Appendix for additional details). Basic statistics of the age distribution, years of work experience, and qualifications are presented in Figure~\ref{fig:basic_hafiz_statistics}. Figure~\ref{fig:hafiz_scaling} shows the linear scaling between the population at the regional level and number of Hafiz recipients. In addition, Figure~\ref{fig:hafiz_scaling_city} illustrates the linear relationship between the number of Hafiz recipients in each city versus the population estimated by the CSDSI. These results suggest the high quality of the data. However, in Figure~\ref{fig:hafiz_bias} we compare the number of Hafiz recipients at a regional level to the unemployment levels as estimated by the Central Department of Statistics (CDS). We observe systematic bias in the census unemployment estimates of the number of female Hafiz recipients. This raises serious questions regarding the declining accuracy and increased bias in unemployment numbers as obtained from surveys.

\subsubsection{Mapping Census Population to Districts}
For the $i^{th}$ TAZ donated by $\tau_i$ we have the population $P_{\tau_i}$ and demographic breakdown (i.e., gender and nationality), as well as housing data (i.e., number of houses, villas, apartments etc.) provided by the ADA. However, the finest resolution for the unemployment data is at the district level. Therefore, for each district $d_i$ we estimate the population $P_{d_i}$ as follows:

$$
P_{d_i} = \sum_{j=0}^{|\tau|} \frac{A_{(d_i \cap \tau_j)} P_{\tau_j}} {A_{\tau_j}}
$$

where $|\tau|$ is the total number of TAZ units, $A_{\tau_j}$ is the area of the $j^{th}$ TAZ unit and $A_{(d_i \cap \tau_j)}$ is the intersection area of $d_i$ and $\tau_j$.

\subsubsection{Mapping Mobile Population to Districts}
For each cell tower $c_j$, we know the total number of different users $T_{c_j}$ with home location (i.e., the tower where a user spends most of the time at night; as in~\cite{Santi2012}) being the $j^{th}$ tower. When one makes a phone call, the network usually identifies nearby towers and connects to the closest one. The coverage area of a tower $c_j$ thus was approximated using a Voronoi-like tessellation. The Voronoi cell associated with tower $c_j$ is denoted by $v_j$. Therefor, we can compute the penetration rate $\sigma_{d_i}$ for district $i$ as follows:

$$
\sigma_{d_i} = \frac{1}{P_{d_i}} \sum_{j=0}^{|v|} \frac{A_{(d_i \cap v_j)} T_{c_j}} {A_{v_j}}
$$

where $|v|$ is the total number of Voronoi cells, $A_{v_j}$ is the area of the $j^{th}$ Voronoi cell (associated with the $j^{th}$ cell tower) and $A_{(d_i \cap v_j)}$ is the intersection area of $d_i$ and $v_j$.

\begin{figure}[h!]
	\centering
	\includegraphics[width=1\columnwidth]{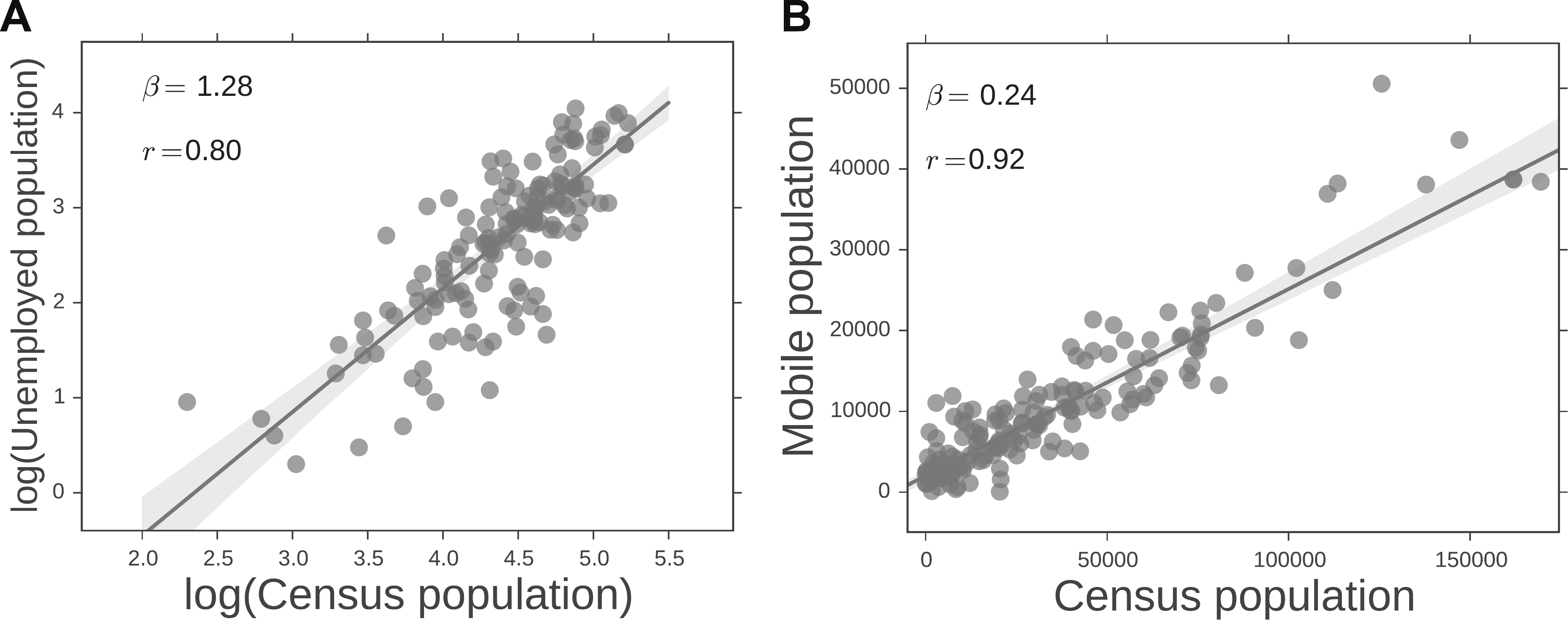}
	\caption{Panel \textbf{(A)} shows the relationship between unemployment and population in 2012, for the 148 districts. Panel \textbf{(B)} shows the number of homes detected, for the 148 districts versus district census population. Best-fit scaling relations are shown as solid lines and the 95\% confidence intervals are shown as a shaded area.}
	\label{fig:mapping_population}
\end{figure}

Figure~\ref{fig:mapping_population} shows the scaling relationships between the district population versus unemployment rate and also population versus mobile users. These results are consisant with previous studies indicating that scaling with population is indeed a pervasive property of urban organization~\cite{bettencourt2007growth,pan2013urban}.

\subsection{Extracting Behavioral Indicators}
\label{sec:indicators}
The goal of this chapter is to investigate how behavioral indicators from mobile phone meta data can be extracted and then related back to the economical wellbeing of geographical regions (i.e., districts). To this end, we define three groups of indicators that have been widely explored in fields like economy or social sciences. Several of these indicators have been implemented in the bandicoot toolbox~\cite{de2013predicting} and can be found at http://bandicoot. mit.edu/docs/. All the indicators are computed at the individual level and then aggregated and standardized (i.e., scaled to have mean zero and variance one) at the district level.

\subsubsection{Activity patterns}

Activity patterns quantify factors related to the aggregate patterns of mobile usage for each district such as volume (average number of records per user.), timing (the average percentage of night calls -- nights are 7pm-7am), and duration (average duration of calls).

\subsubsection{Social interactions}

The social interaction indicators capture the structure of the individual's contact network. We focus on the egocentric networks around the indivdiaul in order to examine the local structure and signify the types of interactions that develop within their circle~\cite{almaatouq2014twitter}. 

Let $E_i = (\mathcal{C}_i,\mathcal{E}_i)$ be the directed egocentric-graph that represents the topological structure of the $i^{th}$ individual where $\mathcal{C}_i$ is the set of contacts (total number of contacts is $k = \vert \mathcal{C}_i\vert $) and $\mathcal{E}_i$ is the set of edges. A directed edge is an ordered pair $(i,j)$ with associated volume $w_{ij}$ (and/or $(j,i)$  with $w_{ji}$ volume) representing the interaction between the ego $i$ and a contact $ j \in \mathcal{C}_i$. Note that by definition $w_{ij} + w_{ji} \in \mathbb{Z}^+$, must be satisfied, otherwise $j \notin \mathcal{C}_i$. Therefore, the volume $w_{ij}$ is set to $0$ when $(i,j) \notin \mathcal{E}_i$, alternatively $w_{ji} = 0$, if the $(i,i)$ direction does not exist. From this we can compute several indicators for an individual within its egocentric network. We define $I(i)$ and $O(i)$ as the set of incoming interactions to (respectively, initiating from) individual $i$. That is,  
$$I(i) = \{j \in \mathcal{C}_i \vert (j,i) \in \mathcal{E}_i \}, \quad \text{and} \quad  O(i) = \{j \in \mathcal{C}_i \vert (i,j) \in \mathcal{E}_i \} \text{.}$$

\paragraph{Percentage of initiated interaction} is a measure of directionality of communication. We define it as $$\mathcal{I}(i) = \vert I(i)\vert / \left(\vert I(i)\vert +\vert O(i)\vert\right)$$ 

\paragraph{Balance of contacts} is measured through the balance of interactions per contact. For an individual $i$, the balance of interactions is the number of outgoing interactions divided by the total number of interactions.

\begin{equation*}
\beta(i) = \frac{1}{k} \sum_{j \in I(i) \cup O(i)} \frac{w_{ij}}{w_{ji} + w_{ij}}
\end{equation*}

\paragraph{Social Entropy} captures the social diversity of communication ties within an individual's social network, we follow Eagle's et al.~\cite{eagle2010network}   approach by defining social entropy, $D_{social}(i)$, as the normalized Shannon entropy associated with the $i^{th}$ individual communication behavior:

\begin{equation*}
D_{social}(i) = -\sum_{j\in \mathcal{C}_i}log(p_{ij})/{log(k)}
\end{equation*}

Where $p_{ij} = w_{ij}/{\sum_{\ell=1}^k w_{i\ell}}$ is the
proportion of $i$'s total call volume that involves $j$. High diversity scores imply that an individual splits his/her time more evenly among social ties.

\subsubsection{Spatial markers}
The spatial markers captures mobility patterns and migration based on geospatial markers in the data. In this chapter, we measure the number of visited locations, which captures the frequency of return to previously visited locations over time~\cite{song2010modelling} (time in our case is the entire observational period). We also compute the percentage of time the user was found at home.


\subsubsection{Self-Organizing Maps}
\label{som}
We use the standard and well understood form of self-organizing maps (SOMs) as an unsupervised exploratory analysis tool. SOMs use a regular grid of ``units'' (unites are sometimes referred to as nodes) onto which objects are mapped. The SOMs does not try to preserve the distances directly but rather the relations or local structure of the input data. As such, objects that are similar are mapped to close units in the grid topology. As in the classical implementation, one starts by assigning a codebook vector (i.e., weight vector of the same dimension as the input data) to every unit, that will play the role of a typical pattern, a prototype, associated with that unit. Proceeding the training phase, different \textit{classes} of units become \textit{specialized} in different subsets of the data. Each district from the data space is then mapped to the unit with the closest (smallest distance metric) codebook vector. The mathematical details based on Kohonen~\cite{kohonen1998self},which assumes that the following process converges and produces the wanted model:

$$
m_i(t+1) = m_i(t) + h_{ci}(t) \left[ x(t) - m_i(t) \right]
$$ 
 where $\{x(t)\}$ is a sequence the input data (of dimension $m$) representing the districts, $t$ is an integer that signifies a step in the sequence, and ${m_i(t)}$ is
 another sequence of m-dimensional real vectors that represent
 successively computed approximations of model $m_i$. Here $i$ is the spatial index of the grid node with which $m_i$ is associated and $h_{ci}(t)$ resembles the kernel that is applied in usual smoothing processes is called the neighborhood function~\cite{kohonen2013essentials}. The implementation can be found at the \textit{Kohonen} R package~\cite{wehrens2007self}.

\begin{figure}[h!]
	\centering
	\includegraphics[width=1\columnwidth]{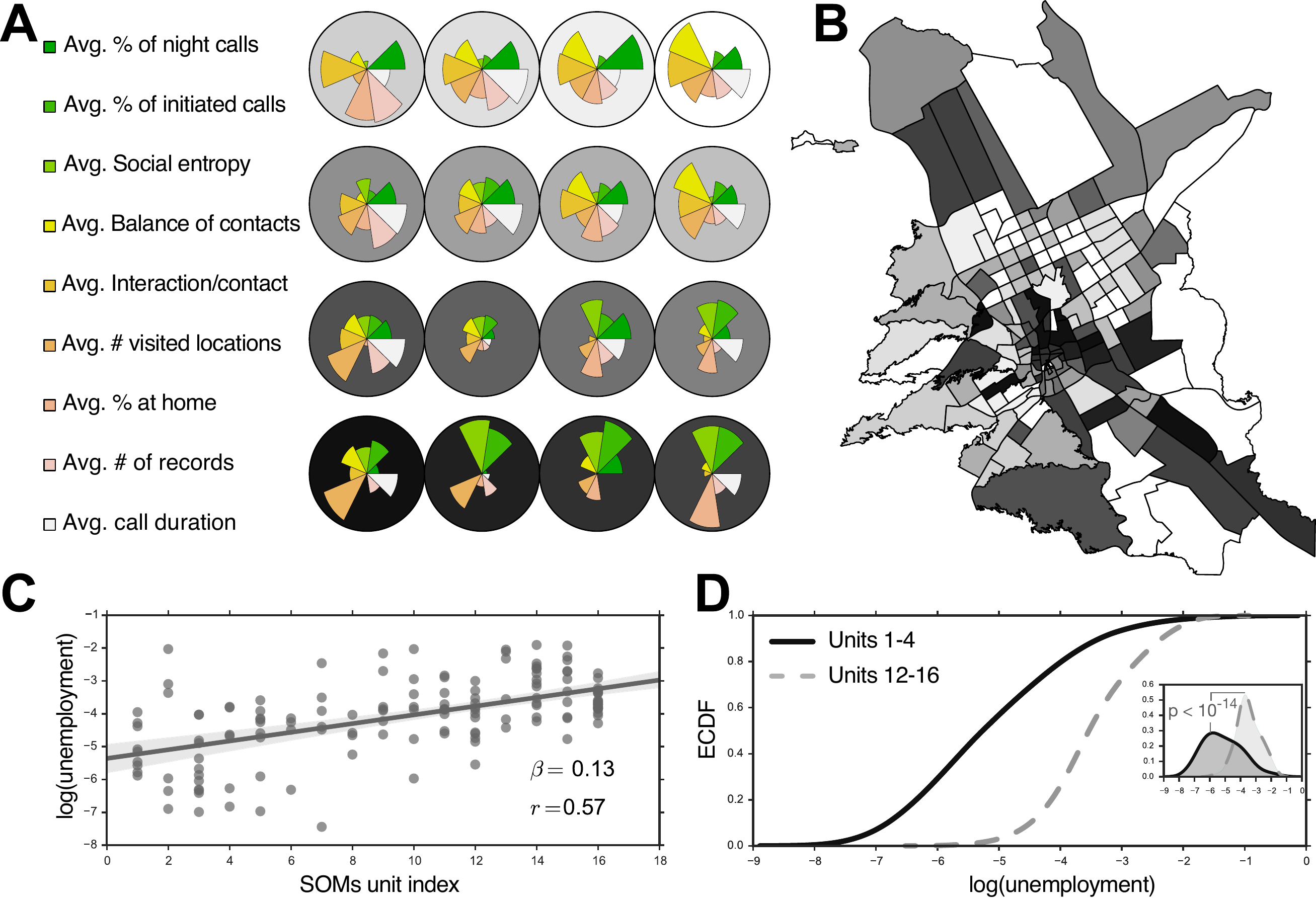}
	\caption{Panel~\textbf{(A)} shows the plot of the color-encoded codebook vectors of the 4-by-4 mapping of the districts behaviors. Panel~\textbf{(B)} shows the combined view of attribute space (i.e., SOMs results) and geographic space (choropleth map). Panel \textbf{(C)} demonstrates the relationship between the clustered behaviors and unemployment rates. Finally, panel \textbf{(D)} shows that the Empirical Cumulative Distribution Function (ECDF) of the unemployment rates for contrasting groups behaviors.}
	\label{fig:som}
\end{figure}

In Figure~\ref{fig:som}A, the codebook vectors from the resulting SOMs are shown in a segments plot, where the gray-scale background color of a unit corresponds to the unit index. Districts having similar characteristics based on the multivariate behavioral attributes are positioned close to each other, and the distance between them represents the degree of behavioral similarity or dissimilarity. High average spatial entropy with small percentage of time being at home, for example, is associated with districts projected in the bottom left corner of the map (i.e., unit index one---black color).  On the other hand, districts with low social entropy, percentage of initiated calls, and balance of contacts are associated with the units at the top column of the map. On the geographic map (see Figure~\ref{fig:som}B), each district is assigned a color, where the meaning of the color can be interpreted from the corresponding codebook vector.
We can see that at the center of the city, most districts are assigned to units with dark backgrounds, and the color gets lighter as we move towards the periphery of the city. As expected from the description of the corresponding codebook vectors, districts projected in the bottom of the map (dark color background) are associated with lower unemployment rates (see Figure~\ref{fig:som}C). It is indeed the case that districts with similar behavioral attributes have similar unemployment rates (see Figure~\ref{fig:som}.D).

\subsubsection{Statistical Correlation}
\label{sec:correlation}

\begin{figure}[h!]
	\centering
	\includegraphics[width=1\columnwidth]{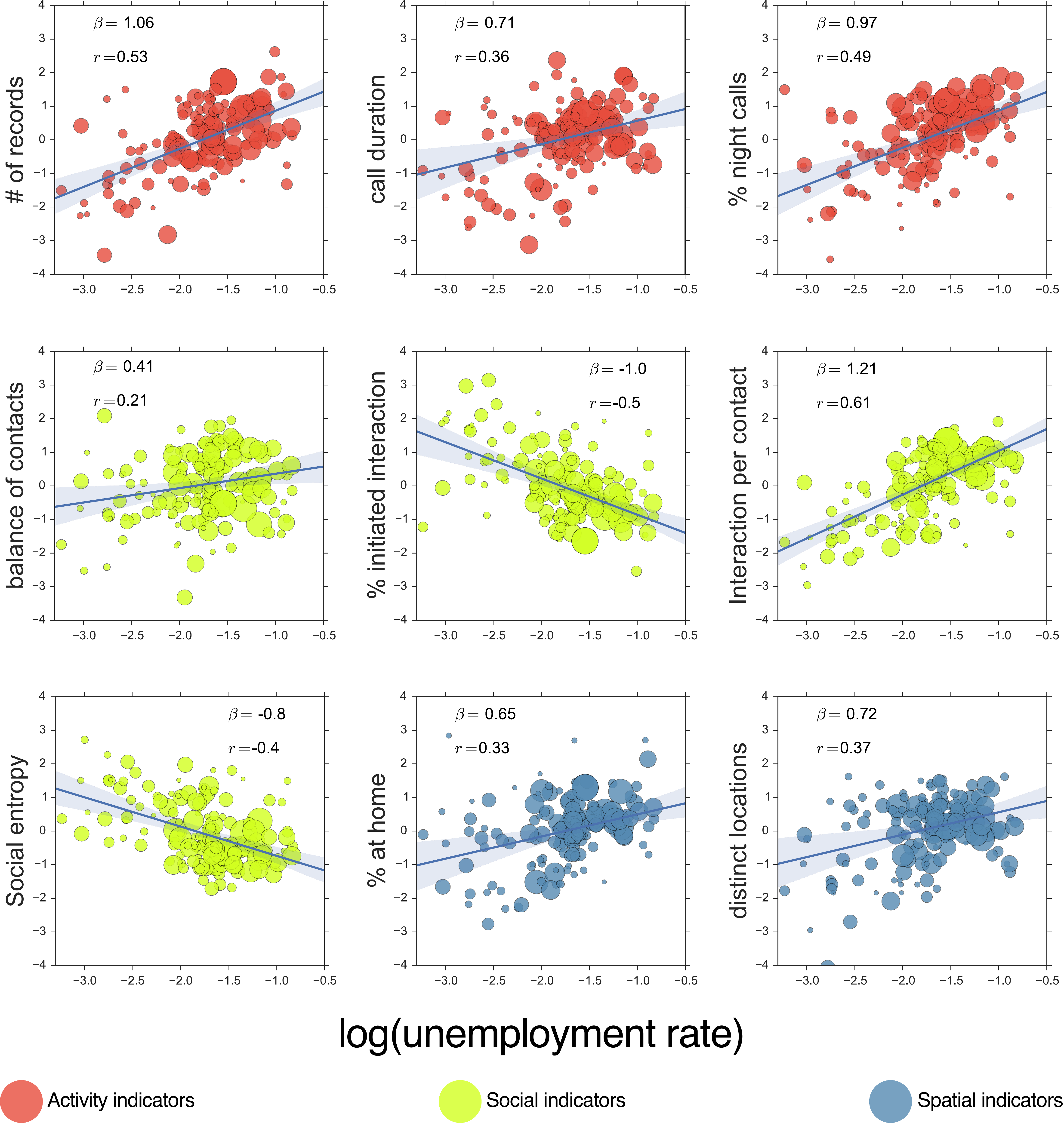}
	\caption{Relations between the mobile extracted indicators (district average) for the 148 districts against its unemployment rate. Size of the
		points is proportional to the population in each district. Solid lines correspond to linear fits to the data and the shaded area represents the 95\% confidence intervals.}
	\label{fig:correlations}
\end{figure}

As we can see in Figure~\ref{fig:correlations}, all the extracted indicators exhibit at least moderate statistical correlations with unemployment. In addition, we find that the indicators relationship with unemployment persists for most indicators even when we include controls for a district's area, population, and mobile penetration rate (see Table~\ref{tab:4}). These results suggest that several of those indicators are sufficient to explain the observed unemployment.
\begin{table}[h!]
	\begin{center}
		\scriptsize
		\begin{tabular}{l D{.}{.}{3.5}@{} D{.}{.}{3.4}@{} D{.}{.}{3.5}@{} D{.}{.}{3.5}@{} D{.}{.}{3.5}@{} D{.}{.}{3.4}@{} D{.}{.}{3.5}@{} D{.}{.}{3.4}@{} D{.}{.}{3.4}@{} }
			\toprule
			& \multicolumn{9}{c}{Dependent variable: log(Unemployment \%)}\\
			\toprule
			& \multicolumn{1}{c}{(1)} & \multicolumn{1}{c}{(2)} & \multicolumn{1}{c}{ (3)} & \multicolumn{1}{c}{(4)} & \multicolumn{1}{c}{(5)} & \multicolumn{1}{c}{(6)} & \multicolumn{1}{c}{(7)} & \multicolumn{1}{c}{(8)} & \multicolumn{1}{c}{(9)} \\

			\midrule
			Population                & 0.15       & 0.16      & 0.12        & -0.07      & 0.09       & 0.12       & 0.12       & 0.16      & 0.16      \\
			& (0.10)     & (0.11)    & (0.11)      & (0.11)     & (0.11)     & (0.11)     & (0.11)     & (0.11)    & (0.11)    \\
			Area                      & 0.07       & 0.15^{*}  & 0.12        & 0.28^{***} & 0.14^{*}   & 0.17^{*}   & 0.11       & 0.20^{**} & 0.19^{**} \\
			& (0.09)     & (0.09)    & (0.08)      & (0.08)     & (0.08)     & (0.08)     & (0.09)     & (0.08)    & (0.08)    \\
			Penetration rate & 0.02       & 0.00      & 0.04        & 0.04       & -0.06      & 0.02       & 0.05       & -0.02     & 0.02      \\
			& (0.11)     & (0.11)    & (0.11)      & (0.10)     & (0.10)     & (0.11)     & (0.11)     & (0.11)    & (0.11)    \\
			\# of records              & 0.31^{***} &           &             &            &            &            &            &           &           \\
			& (0.09)     &           &             &            &            &            &            &           &           \\
			Call duration             &            & 0.17^{**} &             &            &            &            &            &           &           \\
			&            & (0.08)    &             &            &            &            &            &           &           \\
			\% initiated inter.  &            &           & -0.28^{***} &            &            &            &            &           &           \\
			&            &           & (0.09)      &            &            &            &            &           &           \\
			\% night calls            &            &           &             & 0.49^{***} &            &            &            &           &           \\
			&            &           &             & (0.09)     &            &            &            &           &           \\
			\% at home                &            &           &             &            & 0.30^{***} &            &            &           &           \\
			&            &           &             &            & (0.08)     &            &            &           &           \\
			Social entropy            &            &           &             &            &            & -0.18^{**} &            &           &           \\
			&            &           &             &            &            & (0.09)     &            &           &           \\
			Inter. per contact   &            &           &             &            &            &            & 0.27^{***} &           &           \\
			&            &           &             &            &            &            & (0.09)     &           &           \\
			Balance of contacts       &            &           &             &            &            &            &            & -0.01     &           \\
			&            &           &             &            &            &            &            & (0.08)    &           \\
			visited locations        &            &           &             &            &            &            &            &           & 0.14^{*}  \\
			&            &           &             &            &            &            &            &           & (0.08)    \\
			(Intercept)                 & -0.00      & 0.00      & -0.00       & 0.00       & -0.00      & -0.00      & -0.00      & -0.00     & -0.00     \\
			& (0.08)     & (0.08)    & (0.08)      & (0.07)     & (0.08)     & (0.08)     & (0.08)     & (0.08)    & (0.08)    \\
			\midrule
			R$^2$                     & 0.15       & 0.10      & 0.14        & 0.23       & 0.16       & 0.10       & 0.13       & 0.07      & 0.09      \\
			Adj. R$^2$                & 0.13       & 0.08      & 0.12        & 0.21       & 0.13       & 0.07       & 0.11       & 0.05      & 0.07      \\
			BIC                       & 421.97     & 430.53    & 424.03      & 407.25     & 420.75     & 430.65     & 425.16     & 434.73    & 431.86    \\
			Num. obs.                 & 147        & 147       & 147         & 147        & 147        & 147        & 147        & 147       & 147       \\
			\bottomrule
			\multicolumn{10}{l}{\scriptsize{$^{***}p<0.01$, $^{**}p<0.05$, $^*p<0.1$. Standard errors in parentheses.}}
		\end{tabular}
		\vspace{2mm}
		\caption{Regression table explaining the districts' unemployment rate as a function of the activity patterns, social interactions, and spatial markers, with the inclusions of controls for a district's area, population, and mobile penetration rate.}
		\label{tab:4}
	\end{center}
\end{table}
For instance, we find that the percentage of night calls to have the highest effect size and explanatory power ($R^2 = 23\%$; model 4). This is expected, as regions with very different unemployment patterns should exhibit different temporal activities. Since working activities usually happen during the day, we would expect that districts with high levels of unemployment will tend to have higher proportion of their activities during the night.

Previous study~\cite{schneider2013unravelling} have found that the duration spent at either home or work is relatively flat distributed with peaks around time spans of 14 hours at home and 3.5-8.6 hours at work. Therefore, we hypothesize that the lack of having a work location for the unemployed would lead to an increase in the duration spent at home (i.e., \% home), and/or reduce the tendency for revisiting locations (i.e., higher visited locations). We indeed find that the percentage of being home and number of visited locations to be associated with unemployment in our dataset.

We also find the percentage of initiated interactions to be negatively correlated with unemployment. This indicators has been shown to be predictive of the Openness (i.e, the tendency to be intellectually curious, creative, and open to feelings) personality trait~\cite{john1999big,de2013predicting}, which in return is predictive of success in job interviews~\cite{caldwell1998personality}.

As in~\cite{eagle2010network,llorente2014social}, we find that districts with high unemployment rates have less diverse communication patterns than areas with low unemployment. This translates in a  negative coefficient for social entropy and positive coefficient for the interaction per contact indicator. The balance of contacts factor was not found to be significant ($p>0.1$).

\subsection{Predictive Model}

Here we are interested in the predictability of unemployment rates of microregions based on the mobile phone extracted indicators and independently of additional census information such as population, gender, income distribution, etc.
Such additional information is often  unavailable in developing nations, which by itself represents a major challenge to policy-makers and researchers. Therefore, it is of utmost importance to find novel sources of data that enables new approaches to demographic profiling.

We analyze the predictive power of the indicators using Gaussian Processes (GP) to predict unemployment based on mobile phone indicators solely. We train and test the model in K-fold-cross validation ($K=5$) and compute the coefficient of determination $R^2$ as a measure of quality for each category of indicators (i.e., activity, social, and spatial) and also for the full indicators (involving all mobile extracted indicators presented in this chapter).

\begin{figure}[h!]
	\centering
	\includegraphics[width=1\columnwidth]{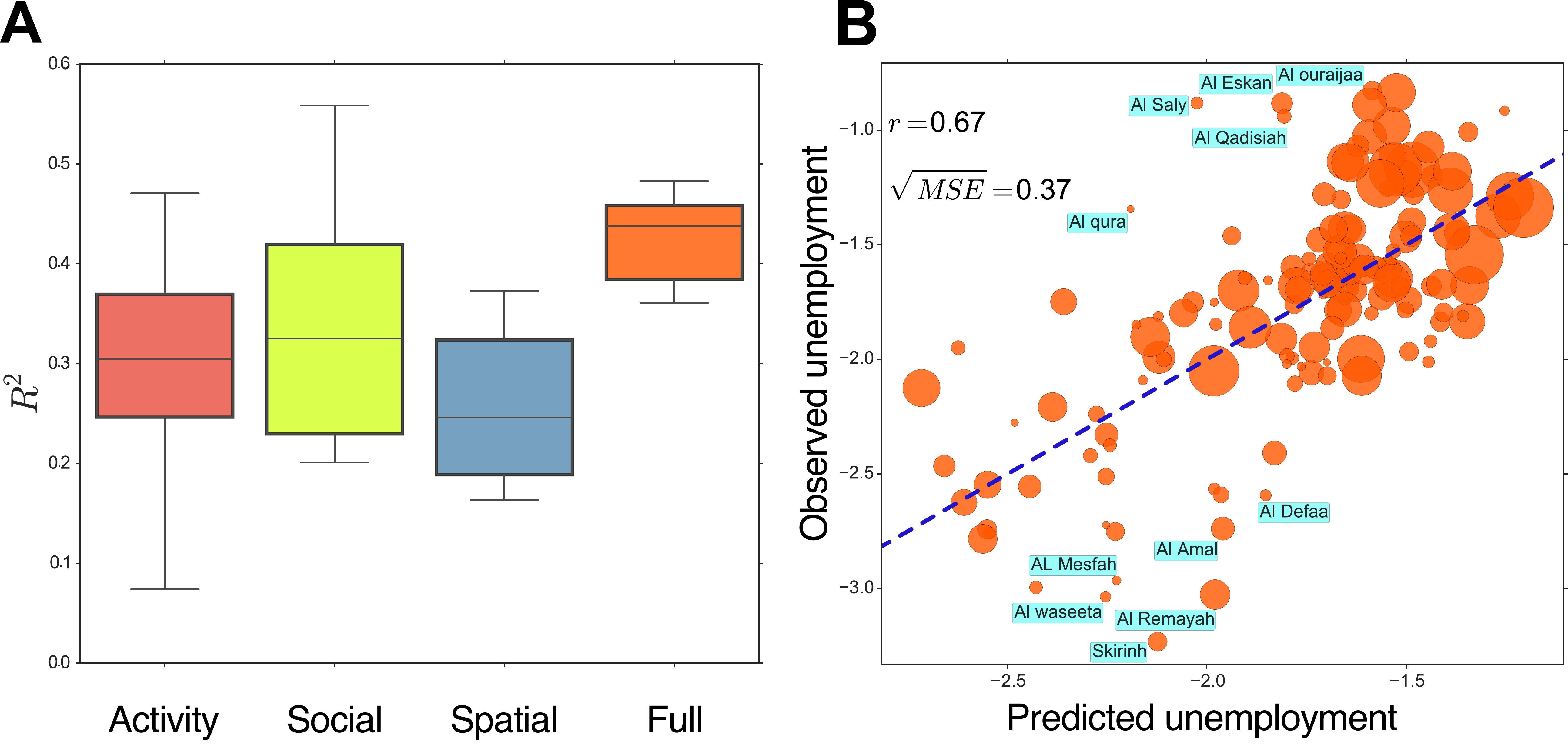}
	\caption{Panel \textbf{(A)} shows the average performance of each indicator category in predicting unemployment. Panel \textbf{(B)} depicts the cross-validated prediction of unemployment rates versus the observed ones, $r= 0.68$. The predicted values are based on the prediction that was obtained for that district when it was in the test set. Dashed line correspond to the equality line.}
	\label{fig:prediction}
\end{figure}

\subsubsection{Mathematical Formulation}
The advantage for using Gaussian Processes (GP) to regress unemployment rates is that the model produces probabilistic (Gaussian) predictions so that one can compute empirical confidence intervals and probabilities that might be used to refit (online fitting, adaptive fitting) the prediction in some region of interest.

The GP works as follows: Given $\mathbf{X}$ an $N \times M$ matrix, each data sample (or a district) corresponds to a vector $\mathbf{x}_n \in R^M, n = 1, ... , N$, where $M$ is the number of district features. Now we want to learn a model $g(X)$ such that:
\begin{align*}
g: & \mathbb{R}^{M} \rightarrow \mathbb{R} \\
& X \mapsto y = g(X)
\end{align*}

The GP model starts by assuming that this function is a conditional sample path of a Gaussian process $G$ which reads as:
$$
G(X) = f(X)^T \beta + P(X)
$$
where $f(X)^T \beta$ is a linear regression model, and $P(X)$ is a 0 mean Gaussian process with a fully stationary covariance function. From this formulation, GP is nothing but an extension of a basic least squares linear regression problem (the addition of $P(X)$). By deriving the best unbiased linear prediction of the sample path $g$ conditioned on the observation, we need to solve the following constrained optimization problem:

\begin{align*}
a(X)^* = \arg \min\limits_{a(X)} & \; \mathbb{E}[(G(X) - a(X)^T y)^2] & \text{[best-fit in the Mean Squared Error sense
	]} \\
{\rm s. t.} & \; \mathbb{E}[G(X) - a(X)^T y] = 0 & \text{[It is unbaised]}
\end{align*}

This optimization problem has a closed form expression solution that can be obtained from the Lagrangian form and satisfying the first order optimality conditions. The complete proof can be found in~\cite{welch1992screening} and implementation in~\cite{nielsen2002dace}. In the end, the solution is shown to be a Gaussian random variate with mean:
$$
\mu_{\hat{Y}}(X) = f(X)^T\,\hat{\beta} + r(X)^T\,\gamma
$$
and variance:
$$
\sigma_{\hat{Y}}^2(X) = \sigma_{Y}^2\,
( 1
- r(X)^T\,R^{-1}\,r(X)
+ u(X)^T\,(F^T\,R^{-1}\,F)^{-1}\,u(X)
)
$$
where $R$ is the correlation matrix, $r$ is the vector of cross-correlations between the point where the prediction, $F$, is the regression matrix (i.e., the Vandermonde matrix), and $u(X) = F^T\,R^{-1}\,r(X) - f(X)$.

\subsubsection{Prediction Results}
In Figure~\ref{fig:prediction}A we find that the social interaction indicators to be very predictive of unemployment with an average $R^2 = 0.43$ (95\% CI: 0.37 -- 0.48), which is more predictive than the activity pattern indicators $R^2 = 0.29$ (95\% CI: 0.15 -- 0.39) and the spatial indicators $R^2 = 0.26$ (95\% CI:0.19 -- 0.33). It is worth mentioning, that the composite model performed significantly better than single category models with an average $R^2= 0.43$ (95\% CI: 0.37 -- 0.48). Figure~\ref{fig:prediction}.B compares the predicted and observed unemployment rate for each based on the prediction that was obtained for that district when it was in the test set. 
\section{The Role of Social Cohesion in Effectiveness of Economic Incentives}
\label{reciprocity}
We investigate the role of social cohesion in increasing the effectiveness of incentives. Incentives are of utmost important in promoting behaviors that are likely to gain job seekers employment and succeed in their chosen occupations. This benefits not only the job seekers themselves, but also the workforce, the community and the economy. In particular, we present evidence that social cohesion---defined as the density of social ties---can amplify the effect of incentives through the mechanism of social pressure. This work has been published in~\cite{almaatouq2016b}.

Traditional policies rely on Pigouvian taxation or subsidies as economic incentives that change individuals' behaviors by internalizing the externalities they produce (e.g., reducing energy consumption, improving the work environment, motivating the unemployed to find jobs, cleaning up the neighborhood, etc). Ideally, economic incentives would lead to long-lasting behavioral change and allow us to eradicate some of the most important problems in modern society, such as pollution, global warming, and rising health care and insurance costs ~\cite{mani2013inducing}. However, changing the behavior of self-interested individuals is extremely difficult~\cite{dietz2003struggle}. A fundamental assumption of Pigouvian taxation or subsidies is that the individual is the correct unit of incentives that promote changes in human behavior. We present evidence that this assumption may have limited utility, that the social networks containing the individuals are an important additional unit for incentives, and that these ''network incentives`` can amplify the effect of incentives on behavioral change~\cite{mani2013inducing,pentland2007collective}. These findings support the hypothesis that social capital, defined as the density of social ties---particularly reciprocal ties---is what increases the effect of economic incentives. It accomplishes this through the mechanism of social pressure.

In this section, we investigate whether \textit{persuasion} and \textit{peer pressure} can amplify the effect of economic incentives. Persuasion is at the core of norm creation, emergence of collective action, and solutions to `tragedy of the commons' problems.
Here, we show that the directionality of friendship ties affects the extent to which individuals can influence each others' behavior.
Moreover, we find that people are typically poor at perceiving the directionality of their friendship ties and this can significantly limit their ability to engage in cooperative arrangements. Through spreading process simulations, we find that although unilateral ties diffuse behaviors across communities, reciprocal ties play a more important role in the early stages of the diffusion process.
This may lead to individuals within the network failing to establish compatible norms, act together, find compromise solutions, and persuade others to act.
We then suggest strategies to overcome this limitation by using two topological characteristics of the perceived friendship network.
The findings of this paper have significant consequences for designing interventions that seek to harness social influence for collective action. 

Studies of collective action have shown that social influence can greatly facilitate mobilization around collective goods, predict success of the action, and guide its design~\cite{margetts2012social}.
For instance, social influence is found to be a key factor in political uprisings~\cite{gray2000declining,marwell1993critical}, strike movements~\cite{marwell1993critical}, and other types of participation~\cite{cialdini2004social}.
Social influence in networks is also a well-recognized key factor in the diffusion of new behaviors~\cite{centola2010spread,centola2011experimental,christakis2007spread,kim2015social}, new ideas~\cite{kempe2003maximizing} and new products~\cite{aral2012identifying,aral2011creating} in society.

Moreover, in recent years, peer-support programs are emerging as highly effective and empowering ways to leverage peer influence to support behavioral change of people~\cite{ford2013systematic}.
One specific type of peer-support programs is the ``buddy system,'' in which individuals are paired with another person (i.e., a buddy) with the responsibility to support their attempt to change their behavior.
Such a system has been used to shape people's behavior in various domains including smoking cessation~\cite{may2000social}, weight loss~\cite{stock2007healthy}, diabetes management~\cite{rotheram2012diabetes} or alcohol misuse. 

Consequently, the need to understand the factors that impact the level of influence individuals exert on one another is of great practical importance.
Recent studies have investigated how the effectiveness of peer influence is affected by different social and structural network properties, such as clustering of ties~\cite{centola2010spread}, similarity between social contacts~\cite{centola2011experimental}, and the strength of ties~\cite{aral2013tie}. However, how the effectiveness of social influence is affected by the reciprocity and directionality of friendship ties is still poorly understood.

Individuals commonly assume their affective relationships to be reciprocal by default~\cite{laursen1993adolescents,hartup1993adolescents}.
For instance, when one considers another individual as ``friend'', the common expectation is that this other individual also thinks of them as friends.
Moreover, reciprocity is implicitly assumed in many scientific studies of friendship networks by, for example, marking two individuals as being friends or not being friends (e.g.,~\cite{newcomb1995children,eagle2006reality,dong2012graph,dong2012modeling}).
Despite this common expectation, in reality not all friendships are reciprocal~\cite{Vaquera2008}.
When analyzing self-reported relationship surveys from several experiments, we find that the vast majority of friendships are expected to be reciprocal, while in reality, only about half of them are indeed reciprocal.
These findings suggest a profound inability of people to perceive friendship reciprocity, perhaps because the possibility of non-reciprocal friendship challenges one's self-image. 

We further show that the asymmetry in friendship relationships has a large effect on the ability of an individual to persuade others to change their behavior. 
Moreover, we show that the effect of directionality is larger than the effect of the self-reported strength of a friendship tie~\cite{aral2013tie} and thus of the implied `social capital' of a relationship. 
Our experimental evidence comes through analysis of a fitness and physical activity intervention, in which subjects were exposed to different peer pressure mechanisms, and physical activity information was collected passively by smartphones.
In this experiment, we find that effective behavioral change occurs when subjects share reciprocal ties, or when a unilateral friendship tie exists from the person applying the peer pressure to the subject receiving the pressure, but not when the friendship tie is from the subject to the person applying peer pressure.

Our findings suggest that this misperception of friendships' character for the majority of people may result in misallocation of efforts when trying to promote a behavioral change.
For instance, in the smoking cessation example mentioned above, our results suggest that the reciprocity and directionality of the friendship relationship between the smoker and the buddy can greatly affect the success of the intervention, and therefore the intervention designers cannot rely on how the smoker perceive his/her relationships with the buddies.

To overcome this limitation, we show that two topological features of the perceived friendship network---social embeddedness and social centrality---can alone effectively identify the most likely targets for effective behavioral change.
As a consequence, people seeking to shape the behavior of others could become more effective by relying on these features rather than on the perceived character of the relationship itself.
Revising our smoking cessation example, the intervention designers can leverage some topological features of the perceived friendship network of the smokers to select buddies that would induce more effective peer-pressure.

\subsection{Friendship Directionality}
We provide a quantitative assessment of people's expectation on the reciprocity of their friendship relationships through a self-reported survey that we collected among 84 students of an undergraduate course.
Similarly to other self-reported friendship surveys (see for example \cite{aharony2011social,madan2012sensing,de2014strength}), we asked each participant to score every other participant on a 0--5 scale, where 0 means ``I do not know this person'', 3 means ``Friend" and 5 means ``One of my best friends.''
In addition, participants were also asked to `predict' how other participants would score them. Participants in the experiment were early career (age 23-38) adults taking a university course in applied management. Gender balance was 40\% male and 60\% female. This age range and gender balance is similar to participants in the FunFit experiment~\cite{aharony2011social}. The study was approved by the Institutional Review Board (IRB) and conducted under strict protocol guidelines.

\begin{figure}
	\centering
	\includegraphics[width=1\columnwidth]{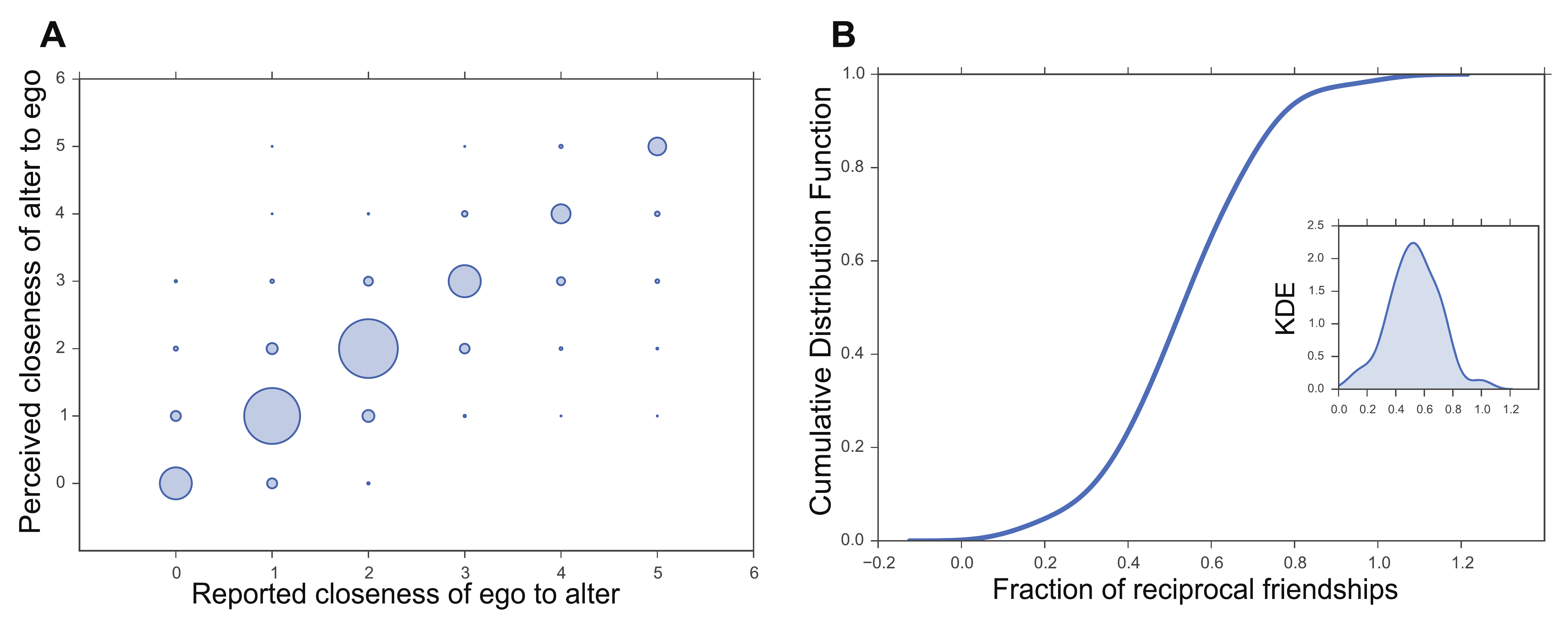}
	\caption{
		\textbf{(A)} The perceived closeness score of nominated alter to nominating ego vs. the reported closeness score of nominating ego to nominated alter. 
		The size of each circle represents the number of edges with the specific perceived and reported closeness scores.
		\textbf{(B)} Distribution (Estimated Cumulative Distribution Function and Kernel Density Estimate) of the fraction of reciprocal friendships at the single participant level.  }
	\label{fig1}
\end{figure}

Examining the relationship between how each subject scored the other subjects and his/her perception of how the other subjects would have scored him/her reveals a very strong and significant correlation ($r = 0.95, p = 0$). 
Fig.~\ref{fig1} shows this high expectation for reciprocity in friendship scores.
In fact, in 94\% (1273 out of 1353) of the cases in which a subject nominated another subject as a friend (i.e., closeness score $>2$), he/she also expected the other subject to nominate him/her back as a friend.

In contrast to the high expectations of reciprocity among the participants, we find that almost half of the friendships are actually non-reciprocal.
We show this by constructing a directional friendship network based on explicit friendship nominations (i.e., closeness scores $>2$).
In this network, we consider a friendship tie to be ``reciprocal'' when both participants identify each other as friends.
Alternatively, the tie is ``unilateral'' when only one of the participants identifies the other as a friend.
The final directed friendship network consists of 84 nodes (i.e., participants) and 775 edges (i.e., explicit friendships).
Examining the relationship between the reported friendship scores on the two sides of these edges reveals a relatively weak correlation ($r = 0.36, p = 0$).
Furthermore, only half (i.e., 53\%) of these edges are indeed reciprocal (413 out of 775).

In order to rule out the possibility that there were a few participants that nominated a large number of other participants as friends (high out-degree) and were not nominated back as friends by many others (low in-degree) and therefore skewed the global fraction of reciprocal friendships, we further calculated the fraction of reciprocal friendships at the single participant level. We find the distribution of these fractions for all participants to be normal and centered around 0.5 (see Fig.~\ref{fig1}). Therefore the overall result is not skewed by a few outlier individuals.

\begin{figure*}[h!]
	\centering
	\includegraphics[width=1\columnwidth]{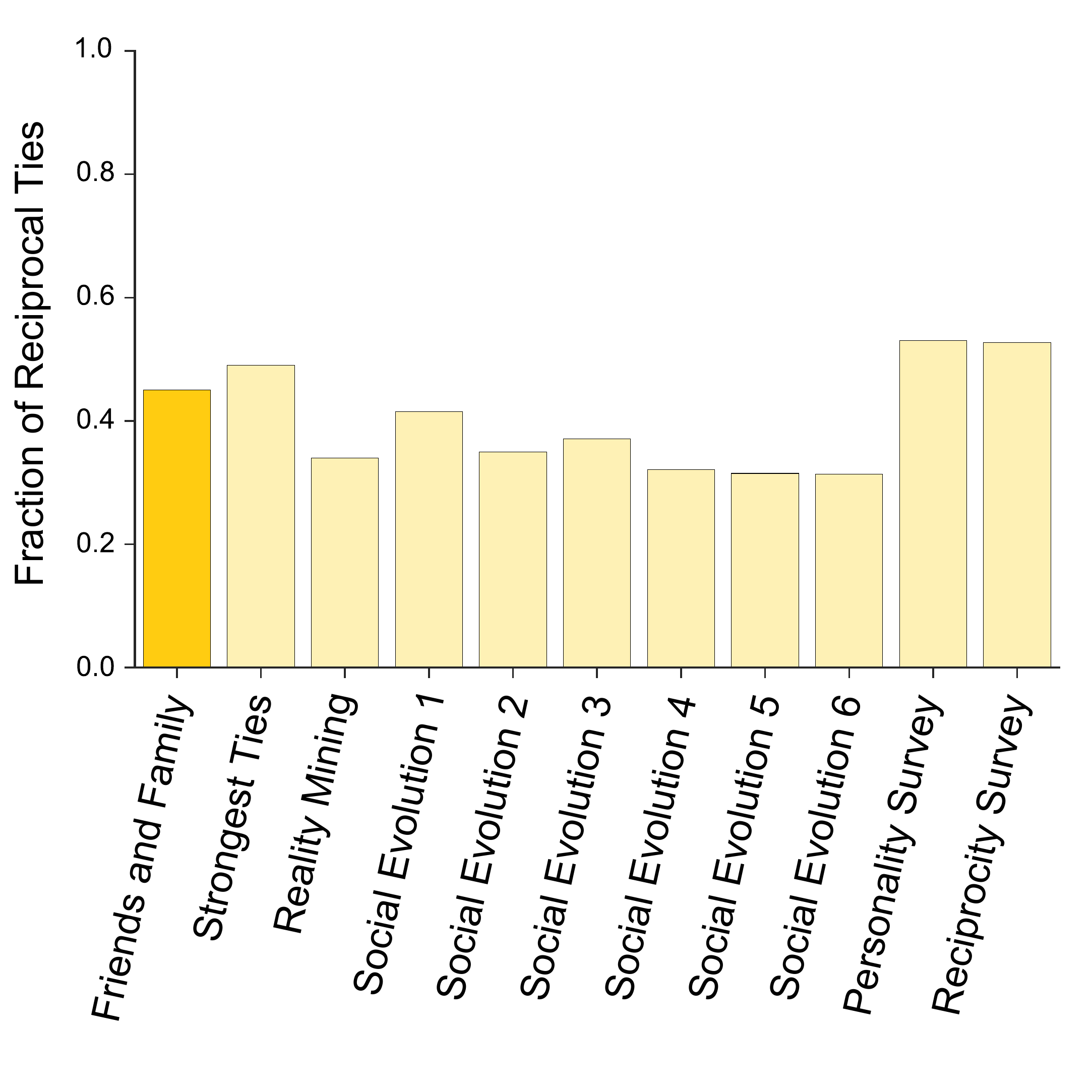}
	\caption{Fraction of reciprocal ties in additional datasets. The fraction of reciprocal ties for the Friends and Family study is reported (dark blue)
		next to the fraction observed in additional datasets (light blue): \textit{Reality Mining}, \textit{Strongest Ties}, \textit{Social Evolution} study,
		a \textit{Personality Survey}, and our \textit{Reciprocity Survey}.
		The \textit{Social Evolution} study is split into six temporal slices, and we report each of them as a separate dataset. The average percentage of reciprocal tie for the entire \textit{Social Evolution} study is 35\%.}
	\label{fig:additional_ds_reciprocal_percentage}
\end{figure*}

We find this result to be consistent across many self-reported friendship networks that we have analyzed: only 45\% (315 out of 698) of friendships are reciprocal in the Friends and Family dataset~\cite{aharony2011social}, 34\% (28 out of 82) in the Reality Mining dataset~\cite{eagle2006reality}, 35\% (555 out of 1596) in the Social Evolution dataset~\cite{madan2012sensing}, 49\% (102 out of 208) in the Strongest Ties dataset~\cite{de2014strength}, and 53\% (1683 out of 3160) in the Personality Survey. The first three surveys were collected at an American university, the fourth at a European university, and the latter at a Middle Eastern university. The participants of all the considered studies were asked to indicate their friendship relationships and the closeness with other participants through self-reported surveys. Fig.~\ref{fig:additional_ds_reciprocal_percentage} shows that for the additional datasets, as well as in the Friends and Family study, the percentage of reciprocal ties is below 55\%.
This supports our hypothesis that generally not all friendships are reciprocal.

Similarly, a previous study~\cite{Vaquera2008} in which adolescents were asked to nominate at most 10 of their best school friends (5 male and 5 female) found that only 64\% of the reported friendships were indeed reciprocal.
Our findings reinforce this finding by investigating multiple datasets from three continents, and by using complete nomination networks (in which each participant is asked about every other participant), resulting in an even more prominent lack of reciprocity.The phenomenon of frequent unreciprocated friendships may be tied to the prevalence of social status and power hierarchy. This suggests that many of the non-reciprocal friendships are aspirational: people want to be friends with higher-status individuals and behave in ways that indicate friendship (e.g., naming them as friends), but higher-status individuals have greater choice in which friendships to reciprocate and choose to only behave as a friend to a subset of the friendships offered to them. In the study of reciprocal friendships in US high schools, the first-generation and black children were found to have many fewer reciprocal friendships, which may indicate that these social groups have stronger hierarchical social structure~\cite{whyte2012street,klein2006cultural}.
Moreover, to the best of our knowledge, this is the first study to analyze the expectations that individuals have from their friendship relationships and to compare the expected and actual relationships.

\subsection{Directionality and Induced Peer-Pressure}
\label{sec:peer_pressure}

Social scientists have long believed that reciprocal friendships are more intimate~\cite{buhrmester1990intimacy,Vaquera2008,gershman1983differential}, provide higher emotional support~\cite{wenz1997importance,gershman1983differential,tur1999reciprocal,stanton2005adolescent}, and form a superior resource of social capital~\cite{lazega2001social,nahapiet1998social,Vaquera2008} when compared to those that are not reciprocated.
This holds whether or not any party of the dyad is aware of the reciprocity status of  their relationship~\cite{Vaquera2008}. 

In order to understand the effect of reciprocal ties on peer-pressure, we turn to the Friends and Family study.
In addition to surveys that were used to determine the closeness of relationships among participants (similarly to the Reciprocity Survey above), it included a fitness and physical activity experimental intervention. The study (Approval\#: 0911003551) was reviewed and approved by the Committee on the Use of Humans as Experimental Subjects (COUHES) at MIT.All participants provided a written consent to participate in this study and COUHES approved the consent procedure.

\begin{figure}
	\centering
		\includegraphics[width=1\columnwidth]{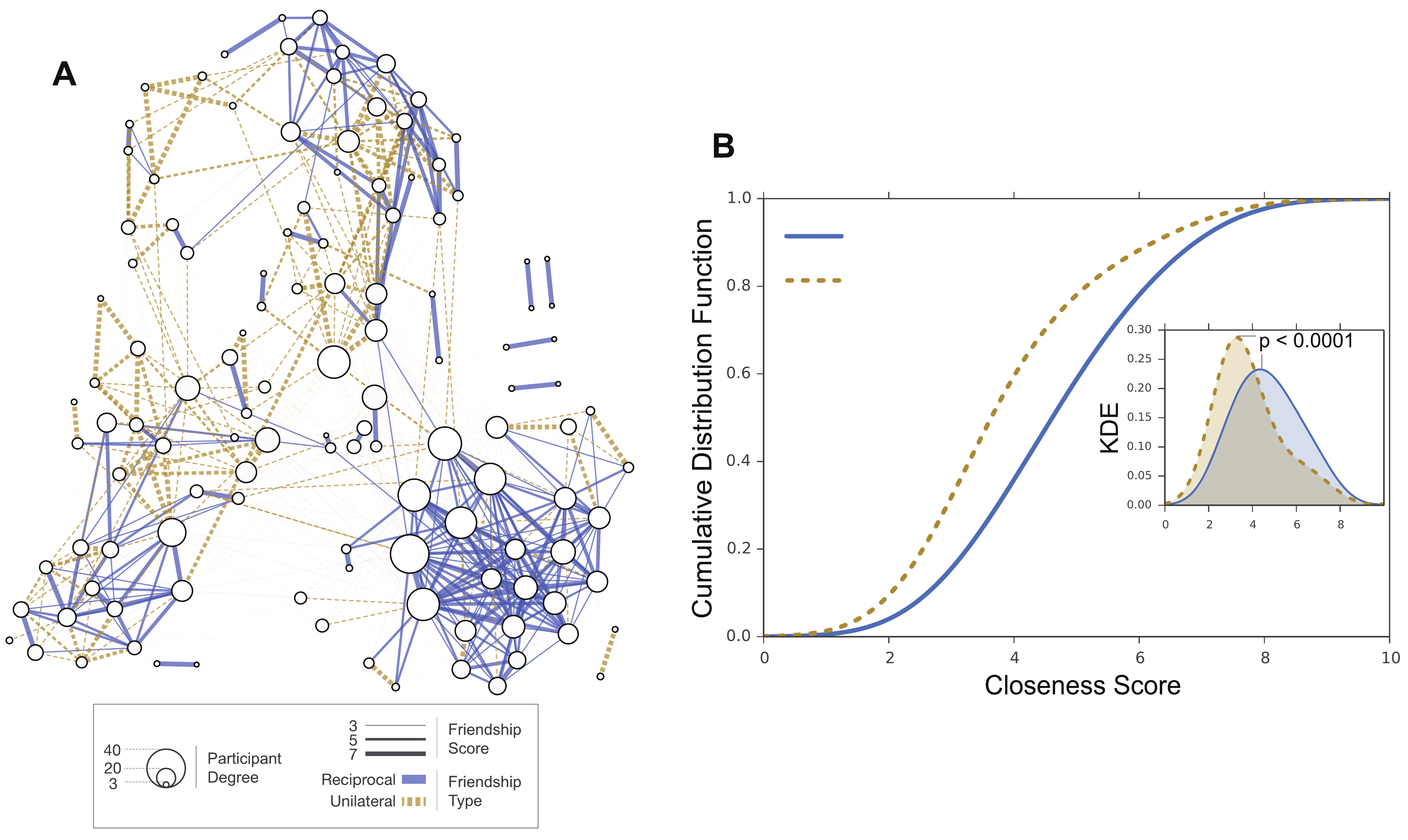}

		\caption{
			\textbf{(A)} The undirected friendship nomination network in the Friends and Family study, where nodes represent participants and edges represent explicit friendship ties (only ties with a closeness score $>2$ are considered).
			The size of a node is proportional to its degree.
			The width of an edge represents the closeness score for unilateral ties and the average closeness score for reciprocal ties (average scores are used for visualization purposes only, for the analysis, reciprocal ties are considered as two separate ties).
			The line style of an edge represents the type of the edge, where reciprocal ties are solid lines and unilateral ties are dashed lines. \textbf{(B)}
			The Cumulative Distribution Function (CDF) and Kernel Density Estimate (KDE) of closeness scores are computed for unilateral ties (dashed line) and reciprocal ties (solid line). Note that due to the nature of the Gaussian KDE process, it is possible that the estimation extends past the largest and smallest values in the dataset.}
		\label{fig2}
	\end{figure}

As part of the closeness surveys, each participant scored other participants on a 0--7 scale, where a score of 0 meant that the participant was not familiar with the other, and 7 that the participant was very close to the other.
Analyzing the distribution of closeness scores associated with the two types of ties in the Friends and Family friendship network (see Fig.~\ref{fig2}) reveals that participants that share a reciprocal friendship tend to score each other higher (average of 4.7) in terms of closeness when compared to participants that share unilateral friendship (average of 3.9) ($p < 10^{-4}$, two-sample t-test).

	However, we hypothesize that `reciprocity' and `directionality' of friendships may be critical factors in promoting peer influence, beyond the mere effect of the total tie `strength' bound up in the relationship. To support our hypothesis, we investigate the FunFit experiment---a fitness and physical activity experimental intervention---conducted within the Friends and Family study population during October to December of 2010. 	The experiment was presented to participants as a wellness game to help them increase their daily activity levels.  Subjects received an `activity app' for their mobile phone which passively collected their physical activity data and showed the participants how their activity level had changed relative to their previous activity level, and the amount of money they had earned by being more active.
\subsubsection{Measuring Physical Activity \& Calculating Rewards}
	Physical activity measurement was based on accelerometer readings from the subjects' smartphones.
	Accelerometer scans were sampled in a duty cycle of 15 seconds every 2 min.
	During the 15 seconds, raw 3-axis accelerometer measurements were sampled at 5 Hz rate and combined to compute the vector magnitude for each sample.	The variance of the magnitude in each one-second block was then computed \cite{eston1998validity}.
	The score was calculated by giving one point for every second, thresholded to three states (i) \emph{still}, (ii) \emph{moderate activity}, and (iii) \emph{high activity}, where the two active levels were combined into a single active level.
	Participants were not constrained in the way they should carry the phone, but were told that the more they carry the phone on their body, the more of their activity would be accounted for their game score.
	
	For analysis purposes, activity levels were normalized to the span of a single sample.
	For example, a normalized \emph{daily average activity} is calculated by summing all accelerometer samples for the day and then dividing by the total count of accelerometer readings for the day.
	This gives the average activity level per reading for that day, and allows to easily do things like comparing between normalized average activity levels in different times of the day.
	It is trivial to convert a normalized value to actual time: for example, a normalized daily average value of 1 for an experimental group represents an average activity of 96 minutes per member.
	
	\textit{Game rewards were calculated every three days}, using a reference window of the seven days preceding the current 3-day bin.
	Average and variance for daily activity count are calculated for the reference window, as well the daily average for the current 3-day bin.
	Reward depended solely on an individual's performance, and was mapped to be between \$0.50-\$5, in \$0.50 increments between one standard deviation above and below the reference average value.
	Values above or below the bounds received max or min value, respectively.
	To avoid discouragement due to zero reward, participants were not given less than 50 cents per reward period.
	
\subsection{Analyzing the Change of Behavior }
	108 out of the 123 active Friends and Family subjects at that time elected to participate and were allocated into three experimental conditions, allowing us to isolate different incentive mechanisms varying monetary reward, the value of social information, and social pressure/influence:
	\begin{itemize}
		\item \textbf{Control:} subjects were shown their own progress and were given a monetary reward based on their own progress in increasing physical activity relative to the previous week.
		\item \textbf{Peer See:} subjects were shown their own progress and the progress of two ``buddies,'' in the same experimental group, and were given a monetary reward based on their own progress in increasing physical activity relative to the previous week.
		\item \textbf{Peer Reward:} subjects were shown their own progress and the progress of two ``buddies,'' in the same experimental group, but their rewards depended only on the progress of the two ``buddies''. This condition realizes a social mechanism based on inducing peer-to-peer interactions and peer pressure~\cite{mani2013inducing}.
	\end{itemize}
	However, for the purpose of our analysis in this section, we combine the samples from the two peer pressure treatments and omit the control group.
	
	During the initial 23 days of the experiment (Oct 5 - Oct 27), denoted as P1, the baseline activity levels of the subjects were collected. The actual intervention period is denoted as P2.
	During the intervention period, the subjects were given feedback on their performance in the form of a monetary reward.
	The monetary reward was calculated as a function of the subject's activity data relative to the previous week and was divided according to the subject's experimental condition (i.e., Peer See and Peer Reward).
	Noting that the physical activity was measured passively by logging the smartphone accelerometer (as opposed to self-reported surveys) and the game was not designed as a competition, every subject had the potential to earn the maximal reward.
	That is, a previously non-active participant could gain the same reward as a highly active one, while the highly active person would need to work harder.
	
	The results in~\cite{aharony2011social} show that the two social conditions (i.e. Peer See and Peer Reward) do significantly better than the control group.
	Furthermore, the results suggest that there is a complex contagion effect~\cite{centola2007complex}, due to the reinforcement of the behavior from multiple social contacts~\cite{centola2010spread,centola2007complex}, related to pre-existing social ties between participants.
	Our analysis here focuses on the role of reciprocity and directionality of friendship ties in this contagion process.
	
	In order to investigate the role of reciprocity and directionality of friendship ties in the contagion process, we performed a regression analysis in which the dependent variable was the change in physical activity between the post-intervention phase and the pre-intervention phase (i.e., the average daily physical activity in P2 divided by the average daily physical activity in P1).
	
	For our study, we refer to a participant whose behavior is being analyzed as ``ego'', and participants connected to the ego (i.e., experimental ``buddies'') are referred to as ``alters''.
	Because friendship nominations are directional, we studied the three possible types of friendships (from the prospective of the ego) as independent variables:
	an ``ego perceived friend'', in which an alter identifies an ego as a friend (i.e., incoming tie);
	an ``alter perceived friend'' in which an ego identifies an alter as a friend (i.e., outgoing tie);
	and a ``reciprocal friend,'' in which the identification is bidirectional (i.e., reciprocal tie).
	Finally, we also included the tie strength (i.e., the sum of the closeness scores between an ego and his or her alters) as a control variable, which has been previously investigated as a moderator of the effect of social influence~\cite{aral2013tie}.

\begin{figure}[h!]
	\includegraphics[width=1\columnwidth]{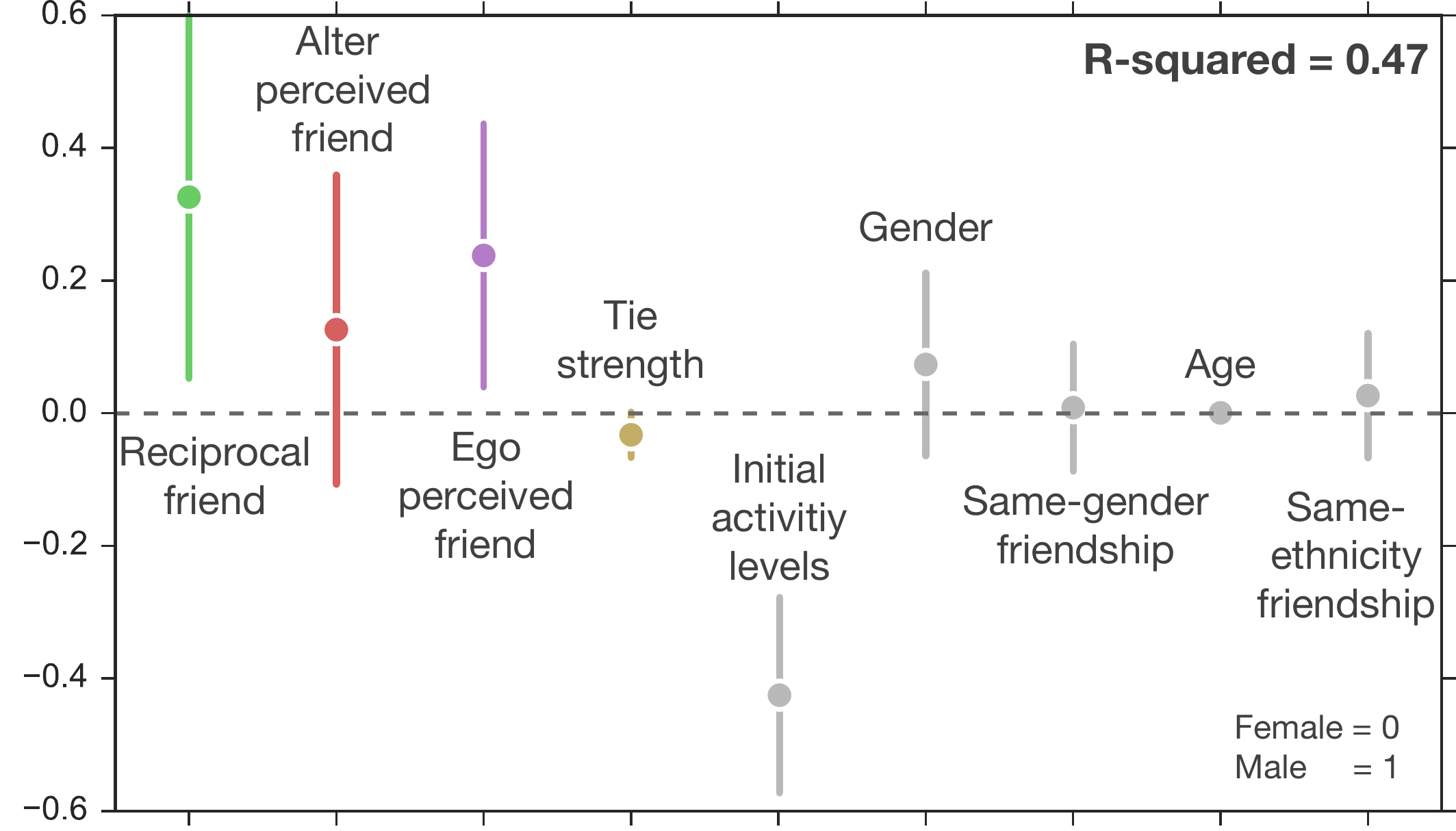}
	\caption{Change in physical activity under experiment conditions shows that the type of friendship is relevant to the effectiveness of the induced peer pressure while controlling over several covariates (gray lines and circles). The plot shows the mean effect size of the covariates (solid circles) and the 95\% confidence intervals (bars).}
	\label{fig:peer_pressure_si}
\end{figure}

Our central hypotheses concern the relationship between the type of relationship subjects share with their buddies, on
the susceptibility of increased physical activity under the experiment conditions.
Several multivariate analyses are presented in table 2.2, but the main story emerging from them is made clear in Fig.~\ref{fig:peer_pressure_si}.
In table~\ref{tab:4} our dependent variable is the change in physical activity between the post-intervention phase (P2+P3) and the pre-intervention phase (P1) and the covariates consist of:
(1) reciprocal friend - the number of buddies with whom the subject has a reciprocal friendship relationship (values 0 - 2);
(2) alter-perceived friend - the number of buddies who were reported by the subject as friends but did not report the subject as friend (values 0 - 2);
(3) ego-perceived friend - the number of buddies who reported the subject as friend but were not reported by the subject as friends (values 0 - 2);
(4) tie strength - which is the sum of the nomination scores between the subject and the buddies;
(5) Initial activity - which controls over the subject's pre-intervention activity levels. This can be described as follows:
(6) gender (Male or Female);
(7) same-gender - the number of buddies who had the same gender as the subject (values 0 - 2);
(8) age; and
(8) same-gender - the number of buddies who had the same ethnicity as the subject (values 0 - 2);

\[
Y_i  = \log \bigg(\frac{\text{Activity($P2+P3$)}}{\text{Activity($P1$)}}\bigg) \sim N(\beta X_i, \sigma^2)
\]

Where $X_i$ represents the vector of demographic and experimental conditions of $i$ with a constant variance $\sigma^2$.

The full specification model in Table 2.2 shows the results of the linear model analysis using a fairly wide specification of independent variables discussed above.
We find that reciprocal ties are consistently associated with higher improvement in activity change and that this relationship persists even when we include detailed controls.
\begin{table}
	\footnotesize
	\begin{center}
		\caption{Regression Coefficients for the change in activity under different experiment conditions}
		\label{tab:4_1}
		\begin{tabular}{l D{.}{.}{2.7}@{} D{.}{.}{2.7}@{} D{.}{.}{2.7}@{} D{.}{.}{2.7}@{} D{.}{.}{2.7}@{} D{.}{.}{2.7}@{} D{.}{.}{2.4}@{} D{.}{.}{2.7}@{} }
			\toprule
			& \multicolumn{1}{c}{Model 1} & \multicolumn{1}{c}{Model 2} & \multicolumn{1}{c}{Model 3} & \multicolumn{1}{c}{Model 4} & \multicolumn{1}{c}{Model 5} & \multicolumn{1}{c}{Model 6} & \multicolumn{1}{c}{Model 7} & \multicolumn{1}{c}{Model 8} \\
			\midrule
			Constant                   & 0.05          & 0.03          & -0.02         & 0.10          & 0.63^{***}    & 0.06          & -0.01     & 0.56^{*}      \\
			& (0.08)        & (0.08)        & (0.09)        & (0.09)        & (0.11)        & (0.29)        & (0.09)    & (0.27)        \\
			Reciprocal friend              & 0.44^{**}     & 0.45^{**}     & 0.45^{**}     & 0.44^{**}     & 0.30^{*}      & 0.44^{**}     & 0.45^{**} & 0.33^{*}      \\
			& (0.16)        & (0.16)        & (0.16)        & (0.16)        & (0.13)        & (0.16)        & (0.16)    & (0.14)        \\
			Alter perc. friend     & 0.15          & 0.15          & 0.18          & 0.16          & 0.12          & 0.15          & 0.16      & 0.13          \\
			& (0.13)        & (0.14)        & (0.14)        & (0.14)        & (0.11)        & (0.14)        & (0.13)    & (0.12)        \\
			Ego perc. friend       & 0.31^{*}      & 0.30^{*}      & 0.31^{*}      & 0.29^{*}      & 0.24^{*}      & 0.31^{*}      & 0.32^{**} & 0.24^{*}      \\
			& (0.12)        & (0.12)        & (0.12)        & (0.12)        & (0.10)        & (0.12)        & (0.12)    & (0.10)        \\
			Tie Strength               & -0.04^{\cdot} & -0.04^{\cdot} & -0.04^{\cdot} & -0.04^{\cdot} & -0.03^{\cdot} & -0.04^{\cdot} & -0.04^{*} & -0.03^{\cdot} \\
			& (0.02)        & (0.02)        & (0.02)        & (0.02)        & (0.02)        & (0.02)        & (0.02)    & (0.02)        \\
			Gender                     &               & 0.05          &               &               &               &               &           & 0.08          \\
			&               & (0.08)        &               &               &               &               &           & (0.07)        \\
			Same-gender         &               &               & 0.07          &               &               &               &           & 0.01          \\
			&               &               & (0.06)        &               &               &               &           & (0.05)        \\
			Interv. group         &               &               &               & -0.08         &               &               &           & -0.01         \\
			&               &               &               & (0.08)        &               &               &           & (0.07)        \\
			Pre-interv. activity &               &               &               &               & -0.43^{***}   &               &           & -0.42^{***}   \\
			&               &               &               &               & (0.07)        &               &           & (0.07)        \\
			Age                        &               &               &               &               &               & -0.00         &           & 0.00          \\
			&               &               &               &               &               & (0.01)        &           & (0.01)        \\
			Same-ethnicity             &               &               &               &               &               &               & 0.08      & 0.03          \\
			&               &               &               &               &               &               & (0.05)    & (0.05)        \\
			\midrule
			R$^2$                      & 0.16          & 0.16          & 0.18          & 0.17          & 0.46          & 0.16          & 0.18      & 0.47          \\
			Adj. R$^2$                 & 0.11          & 0.10          & 0.12          & 0.11          & 0.42          & 0.10          & 0.12      & 0.39          \\
			Num. obs.                  & 76            & 76            & 76            & 76            & 76            & 76            & 76        & 76            \\
			\bottomrule
			\multicolumn{8}{l}{\scriptsize{ Gender (Male = 1, Female = 0); dependent variables significance levels: $^{***}p<0.001$, $^{**}p<0.01$, $^*p<0.05$, $^{\cdot}p<0.1$}}
		\end{tabular}
	\end{center}
\end{table}

Both the gender of the subject and whether the buddies are from opposite gender had no significant association with the increase in activity.
Controlling for a subject's age showed no effect on the change of activity as well.
This was expected due to the highly homogeneous group of participants (e.g., age $mean = 33$, $SD = 5$, both in years).
A recent study by Aral et al., showed the importance of tie strength in moderating the effect of peer pressure~\cite{aral2013tie}.
In this work, we find the strength of the tie to be significant at $1\%$ only, which highlights the importance of the tie `type' rather than the tie `strength'.
Finally, the initial activity level coefficient is negative and statistically significant ($P < 0.001$).
This suggests the difficulty of increasing the relative activity levels for already very active users in the post intervention period.

Previous studies have found that passive exposure to peers is sometimes sufficient to adopt a new health related behavior (e.g.,~\cite{madan2010social,centola2010spread}).
This is particularly true when the behaviors to be transmitted are `not difficult' and have low adoption barrier (e.g., registering for a health forum Web site~\cite{centola2010spread} or signing up for an Internet-based diet diary~\cite{centola2011experimental}).

We hypothesize that transmitting more complex behaviors (e.g., getting a vaccination, improving a diet, using condoms, or committing to physical exercises) might require people to engage in purposeful persuasion and therefore, the effects of interpersonal relationships will become more significant.
Indeed, we see this as a trend in the current analysis, where the participants in the passive peer-see condition show no significant directionality effect while participants in the active peer-reward condition show a large directionality effect (see Table~\ref{tab:see_pay_only}).

\begin{table}
	\begin{center}
		\caption{Regression coefficients for the change in physical activity for peer-see and peer-reward intervention groups.}
		\begin{tabular}{l D{)}{)}{11)2}@{} D{)}{)}{11)2}@{} }
			\toprule
			& \multicolumn{1}{c}{Peer-See} & \multicolumn{1}{c}{Peer-Reward} \\
			\midrule
			Constant               & 0.02 \; (0.09)  & 0.11 \; (0.14)     \\
			Reciprocal friend          & 0.33 \; (0.20)  & 0.55 \; (0.27)^{*} \\
			Alter perceived friend & 0.12 \; (0.16)  & 0.20 \; (0.23)     \\
			Ego perceived friend   & 0.24 \; (0.16)  & 0.36 \; (0.20)^{*} \\
			Tie Strength           & -0.03 \; (0.02) & -0.05 \; (0.04)    \\
			\midrule
			R$^2$                  & 0.13            & 0.16               \\
			Num. obs.              & 37              & 39                 \\
			\bottomrule
			\multicolumn{3}{l}{\scriptsize{$^{***}p<0.001$, $^{**}p<0.05$, $^*p<0.1$}}
		\end{tabular}
		\label{tab:see_pay_only}
	\end{center}
\end{table}
	
The strongest effect for both treatment groups in this study was found for the reciprocal factor even when controlling over the strength of the tie (the tie strength is weakly significant $p = 0.07$).
That is, alters in reciprocal friendships have more of an effect on the ego than alters in other types of friendships.

Interestingly, when the ego was perceived as a friend by the alters (i.e., incoming edges from the alters to the ego), the effect was also found to be positive and significant ($p < 0.05$).
On the other hand, no statistically significant effect was found when the alters were perceived as friends by the ego (i.e., outgoing edges from the ego to the alters).
Therefore, the amount of influence exerted by individuals on their peers in unilateral friendship ties seems to be dependent on the direction of the friendship.

Unlike previous works on social contagion effects~\cite{christakis2007spread,fowler2009dynamic}, which were conducted without peer-to-peer incentives, we find that influence does not flow from nominated alter to nominating ego. Surprisingly, alter's perception of ego as a friend would increase alter's ability to influence ego's behavior when ego does not reciprocate the friendship.
We attribute this difference to the fact that there is a peer-to-peer incentive mechanism, and therefore there are likely to be differences in communication when the alters believe the ego to be their friend versus when they do not.


We find these results to be consistent even when including additional detailed controls for the ego's age, gender, whether the buddies are from the same or opposite gender, whether the buddies have the same ethnicity, and their pre-intervention activity levels (Model 2 \& Model 3).
We find the pre-intervention activity to be the only control variable with a significant effect.
The asymmetry we observe would explain why targeting individuals based on the incoming-degree centrality (i.e., picking people receiving the highest number of friendship nominations) is not an efficient intervention strategy and has been found to be no better~\cite{kim2015social}, or worse~\cite{cho2012identification} than random targeting.
\subsection{Reciprocity and Global Adoption}
\label{sec:global_adoption}
In order to understand the effect of reciprocal ties on global behavior adoption, we experimented with a variation of the classic epidemic spreading model, Susceptive-Infected (SI) model.
We refer to this variant as the Bi-Directional Susceptive-Infected (BDSI) model.
Unlike the classic SI in which behavior is transmitted along edges with a constant probability, the proposed BDSI model considers the direction in which behaviors can be transmitted with different probabilities based on the direction and type of edges---i.e. $p_{rec}$ for reciprocal edges and $p_{+}$/$p_{-}$ for the two possible directions of unilateral edges.

\begin{figure}
	\centering
	\includegraphics[width=1\columnwidth]{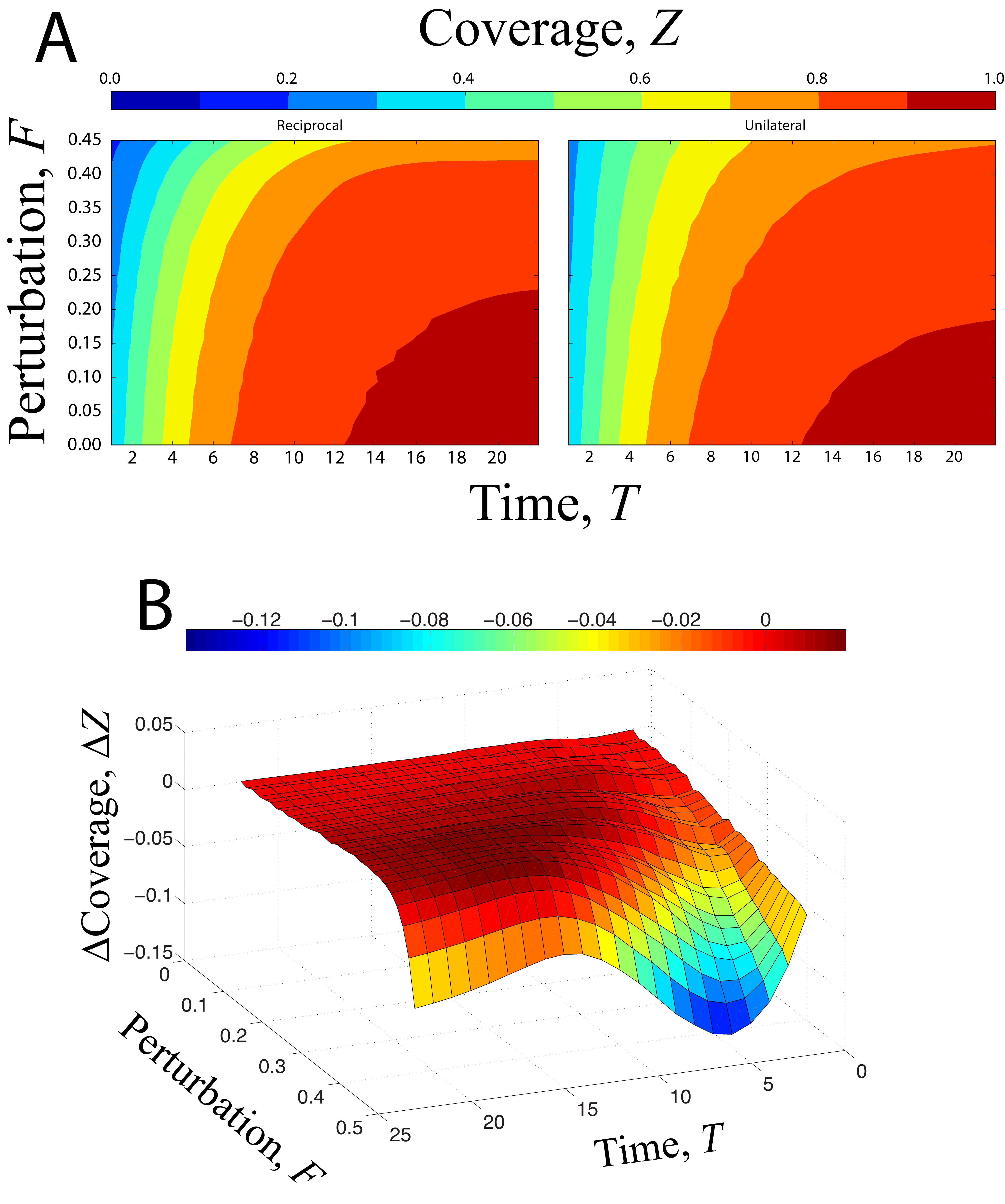}
	\caption{
		 \textbf{(A)} demonstrates the effect of perturbation on the coverage and speed of adoption. \textbf{(B)} illustrates how the coverage and speed decay faster when removing reciprocal edges in comparison with unilateral edges.}
	\label{fig:simulation}
\end{figure}

In order to observe the effects of reciprocal edges on diffusion, we employ an edge percolation process in which we measure the coverage (i.e., number of infected nodes), denoted by $Z$, and time to infect, denoted by $T$, when removing reciprocal and unilateral edges successively (i.e., perturbation $F$).
That is, $F$ is the percentage of edges removed in perturbation.
We find the nature of the simulation results are qualitatively independent of the choice of $p_{rec}$, $p_{+}$ and $p_{-}$ given that $p_{rec} \ge p_{+} \ge p_{-}$.

Figure~\ref{fig:simulation} shows the behavior adoption coverage when simulating the BDSI model on the self-reported friendship network from the Friends and Family dataset.
As can be seen in the figure, the coverage $Z$ decays much faster when removing reciprocal edges (left figure) compared with removing the same amount of unilateral edges (right figure).
Moreover, the difference in coverage $\Delta Z$ is affected remarkably by the removal of reciprocal edges most notably in the early stages of the diffusion process (e.g., $T\in [5,10]$). 
This can be attributed to the rapid diffusion within a single community through reciprocal edges, corresponding to fast increases in the number of infected users in early stages of the diffusion process, followed by plateaus, corresponding to time intervals during which no new nodes are infected the behavior escapes the community (i.e., through the strength of weak tie~\cite{granovetter1973strength}) to the rest of the network through unilateral edges.

\section{Future Work}
\section*{Integrated Transportation Systems}
\begin{figure}[h!]
	\centering
	\includegraphics[width=1\columnwidth]{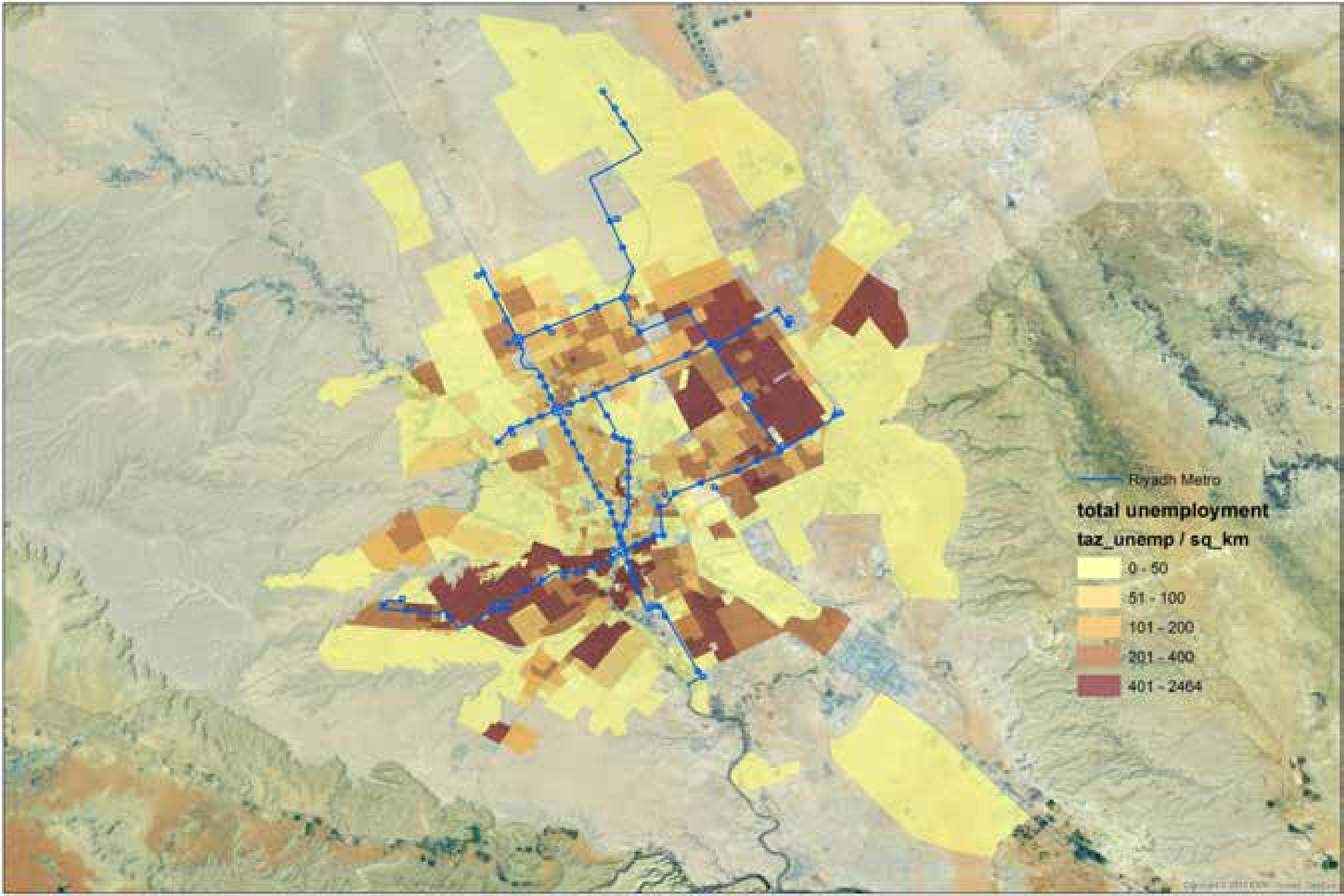}
	\caption{Unemployment population density and the planned Riyadh Metro are shown for Riyadh, the capital of Saudi Arabia.}
	\label{fig:transit}
\end{figure}

In this chapter, we found the relationship between unemployment and social behavior (i.e., exploration and engagement) to be very strong. We also explored how social cohesion can play an important role in promoting behavioral change (i.e., increasing opportunity/information flow). Better transportation infrastructure is another way to lower the cost of social exploration. For example, CDR-inferred sociodemographic indicators can be combined with travel demand estimates in order to test various scenarios for transit system adoption in Riyadh, where major public transit system planning is underway. Such research presents an innovative approach to transit system design in the information age, which proactively targets the maximum environmental and social impact~\ref{fig:transit}

\section*{Investigating the Determinants of Unemployment---Living Lab Approach in KSA}
In our future work, we intend to intersect the Call Detail Records (CDR) and unemployment data derived from the unemployment benefit program at the individual level. This will allow for the study of how the behavioral signature of a single individual can be used to predict that same individual's employment status. This could reveal the key determinants of unemployed people to find a job and allow for designing personalized intervention mechanisms. Earlier studies proved that behavior is highly shaped and constrained by one's social network and demonstrated ways in which individuals can manipulate these networks to achieve specific goals. A great example is the much studied 'strength of weak ties` hypothesis, which states that the strength of a tie between A and B increases with the overlap of their friendship circles, resulting in the importance of weak ties in connecting communities. Mark Granovetter~\cite{granovetter1973strength} first proposed this idea in a study that laid emphasis on the nature of the tie between job changers and the contact person who provided the necessary information in a Boston suburb.  Basically, although those with whom one has strong tie are more motivated to help with job information, the structural position of weak ties played more important role. The intuition is those to whom one is weakly tied are more likely to move in circles different and will thus have access to information different from that which one receives. 

However, we intend to investigate the determinants of unemployment through a Reality Mining/Living  approach. 

Living Labs are a unique, new type of user-centric field research methodology for sensing, prototyping, validating and refining complex solutions in multiple and evolving real life contexts. The Living Labs allow the translation of big data analysis into action, generating new insights by quantifying real world interactions and delivering interventions that change behavior at scale.

The purpose of the proposed Living Lab is to understand the social and behavioral dynamics associated with successful job search. Specifically, investigating the causal mechanisms of job seekers to get jobs through quasi-experimental design. This will allow for developing guidelines and frameworks based on quantitative measures to support job seekers in finding jobs

By combining theoretical models with rich and systematic measurements, it is possible to gain insight into the underlying behavior of complex social systems. The use of pervasive technologies as a proxy for more expensive traditional survey methods is of a particular interest as such methods can provide quick insights and understanding of social patterns, supply, and demand.  This approach is also immune to some of the biases of surveys has and provide much higher time and space resolutions enabling better understanding at unprecedented scales. 

\subsection{Living Lab Research Design}
By combining theoretical models with rich and systematic measurements, it is possible to gain insight into the underlying behavior of complex social systems. The use of pervasive technologies as a proxy for more expensive traditional survey methods is of a particular interest as such methods can provide quick insights and understanding of social patterns, supply, and demand.  This approach is also immune to some of the biases of surveys has and provide much higher time and space resolutions enabling better understanding at unprecedented scales.

\subsection{Quasi-experimental Econometric Analysis}
In medical research, researchers measure the effects of a new medicine by randomly selecting a treatment group to receive the medicine and a control group to receive the placebo. The fact that the medicine/placebo was assigned randomly means that researchers can argue with high confidence that the medicine was solely responsible for any difference in health between control \& treatment groups. In the same way, most quasi-experimental econometric methods use a source of randomness to measure the effects of economic programs. Specifically, we can use a case where different behaviors and job seeking patterns associated with two or more similar groups to test the effects of these patterns on these groups. In other words, this technique will allow us to test how changes in social behaviors (as evidenced by mobile phone CDR's) causes changes in the probability of success of a job seeker, which can guide intervention mechanisms that would cause specific types of people to find employment and/or improve their lives. Specifically, we can test:

\begin{description}
	\item [How] do specific aspects of social network or behavior causes people to connect with jobs, determine duration of unemployment, and explain particular qualities of jobs (higher salaried, etc.)?
	\item [How] do different types of people connect to different types of jobs?
	\item [How] can the employment assistant programs be adjusted to increase the benefit to the people most serious about finding employment? 
\end{description}

\begin{figure}[h!]
	\centering
	\includegraphics[width=0.8\columnwidth]{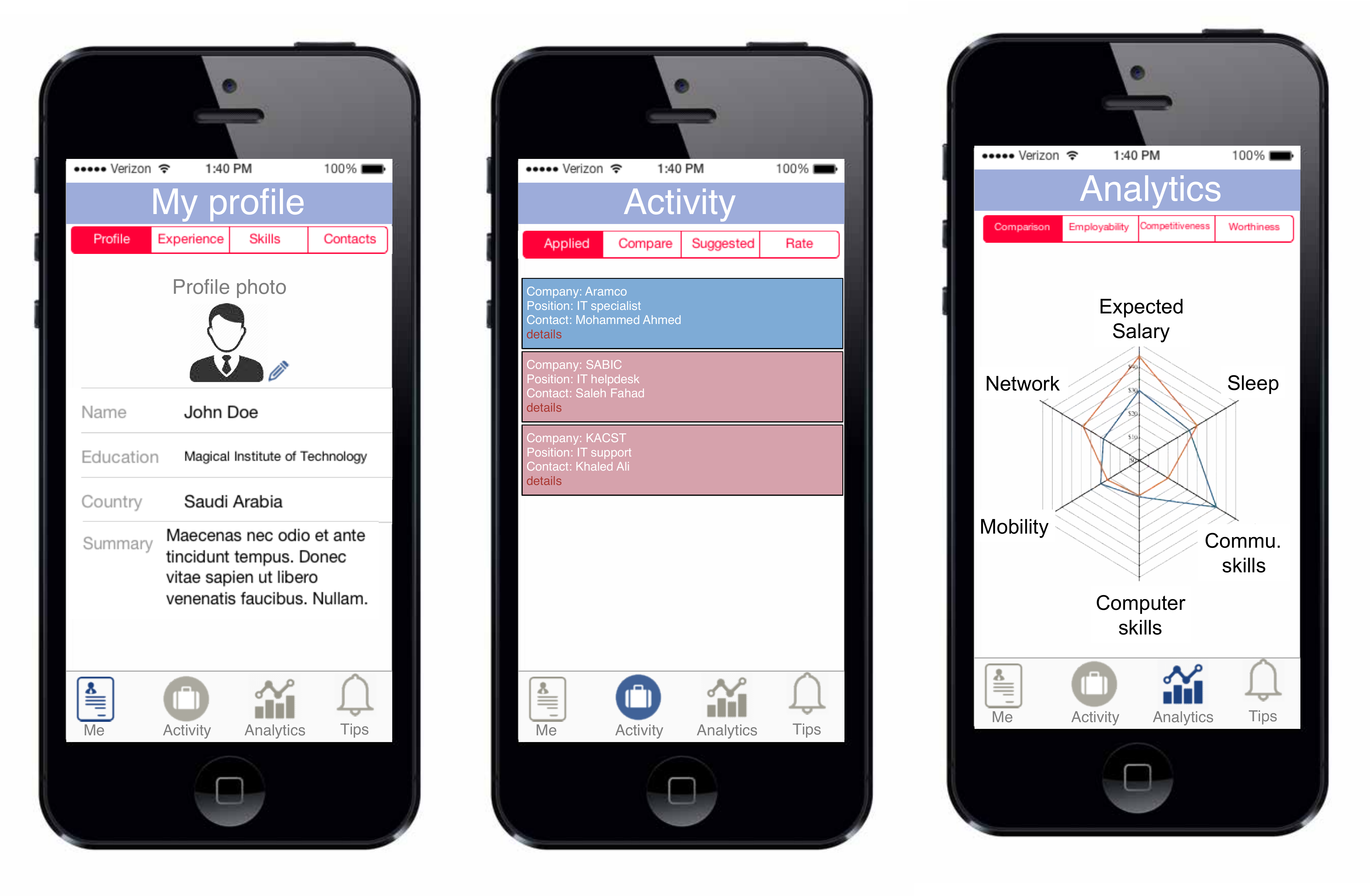}
	\caption{Mockup screens of the potential mobile app that could help in changing the behavior of job seekers in order to increase their employability.}
	\label{fig:mobile}
\end{figure}

As such, this direction will allow to address the behavioral patterns of job seekers in order to improve the employability of individuals in the emerging sectors of the Saudi economy. In particular, we expect to be able to develop a mobile application, `Job Coach', that will provide personalized advice and coaching for job seekers in order to improve their employability~\ref{fig:mobile}.

\section{Discussion and Chapter Summary}
\label{sec:conclusion}

In this chapter we have demonstrated that mobile phone indicators are associated with unemployment rates and that this relationship is robust to the inclusion of controls for a district's area, population, and mobile penetration rate. Following this analysis, we also investigated the predictability of unemployment rates with respect to three categories of indicators---activity patterns, social interactions, and spatial markers. The results of these analyses highlighted the importance of social interaction indicators for predicting unemployment. 

Note that we are not stating a causality arrow between the indicators and the unemployment rate, as we do not have individual-level mapping of unemployment with which to test for individual differences. In this chapter, our goal is to show that aggregate behavioral indicators of the members of a district represent a strong statistical signature that can be used as alternative measuring approach with a real translation in the local economy.

We also have demonstrated the important role that  reciprocity and directionality of friendship ties play in inducing effective social persuasion.
We have also shown that individuals have difficulty in judging the reciprocity and directionality of their friendship ties (i.e., how others perceive them). 
This can be a major factor limiting the success of cooperative arrangements such as peer-support programs.

For instance, in ~\cite{kim2015social,cho2012identification} it was shown that the assumption that individuals with a high number of incoming friendship nominations are ``influencers'' is flawed. Such people are no better and often worse than average people at exerting social influence. Our results suggest that this is because many of those ties are either not reciprocal or flow in the wrong direction for effective persuasion.

Finally, we investigated the predictability of friendship ties in friendship networks with respect to two categories of features, namely, \textit{Social Embeddedness} and \textit{Social Centrality}, and showed that these features can be used to markedly improve judgments about the reciprocity and directionality of friendship ties (see Appendix).

Previous studies have found that people tend to adopt the behaviors of peers that they are exposed to passively, with the friends and intimate social acquaintances they explicitly identify as such playing a peripheral role (e.g.,~\cite{madan2010social,centola2010spread}).
Other studies have shown that passive exposures to peer behavior can increase the chances of becoming obese~\cite{christakis2007spread,madan2010social}, registering for a health forum Web site~\cite{centola2010spread}, signing up for an Internet-based diet diary~\cite{centola2011experimental}, or adopting computer applications~\cite{aral2012identifying}.
However, our results suggest a fundamental difference between how social learning (i.e., passive exposure) and social persuasion (i.e., active engagement) spread behaviors from one person to another.
Indeed we see this as a trend in our data, with participants in the active Peer Reward condition showing more of the directionality effect than those in the Peer See condition. This is consistent with the effect of the incentive mechanism: the Peer See condition induces social pressure only through social comparison, whereas in the Peer Reward condition there is a monetary reward in addition to the pressure induced by social comparison.

Although we demonstrated the ability to predict tie direction and reciprocity from Social Embeddedness and Difference in Social Status, these measures are also likely to constrain influence, independently of directionality, via mechanisms such as social reinforcement, tie strength, and social status (see Appendix). Therefore, these potential confounding effects need to be taken into account when moving from an experimental to field setting.

The findings of this work have significant consequences for designing interventions that seek to harness social influence for collective action. This thesis also has significant implications for research into peer pressure, social influence, and information diffusion, as these studies have typically assumed undirected (reciprocal) friendship networks, and may have missed the role that the directionality of friendship ties plays in social influence.
This suggests that both practitioners and researchers should consider the reciprocity and directionality of friendship relationships when thinking about behavior change due to social influence.

\include{chap2}
\include{chap3}
\chapter{Economic \& Job Complexity: The Basis of the Labor Economy}
In this chapter, we extend the idea of opportunity/information flow beyond job-seekers to clusters of other economic activities. Similar to the previous chapter, we expect the flow within clusters of related activities to be higher when compared to the flow within isolated activities. This captures the intuition that more ``capitals'' are involved and that such activities require similar ``capabilities.'' Therefore, more extensive clusters of economic activities should generate greater growth by exploiting the greater flow of opportunities/information. We quantify the opportunity/information flow using a ``complexity measure'' of two economic activities (i.e. jobs and exports). We find that this measure indeed predicts economic growth in terms of GDP as measured at the city and country levels.

\section{The City: Job Relatedness Networks}\label{ch3:job}
Job markets are one of the main drivers of human society, from fostering innovation, to dictating unemployment, all the way to a having a profound impact on the economy. Understanding the ecosystem of job markets and possible changes within them is key when planning for the future. In this chapter, we introduce and investigate a measure associated with the complexity of jobs. A job with high complexity means that for a city to have significant prominence for this job, it needs to have signficant prominence for many others. This measure captures dependencies or coexistence between jobs. For example, medical doctor is a complex job. It depends on having nurses, lab technicians, health care administrators etc. However, medical doctor is not a fluid job~\footnote{The concept of fluidity is based on the eigenvector centrality of the jobs-skills bipartite network. If a job has a high fluidity score, a person working in this job has skills that will transfer easily to many other jobs. The concept of job fluidity will not be addressed in this thesis.}. In addition, we find this measure to be a strong predictor of a city's economic performance (measured in GDP), and also a good predictor of a particular job's susceptibility to automation\footnote{this chapter is developed in collaboration with Ahmad Alabdulkareem (kareem@mit.edu), whose research focuses on urban growth and change}.
%
%
%


\subsection{The Complexity of Jobs}

In relating jobs to each other, we look at job-city associations using data from the U.S. Bureau of Labor Statistics (BLS), which reports yearly employment data for metropolitan areas in the United States (http://www.bls.gov/data/). We interpret these data as bipartite networks~\footnote{A bipartite graph or network is a set of nodes and links in which nodes can be separated into two partitions, such that links only connect nodes in different partitions} in which cities are connected to the prominent jobs within them~\footnote{The methodology in this chapter is  influenced by the work of Hidalgo et al. in developing the product space, in which they used a network science approach to evaluate the underlying structure of product creation~\cite{hidalgo2007product,hidalgo2009building}}. Mathematically, we represent this network using the adjacency
matrix $M_{cj}$, where $M_{cj} = 1$ if city $c \in C$ has a significant prominence in
job $j \in J$ and 0 otherwise. We consider city $c$ to have a significant
prominence in job $j$ if the percentage of $c$'s population employed in $j$ is greater than the national average. This means that a city $c$ has higher number of employees in a job $j$, as a share of its total workforce, than the ``average'' city. Therefore, the adjacency matrix $M_{cj}$ relating cities to jobs can be defined as:

\begin{equation}
\label{equation:city_job}
M_{cj} =
\begin{cases} 
1  & \text{if}~ \frac{X_{cj}}{\sum\limits_{j\prime\in J} X_{cj\prime}} > 
\frac{\sum\limits_{c\prime\in C} X_{c\prime j}}{\sum\limits_{j\prime\in J} \sum\limits_{c\prime\in C} X_{c\prime j\prime}} \\
0 & \text{otherwise}
\end{cases}
\end{equation}
Where $X_{cj}$ is the number of employees in job $j$ within city $c$. As such, if 1\% is the fraction of US employees that are medical doctors, then a city (e.g., Boston) would need to have more 1\% of its workforce employed as medical doctors to have significant prominence in this job.

\begin{figure}[h!]
	\centering
	\includegraphics[width=0.85\columnwidth]{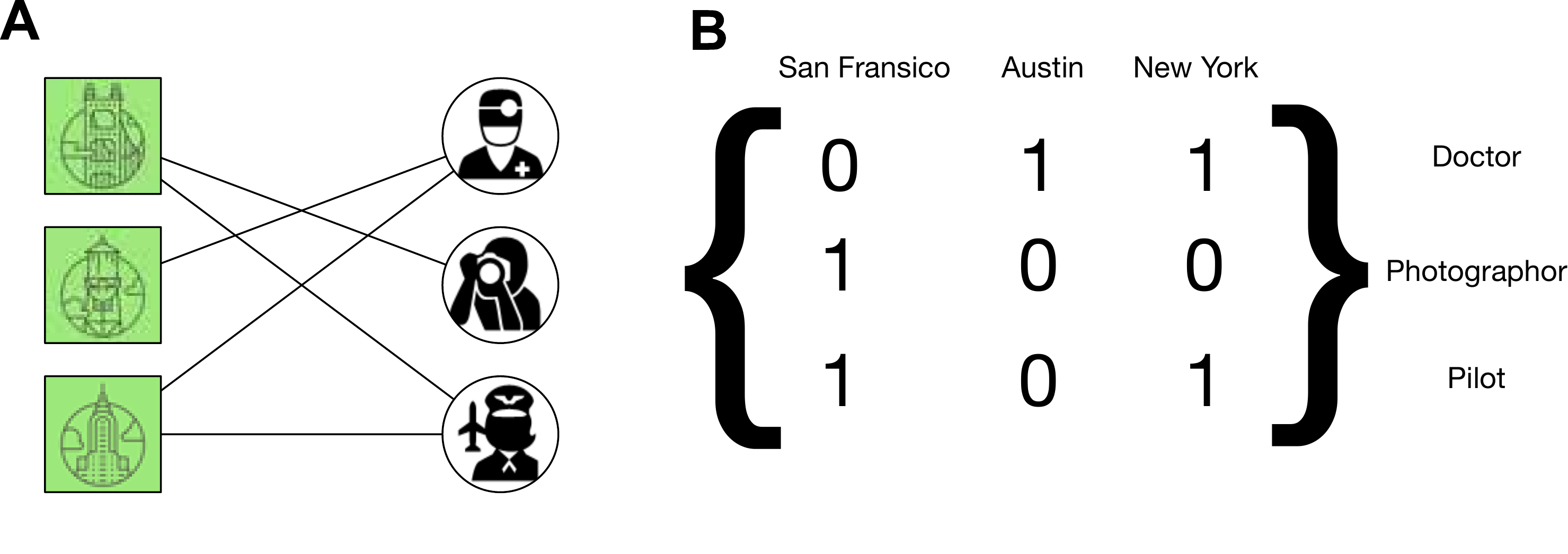}
	\caption{Panel \textbf{(A)} depicts a simple network to exemplify the bipartite
		nature of the matrix connecting cities to jobs, and panel \textbf{(B)} shows the corresponding adjacency matrix $M_{cj}$ computed as in equation~\ref{equation:city_job}.}
	\label{fig:bipartite}
\end{figure}

\subsubsection{The Method of Reflection}
Analogous to the work of Hidalgo et al.~\cite{hidalgo2009building} in building the Economic Complexity Index, we characterize cities and jobs by introducing a family of variables that captures the structure of the network defined by $M_{cj}$.  The method of reflection produces a symmetric set of variables for the two types of nodes in the network (cities and jobs). The methodology consists of iteratively calculating the average value of the previous-level properties of a node's neighbors and is defined as the set of observables:

\begin{equation}
k_{c,N} = \frac{1}{k_{c,0}} {\sum\limits_{j \in J} M_{cj} k_{j,N-1} }
\end{equation}

\begin{equation}
k_{j,N} = \frac{1}{k_{j,0}} {\sum\limits_{c \in C} M_{cj} k_{c,N-1}} 
\end{equation}

\begin{figure}[h!]
	\centering
	\includegraphics[width=0.85\columnwidth]{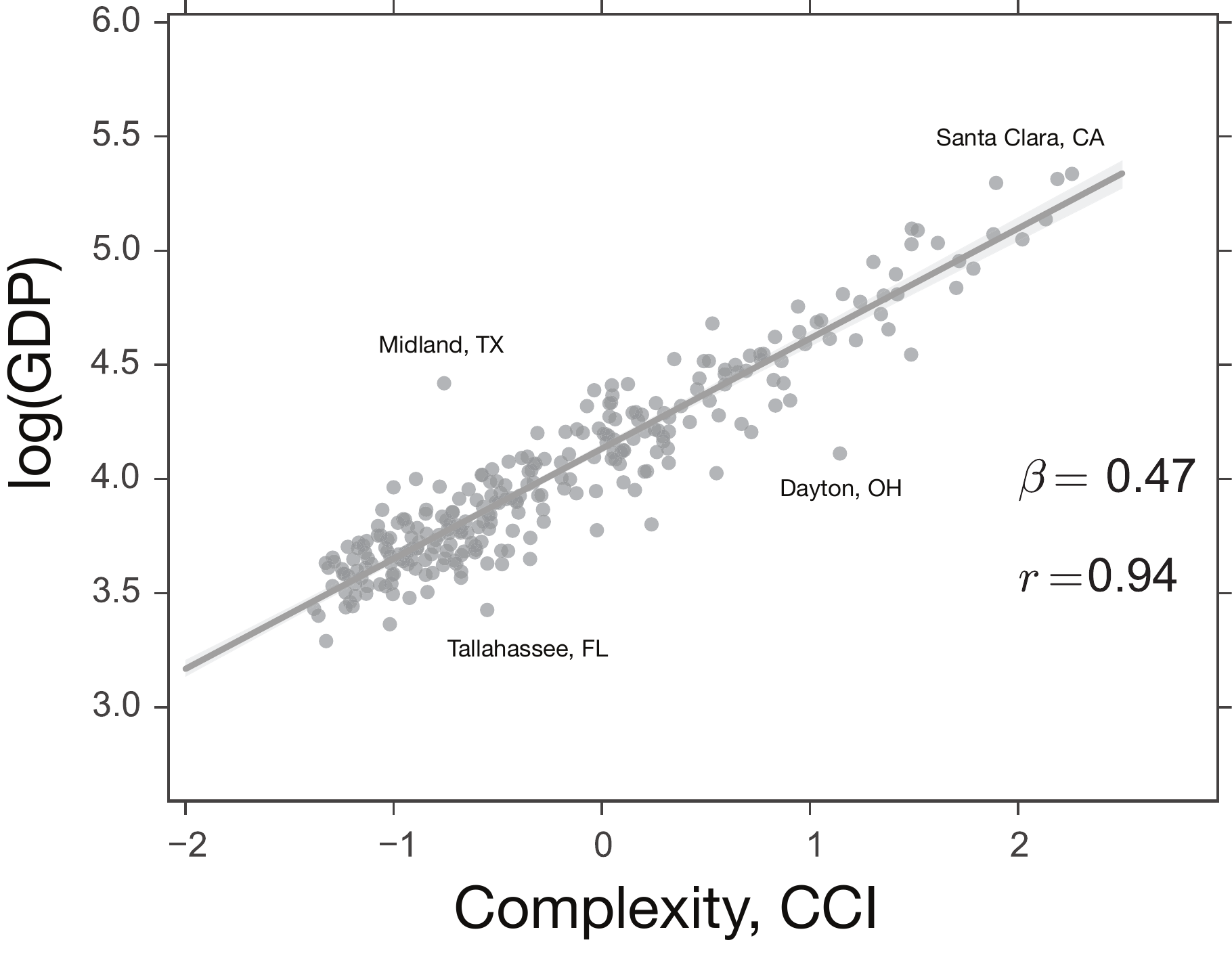}
	\caption{The relationship between the City Complexity Index \textit{(CCI)} and GDP. The best-fit scaling relation is shown as a solid line and the 95\% confidence interval is shown as a shaded area.}
	\label{fig:complexity}
\end{figure}

Initially, $k_{c,0}$ and $k_{j,0}$ represent, respectively, the number of jobs with significant prominence in that city (i.e., city complexity), and the number of cities that have that job with significant prominence (i.e., job complexity), while $N > 0$ is the number of iterations used to define the indicators. Therefore, $k_{c,N}$ converges to City Complexity Index $CCI$ and $k_{j,N}$ converges to the Job Complexity Index $JCI$, for large $N$ values.

\begin{figure}[h!]
	\centering
	\includegraphics[width=0.85\columnwidth]{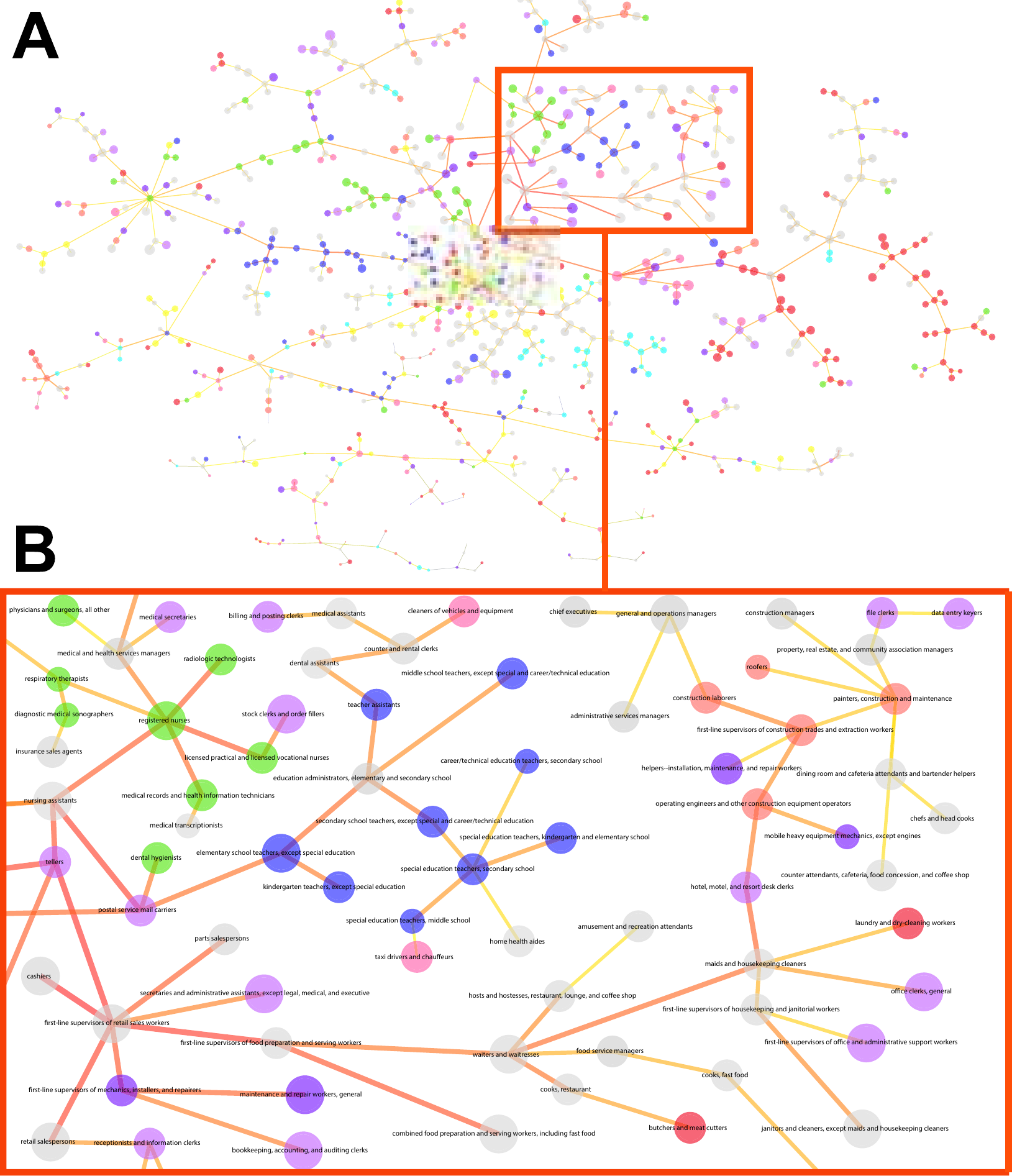}
	\caption{Panel~\textbf{(A)} Network representation of the resulting job relatedness network based on significant prominence. Links are color coded with their proximity value. The sizes of the nodes are proportional to the national number of employees, and their colors indicate the industry cluster that each job belongs to. Panel~\textbf{(B)} provides closer look at a subpart of the network}
	\label{fig:job_network}
\end{figure}

Figure~\ref{fig:complexity} shows the linear relationship between the City Complexity Index \textit{(CCI)} computed on the job data from 2006 and GDP of the city in 2014. We find a very strong correlation between the complexity of the jobs and their corresponding GDP. The network of the 811 jobs is depicted in Figure~\ref{fig:job_network}.

\subsection*{Future Work: Proximity Measure}
We can define our concept of proximity between jobs, in order to project this bipartite network into a simple undirected network relating jobs to each other. Formally, the proximity $\phi_{ij}$ between job $i \in J$ and $j \in J$ is the minimum of the pairwise conditional probabilities of a city having a significant prominence in one job, given that it has significant prominence in the other. The conditional probabilities can be defined as:

$$
P(M_j = 1 \vert M_i = 1) = \frac{\sum\limits_{c \in C} M_{cj} M_{ci} }{\sum\limits_{c \in C} M_{ci}}
 $$

As the network is undirected, we can formally define the proximity as:

\begin{equation}
 \phi_{i,j} = \min\{P(M_j = 1 \vert M_i = 1), P(M_i = 1 \vert M_j = 1)\}
 \end{equation}

This measure of job proximities would allow us to understand the diffusion process of jobs and how they evolve (i.e., which jobs will increase in number and have significant prominence in the future).

\section{The Country: Beyond The Theory of Economic Complexity}\label{ch4:}
Hidalgo and Hausmann~\cite{hidalgo2009building} recently developed the \textit{Method of Reflection (MR)}, which provides an original approach to measure the economic and product complexities based on a network view of international trade data. As a result of their methodology, two measures are presented: $(i)$ the Economic Complexity Index ($ECI$); and $(ii)$ the Product Complexity Index ($PCI$). The main idea is based on the argument that the economies can develop products close in terms of required capabilities (i.e., in terms of human capital, know-hows, job-market, regulations etc.) to those they currently produce~\cite{hidalgo2007product}. This theory accounts for the process by which countries accumulate capabilities; indicating that development efforts should focus on generating the conditions that would allow complexity to emerge to generate sustained growth and prosperity. The authors suggest that these complexity indicators (i.e., $ECI$ and $PCI$) can capture information not only related to the countries' current income, but also can predict future growth.

Several follow-up studies investigated alternative methods, which correct conceptual weaknesses of the $MR$~\cite{caldarelli2012network,cristelli2013measuring,tacchella2013economic} or as a mean for measuring knowledge diffusion~\cite{bahar2014neighbors}. However, rigorous empirical validation of the indicators properties and developing frameworks and guidelines for policy-makers to better design industrial policies, remains an understudied area.

In this work, we investigate the main features of the $ECI$ and highlight the ability of this single measure to capture many information regarding the capabilities of a country. We also find that the $ECI$ index is consistently associated with higher income growth and that this relationship persists even when we include detailed controls for the country's population, life expectancy, years of education (i.e., primary and secondary education) and the economy's size. Therefore, the robustness of the measure should make the indicator more appealing to empirical economists. In addition, we show that the $ECI$ can best predict long-term growth rather than short-term one, which can be relevant to policy-makers.

\subsection*{Preliminaries}
This section explains how the $ECI$ is computed followed the procedure proposed by~\cite{hidalgo2009building} and~\cite{hausmann2014atlas}. We also describe the data used in order to conduct this work.

\subsubsection{Computing the Economic Complexity Index}
The first step in computing the $ECI$ is to use Balassa's revealed comparative advantage~\cite{balassa1965trade}, denoted by $RCA_{c,p}$, which is computed as follows:

\begin{equation}
	RCA_{c,p} = \frac{x_{c,p}}{\sum_P x_{c,p}} \Big{/} \frac{\sum_C x_{c,p}}{\sum_{c,p} x_{c,p}},
\end{equation}

where $x_{c,p}$ is the export value of product $p \in P$ by country $c \in C$. Therefore, the $RCA_{c,p}$ can be seen as the importance of a product $p$ in country $c$ total exports relative to the importance of the product across countries. After computing the $RCA_{c,p}$, it is possible to construct a matrix of countries (rows) and products (columns), denoted by $M_{c,p}$. Hidalgo et al.~\cite{hidalgo2009building} uses $R^* = 1$ as a threshold for the $RCA_{c,p}$, where the matrix entry $M_{c,p} =1$, when $RCA_{c,p} \ge R^*$ and $RCA_{c,p} = 0$ otherwise. This means that we will only consider those products that have a higher or equal weight in the country's export than in the global trade.

From the $M_{c,p}$ matrix, one can compute the complexity of a product $k_p$ and the complexity of a country's economy $k_c$.

\begin{equation}
	\label{eq:products}
	k_{p,n} = \frac{1}{k_{p,0}} \sum_{c=1}^{N_c} M_{c,p} . k_{c,n-1},
\end{equation}

\begin{equation}
	\label{eq:countries}
	k_{c,n} = \frac{1}{k_{c,0}} \sum_{p=1}^{N_p} M_{c,p} . k_{p,n-1},
\end{equation}

where $n$ is the number of iterations used to define the indicators. This recursive relationship would ensure getting rid of distortionary effects. Therefore, $k_{c,n}$ converges to economic complexity index $ECI$ and $k_{p,n}$ converges to the Product Complexity Index $PCI$, for large $n$ values.

\begin{figure}[h!]
	\centering
	\includegraphics[width=1\columnwidth]{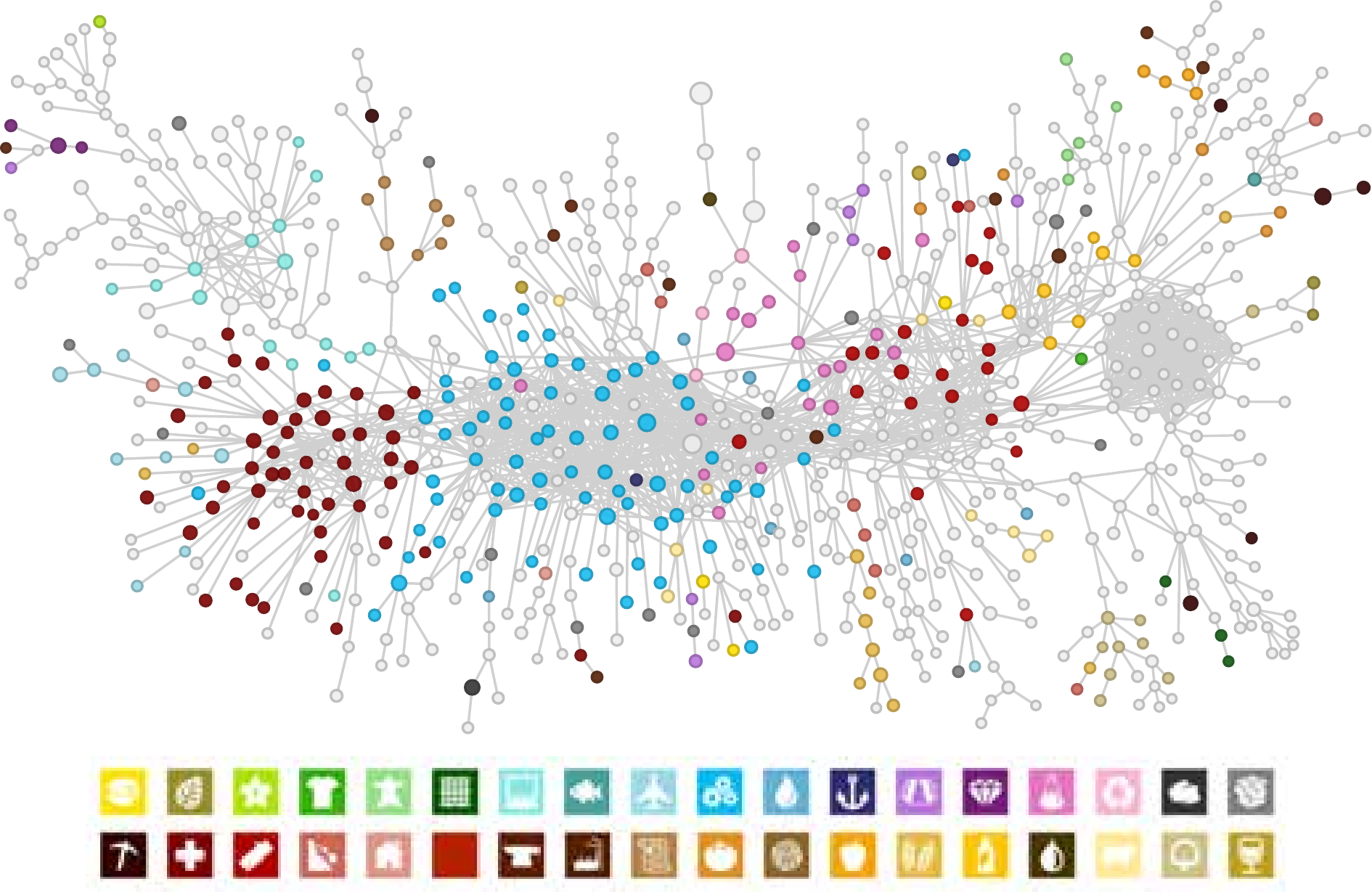}
	\caption[]{Illustration of the product space for the United States connecting 775 products based on their proximity matrix. The color of the node represents the product classification. The visualization is based on the Observatory of Economic Complexity\footnotemark.}
	\label{fig:product_space}
\end{figure}

Conceptually, equation~\ref{eq:products} captures the intuition that the number of countries that can make a product can tell us something about the variety of productive knowledge that a product requires. Similarly, equation~\ref{eq:countries} indicates that the number and type of products that a country makes can reveal something about the productive knowledge that a country has. This insight greatly facilitates the analysis because, while it is difficult to measure productive knowledge directly, it is easy to measure the number of products that a country makes, and how many countries make a given product. Therefore, the recursive definition in equations~\ref{eq:countries}and ~\ref{eq:products} $ECI$ can be simplified and mathematically defined as the eigenvector associated with the second largest eigenvalue a of a matrix connecting countries to countries, which is a projection of the matrix connecting countries to the products they export. Further mathematical derivation and definitions can be found in~\cite{hidalgo2007product,hidalgo2009building,hausmann2014atlas}.  An illustration of the full product proximity network (i.e., product space), which is used to compute the $ECI$ is shown in Fig~\ref{fig:product_space}.
\footnotetext{https://atlas.media.mit.edu; accessed on 20 August 2015}

In this work, we focus on deepening our understanding of the $ECI$ only, and examine its association with short and long term economic growth.

\subsubsection{Datasets}

All auxiliary data comes from the World Development Indicators (WDI) as reported by the World Bank~\footnote{http://data.worldbank.org/; accessed on 20 April 2015}, except data on per capita GDP at Purchasing Power Parity (PPP) for which we use the International Monetary Fond (IMF)~\footnote{http://www.imf.org/external; accessed on 25 March 2015.}.

\subsection{Dimensions of Economic Complexity}
\begin{figure}[h!]
	\centering
	\includegraphics[width=1.0\columnwidth]{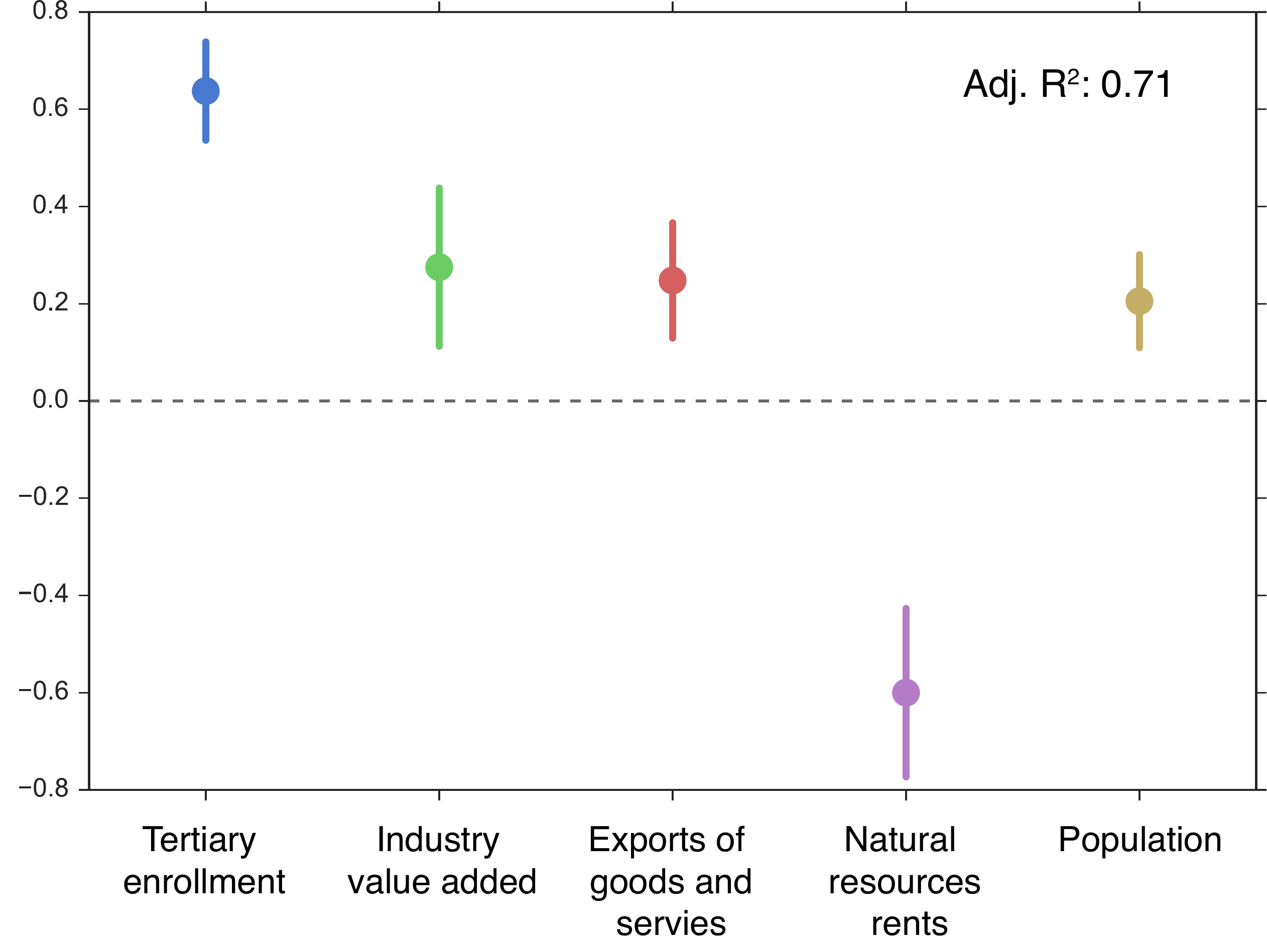}
	\caption{Economic Complexity Index (ECI) captures many dimensions of economic complexity. The plot shows the mean effect size of the covariates (solid circles) and the 95\% confidence intervals (bars). The table of the results is included in the Appendix}
	\label{fig:eci_dimensions}
\end{figure}

Previous work~\cite{ourens2013can, caldarelli2012network} has indicated the difficulty of interpreting the $ECI$ for high iterations. As the MR method used to compute the $ECI$ from the trade data gets rid of distortionary effects and reach an evaluation of economic complexity, one would expect it to be explained by different kinds of capabilities. Figure~\ref{fig:eci_dimensions} shows the results of OLS estimation with $ECI$ as a dependent variable and including covariates that can account for diverse economic dimensions, namely:

\begin{itemize}
	\item \emph{Tertiary enrollment}: the total enrollment in tertiary education, regardless of age, expressed as a percentage of the total population. One would expect that high economic complexity will generate new industries and encourage people to participate in tertiary education.
	\item \emph{Industry value added}: the net output of a sector after adding up all outputs and subtracting intermediate inputs. This would capture the accumulated capabilities  expressed in the country's industrial composition.
	\item \emph{Exports of goods and services}: computed as a percentage of the GDP and includes the value of merchandise, freight, insurance, transport, and other services, such as communication, construction, financial, personal, and government services. The measure excludes compensation of employees and investment income. As the $ECI$ is computed based on export data, one would expect that the exports of goods and services would explain a large proportion of the variance.
	\item \emph{Natural resources rents}: the sum of oil rents, natural gas rents, coal rents (hard and soft), mineral rents, and forest rents~\cite{jarvis2011changing}. Previous work (i.e.,~\cite{hidalgo2009building,hausmann2014atlas}) showed that countries with high $ECI$ tend to have comparative advantage on non-natural resource products~\cite{hidalgo2007product}. Therefore, one would expect negative coefficient on the natural resource rents.
	\item \emph{Population}: Finally, the logarithm of the population as a control for the size of the economy.
\end{itemize}

\begin{table}[h]
	\begin{center}
		\begin{tabular}{l l }
			\hline
			& Model 1 \\
			\hline
			(Intercept)                             & $0.05 \; (0.06)$        \\
			Tertiary education               & $0.64 \; (0.06)^{***}$  \\
			Industry value added            & $0.28 \; (0.10)^{**}$   \\
			Exports of goods and services & $0.25 \; (0.07)^{***}$  \\
			Natural resources rents         & $-0.60 \; (0.10)^{***}$ \\
			Population                        & $0.21 \; (0.06)^{***}$  \\
			\hline
			R$^2$                                   & 0.73                    \\
			Adj. R$^2$                              & 0.71                    \\
			Num. obs.                               & 95                      \\
			\hline
			\multicolumn{2}{l}{\scriptsize{$^{***}p<0.001$, $^{**}p<0.01$, $^*p<0.05$}}
		\end{tabular}
		\caption{Economic Complexity Index (ECI) captures many dimensions of economic complexity. The table shows the effect size of the covariates and Standard Errors in brackets for 95 countries in 2005.}
		\label{table:coefficients}
	\end{center}
\end{table}

We find that the $ECI$ is remarkably capturing rich information regarding an economy's complexity. This result provide evidence supporting the hypothesis that $ECI$ captures many more dimensions of economic complexity since all covariates are significant and yet part of the $ECI$ variation is explained by unobservables. Figure~\ref{fig:eci_dimensions} shows that human capital and manufactured capital (expressed by the tertiary enrollment and industry value added, respectively) are both positive and statistically significant ($p < 0.05$). On the other hand, the natural capital is negative and significant, as one would expect from the $ECI$ theory. This result provides an empirical evidence that the theory captures the specialization a country has. In the traditional theory of growth~\cite{ramsey1928mathematical}, a country that is rich with natural resource (i.e., low $ECI$) has the same potential of economic growth of a country with more industrial reliance. Therefore, a country like Saudi Arabia---one of the world's largest producer of petroleum---should be de-incentivised to diversify its economy. However, as we proceed to show, the type of specialization captured by the $ECI$ is a significant and robust indicator of future growth.

\subsection*{Robustness of ECI}
Hidalgo and Hausmann~\cite{hidalgo2009building} have shown that the ECI is correlated with country's level of income and that deviations from this relationship are predictive of future growth. In this section we study this hypothesis more thoroughly: $(i)$ we test whether the predictive power of ECI is robust to the inclusion of different control variables, and $(ii)$ we perform matching on country's background characteristics to eliminate model dependence and shed some light on a possible casual relationship.

\subsection*{Simple Regression}
We start by replicating Hidalgo and Hausmann's findings, focusing on four 5-year periods between 1985 and 2005 (85-90, 90-95, 95-00, 00-05). We regress the log of the income growth on the ECI, controlling for the initial level of GDP per capita:
\[ \log \bigg(\frac{\text{GDP($t + \Delta t$)}}{\text{GDP($t$)}}\bigg) = \beta_0 + \beta_1 \log(\text{GDP($t$)}) + \beta_2 \text{ECI($t$)} \]
We find that ECI is highly significant predictor of the rate of economic growth ($p \ll 0.01$) and that larger ECI corresponds to higher growth rate (Table~\ref{tab:regressions}, Model 1). This is consistent with \cite{hidalgo2009building}. 

\subsubsection*{Including Control Variables.}
Next, we test whether ECI is still a robust predictor of economic growth when controlling for different background characteristics of a country. We consider: population size, life expectancy, and years of education. Table~\ref{table:vars} shows the distributions and summary statistics of these variables. Again, we regress the log of the income growth on the ECI, now controlling for the initial values of all variables: 
\begin{align*} 
\log \bigg(\frac{\text{GDP($t + \Delta t$)}}{\text{GDP($t$)}}\bigg) &= \beta_0 + \beta_1 \text{ECI($t$)} + \beta_2 \log(\text{GDP($t$)}) \\
& + \beta_2 \log(\text{population($t$)}) + \beta_3 \text{life exp($t$)} \\
&+ \beta_4 \text{years of education($t$)}.
\end{align*}
ECI is still a significant predictor of the rate of economic growth ($p = 0.03$), with a positive coefficient (Table~\ref{tab:regressions}, Model 2). Also, life expectancy is a significant predictor. At the end of this section, we show how that influences the quantity of interest.

\subsubsection*{Matching}
\emph{CEM.} 
To control for the confounding influence of pretreatment control variables we perform Coarsened Exact Matching~\cite{iacus2011causal}. CEM rest on the observation that continues variables have natural breaking points and that they can be coarsened so that substantively indistinguishable values are grouped and assigned the same numerical value. Then, exact matching (i.e., matching treatment and control units that have the exact same values of the covariates) is applied to the coarsened data to determine the matches and to prune unmatched units. Finally, the coarsened data are discarded and the original (uncoarsened) values of the matched data are retained.

\emph{Coarsening.}
We first choose a reasonable coarsening for each variable. Table~\ref{table:vars} column distribution shows the distribution of each variable, where the colors correspond to coarsened regions. We divide ECI (the treatment variable) in there equal frequency bins: low [-2.8, -0.6), medium [-0.58, 0.4), and high [0.4, 2.4). For GDP (log) we use the following cutpoints: [5.18, 7.58), [7.58, 8.73), [8.73, 10.92). We divide the population size (log) in three bins: [12.8, 15.6), [15.6, 16.8), [16.8, 21.0). For life expectancy we use: [40.8, 64.3), [64.3, 73.4), [73.4, 81.1). Finally, we divide years of education in two bins: (9, 11] and (12, 13]. The initial multidimensional imbalance is $\mathcal{L}_1=0.906$. 

\begin{table*}[t!]
	\centering
	\begin{tabular}{ l c c c l }
		\toprule
		\textbf{Variable} & \textbf{Min} & \textbf{Median} & \textbf{Max} & \textbf{Distribution} \\
		\midrule
		ECI & -2.78 & 0.04 & 2.41& \includegraphics[width=0.30\columnwidth, trim=10mm 26mm 10mm 19mm, clip=true]{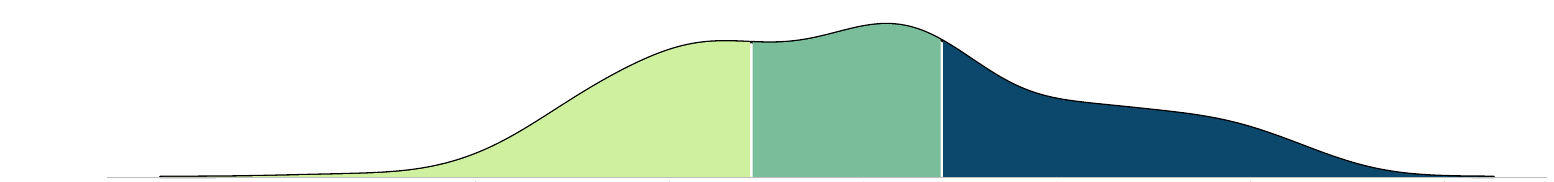} \\
		GDP-PPP (log) \qquad \qquad \qquad \qquad& 5.18 & 8.41 & 10.92 & \includegraphics[width=0.30\columnwidth, trim=10mm 26mm 10mm 19mm, clip=true]{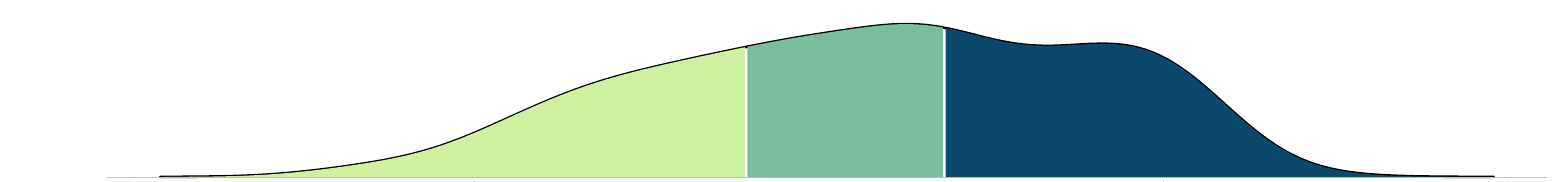} \\
		Population (log) & 12.8 & 16.1 & 21.0 & \includegraphics[width=0.30\columnwidth, trim=10mm 26mm 10mm 19mm, clip=true]{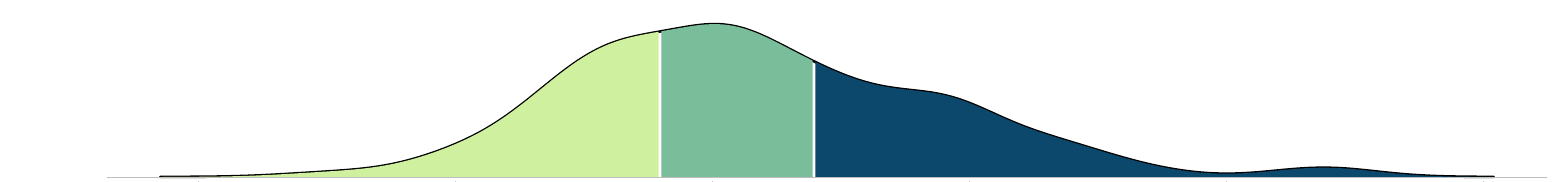} \\
		Life expectency & 40.8 & 69.4 & 81.1 & \includegraphics[width=0.30\columnwidth, trim=10mm 26mm 10mm 19mm, clip=true]{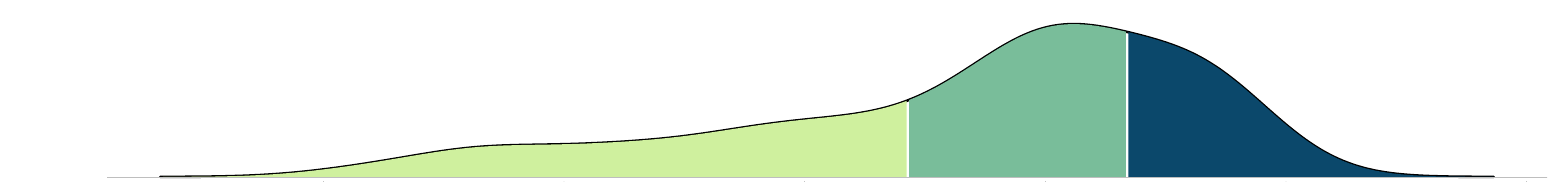} \\
		Years of education & 9.0 & 12.0 & 13.0 & \includegraphics[width=0.30\columnwidth, trim=16mm 26mm 15mm 22mm, clip=true]{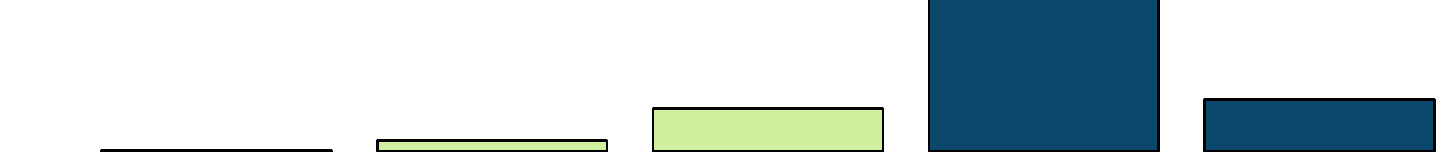} \\
		\bottomrule
	\end{tabular}
	\caption{Statistics of the Economic Complexity Index (treatment variable) and four control variable capturing the background characteristics of a country. The different colors in the distribution plots corresponded to the coarsened values used for matching. The data consist of four 5-year increments between 1985 and 2005, resulting in total of 446 observations.}
	\label{table:vars}
\end{table*}

After running CEM with the above coarsening, we obtain $m_{\text{ECI=low}} = 54$ units with low ECI, matched with $m_{\text{ECI=medium}} = 107$ units with medium ECI, and $m_{\text{ECI=high}} = 67$ units with high ECI. This results in total of 228 matched and 218 unmatched units. The multidimensional imbalance after matching is $\mathcal{L}_1=0.895$, corresponding to 1.23\% imbalance reduction.

\emph{Estimation.}
Next, we estimate the causal effect under CEM. We compute the FSATT (Feasible Sample Average Treatment on the Treated), using the CEM-matched units via linear regression using all variables in the data set. We obtain the following estimates: $\tau_{low \rightarrow medium} = 0.104$ (i.e., 10.4\% growth), $\tau_{medium \rightarrow high} = 0.0376$ (i.e., 3.76\% growth), and $\tau_{low \rightarrow high} = 0.144$ (i.e., 14.4\% growth). Table~\ref{tab:regressions} (Model 3), shows the results of the regression; both ECI-medium and ECI-high are statistically significant.

\begin{table}[h!]
	\begin{center}
		\scalebox{1}{
			\begin{tabular}{l D{)}{)}{13)3}@{} D{)}{)}{13)3}@{} D{)}{)}{13)2}@{} }
				\toprule
				& \multicolumn{1}{c}{Model 1} & \multicolumn{1}{c}{Model 2} & \multicolumn{1}{c}{Model 3} \\
				\midrule
				(Intercept)          & 0.467 \; (0.068)^{***}  & 0.317 \; (0.182)        & 0.073 \; (0.356)       \\
				GDP per capita (log)             & -0.031 \; (0.008)^{***} & -0.061 \; (0.011)^{***} & -0.049 \; (0.017)^{**} \\
				Population (log)          &                         & 0.005 \; (0.005)        & 0.020 \; (0.008)^{*}   \\
				Life Expectancy     &                         & 0.008 \; (0.001)^{***}  & 0.008 \; (0.003)^{**}  \\
				Years of Education &                         & -0.017 \; (0.010)       & -0.030 \; (0.018)      \\
				ECI                  & 0.050 \; (0.010)^{***}  & 0.023 \; (0.011)^{*}    &                        \\
				ECI Medium           &                         &                         & 0.083 \; (0.026)^{**}  \\
				ECI High           &                         &                         & 0.089 \; (0.029)^{**}  \\
				\midrule
				R$^2$                & 0.058                   & 0.131                   & 0.204                  \\
				Adj. R$^2$           & 0.054                   & 0.121                   & 0.183                  \\
				Num. obs.            & 446                     & 446                     & 228                    \\
				\bottomrule
				\multicolumn{4}{l}{\scriptsize{$^{***}p<0.001$, $^{**}p<0.01$, $^*p<0.05$}}
			\end{tabular}
		}
		\caption{Robustness of the Economic Complexity Index, where the dependent variables is the log of the economic growth.}
		\label{tab:regressions}
	\end{center}
\end{table}

\begin{figure}[t!]
	\centering
	\includegraphics[width=1.0\columnwidth]{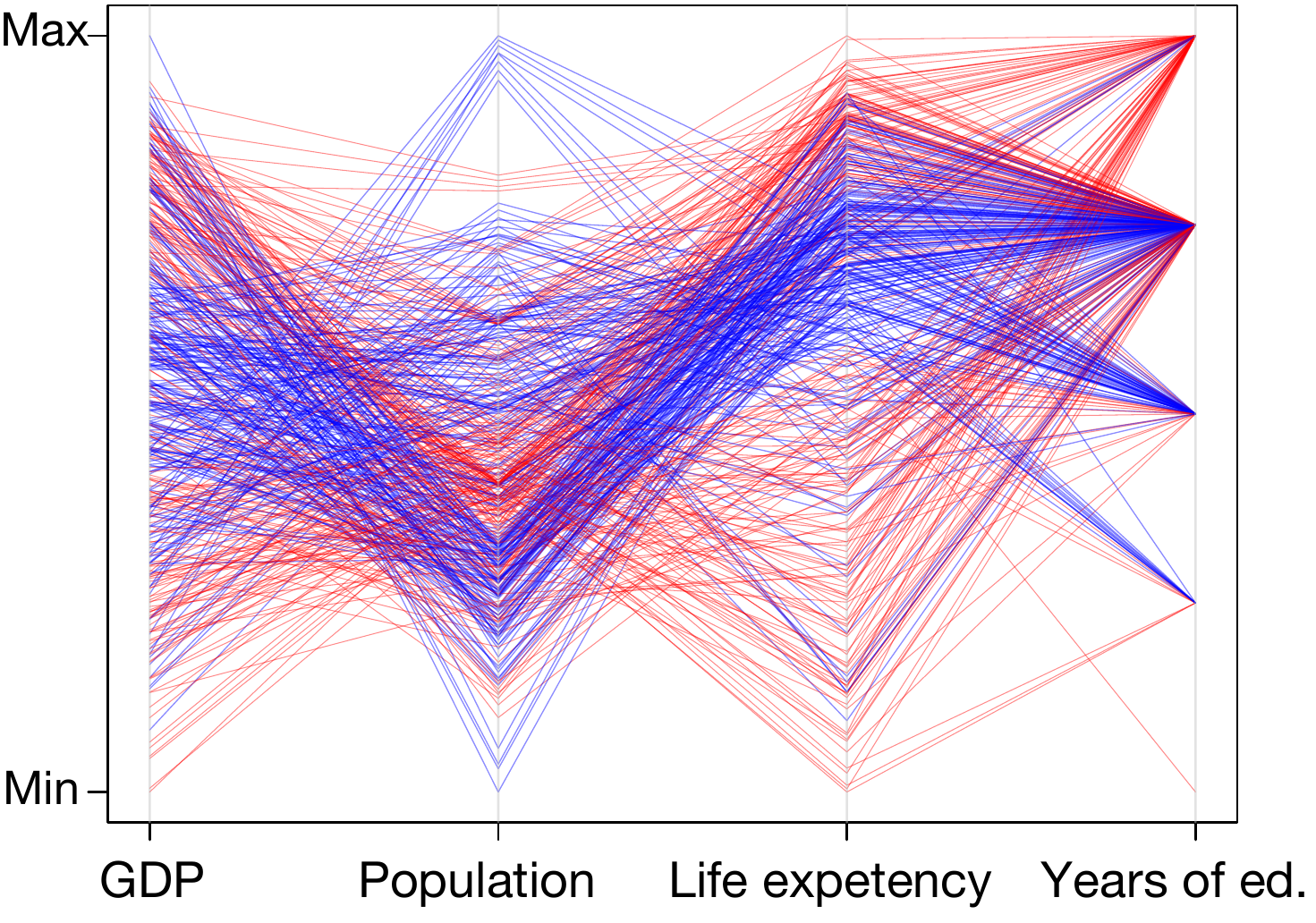}
	\caption{Parallel plot of numerical variables for CEM matched units (blue) against unmatched units (red).}
	\label{fig:parallel_plot}
\end{figure}

\emph{Defining the FSATT.}
Following~\cite{iacus2011causal}, we define the estimands via a parallel plot (Figure~\ref{fig:parallel_plot}). We represent each observation in the original data set by a single line that traces out its values on each of the continuous variables (horizontally) between its minimum and maximum values (vertically). The matched units are colored blue, while the unmatched units are colored red. The estimands are the average treatment effects for countries with different levels of GDP, population size, and years of education, but mostly high levels of life expectancy. 

\begin{figure}[t!]
	\centering
	\includegraphics[width=1.0\columnwidth]{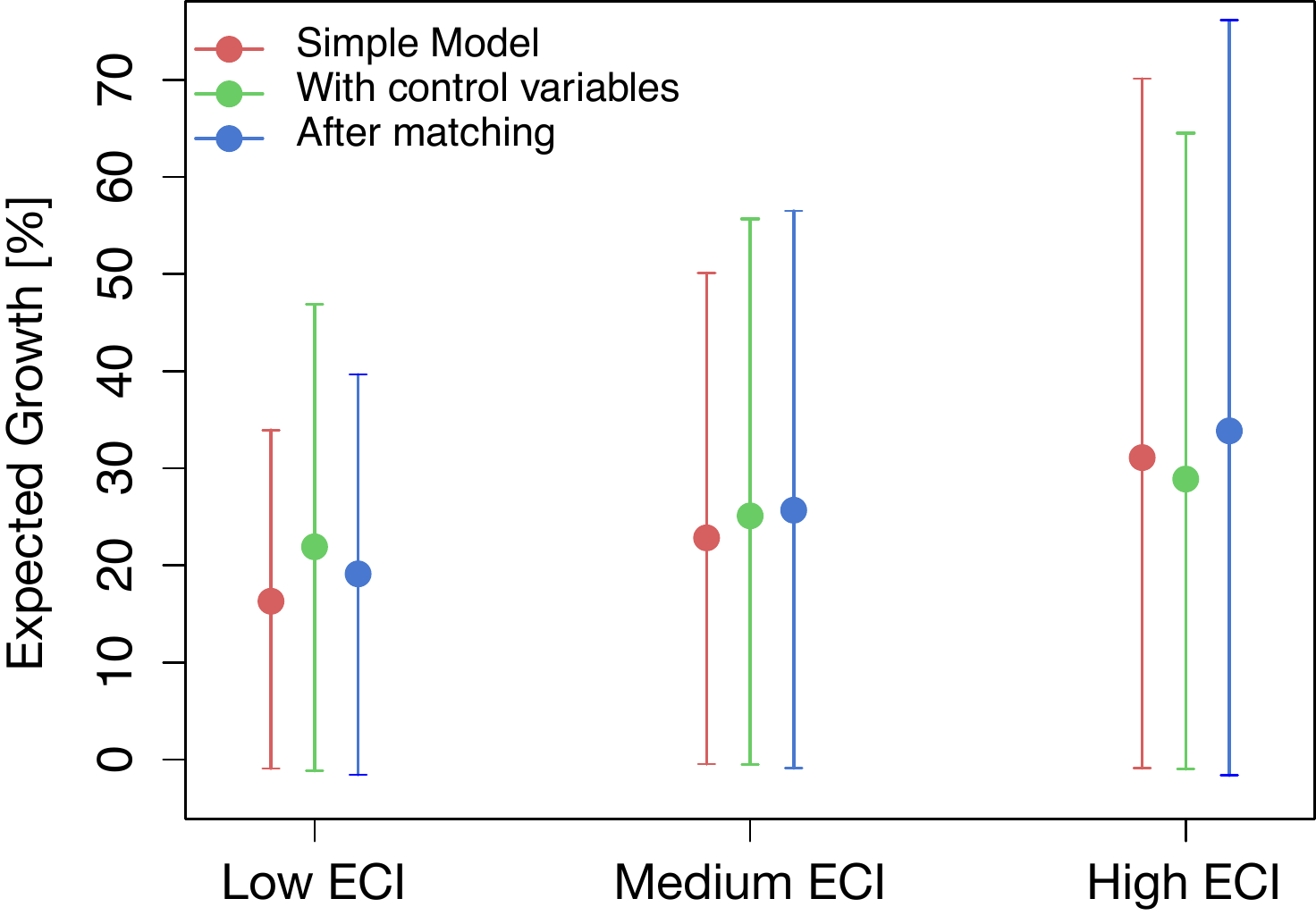}
	\caption{Expected growth of a median country with low, medium, and high ECI. 
The different lines correspond to different models. Red: Growth ~ ECI + GDP (Simple Model). Blue: Growth ~ ECI + GDP + Population + Life expectancy + years of education (With control variables). Red: applying the the same model, but using Coarsened Exact Matching as a preprocessing step (After matching).}
	\label{fig:matching_sims}
\end{figure}

\emph{Comparison between models.}
We compare the estimates of expected growth for a median country produced by the different models. 
We set $x$ to the median values for GDP, population, life expectance and year of education, and we vary the ECI, setting it to the mean value of each category (low: -1.13, medium: -0.05, high: 1.22). 
We perform 1000 simulations to incorporate the estimation and fundamental uncertainty. Figure~\ref{fig:matching_sims} shows the expected values and the 5\% confidence intervals. 
There are two things to note: $(i)$ the estimates of all models are very similar, even without including control variable and matching we obtain reliable estimates; 
$(ii)$ in all cases larger ECI results in increased economic growth, confirming our findings in the previous section.

\subsection*{Out of Sample Prediction of Growth}
In the Section ``Robustness of ECI,'' we have shown that the ECI is a significant predictor of the economic growth in the subsequent five years, even when controlling for different variables and matching. In this section, we test the out-of-sample predictive power of the ECI over different time increments. 

\begin{figure*}[t!]
	\centering
	\includegraphics[width=1\columnwidth]{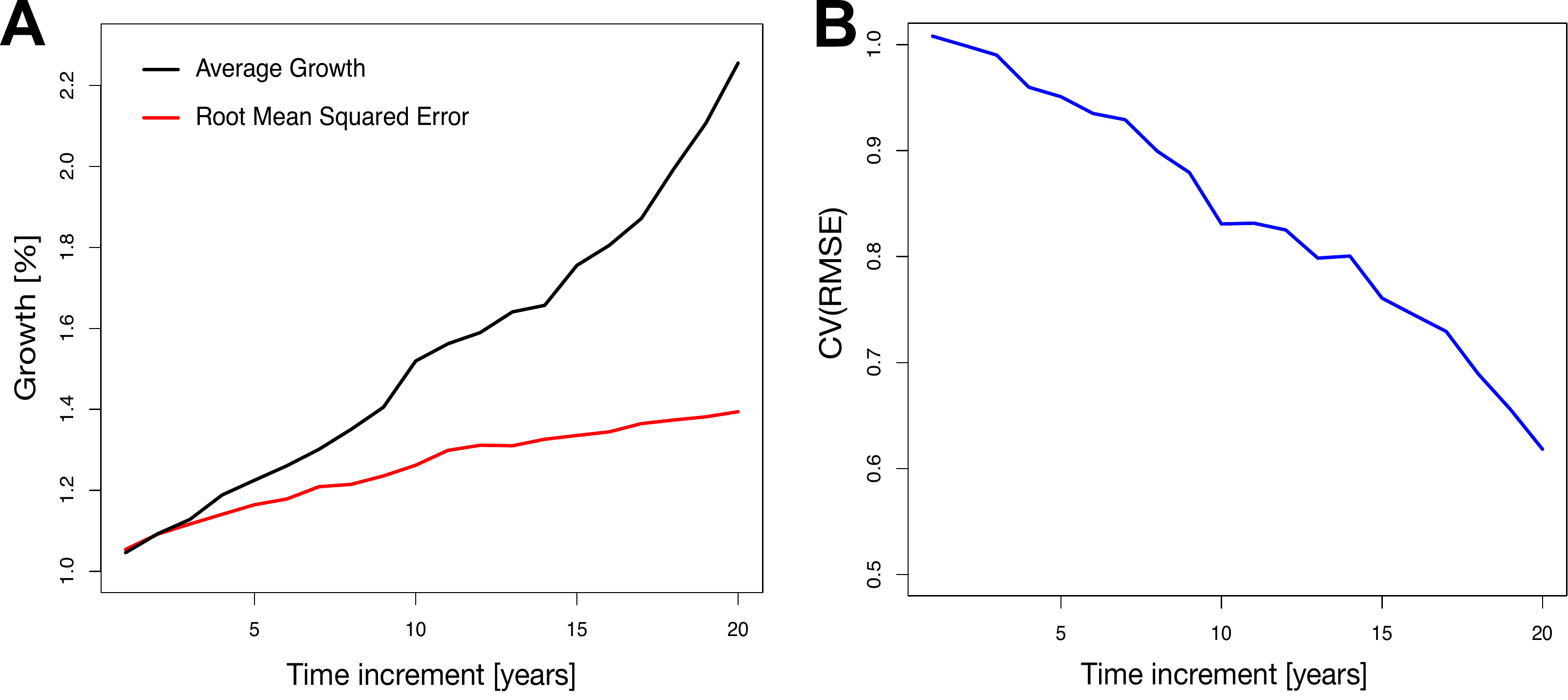}
	\caption{Panel~\textbf{A} shows the Root Mean Squared Error (RMSE) and Average Growth. Panel~\textbf{B} depicts the Coefficient of Variation of the RMSE (i.e., RMSE / Average growth) for different time increments. The CV(RMSE) measure the out-of-sample predictive power of ECI and is comparable across time increments.}
	\label{fig:rmse}
\end{figure*}

We consider time increments between one and twenty years in the period from 1985 to 2005. For each increment we create a dataset of all (non-overlapping) occurrences of the increment. For instance, to build the dataset for ten year increments, we consider the growth between 1985-1995, and between 1995-2005.
We perform 10-fold cross-validation on each of these datasets as follows. $(i)$ We randomly partition the dataset in ten equal size subsets, we pick one subset and we set it aside, $(ii)$ using the remaining nine subsets, we build a linear regression model of the economic growth (between the start and the end of the period) as a function of the ECI at beginning of the period; $(iii)$ we use the model to predict the remaining subset and we measure the prediction error, repeating this procedure for every subset.

To measure the prediction error, we use \textit{Root Mean Square Error}:
\[ \text{RMSE} = \sqrt{\frac{\sum_{i=1}^{n} (p_i - y_i)^2}{n}}, \]
where $p_i$ is the predicted growth, $y_i$ is the observed growth, and $n$ is the number of observations.
However, RMSE is not comparable across time increments, due to differences in magnitude of growth. For instance, the average growth of a 20-year increments is much higher then the 5-year increments, and although the RMSE of the 5-year increments maybe smaller, the error maybe more significant relative to the average growth. Thus, to compare the errors between different time increments, we compute the coefficient of variation (CV) of the RMSE, i.e., we normalize by the average growth of the increment:
\[ \text{CV(RMSE)} = \frac{\text{RMSE}}{\bar{y}} \]
where $\bar{y}$ is the average growth.

We find that the ECI is more predictive of the economic growth in the long run and has limited predictive power for smaller time increments (Figure~\ref{fig:rmse}, left). 
When predicting the growth in the next year the ECI maybe off by 5\% (Figure~\ref{fig:rmse}, right), as much as the average growth for one year increments. 
In contrast the prediction error for five and ten year increments is 16\% and 26\%, or 95\% and 83\% of the average growth, respectively. 
Finally, ECI is most predictive on the twenty year increments being off by only 39\% or 62\% of the average growth.

Note that the errors reported are obtained using only the ECI as the explanatory variable.


\subsection*{Return on Investment} 
Policymakers are often faced with the choice of which products or industries their country should invest in to generate sustained economic growth and prosperity. While, it is easy to predict how investing in a new product will increase the ECI of a country, it is unclear how this investment will influence its future growth. In this section, we perform simulations to help policymakers understand how increase in ECI affects both long and short term economic growth. 

We collect data on the economic complexity and background characteristics of all countries in the period between 1985 and 2005 (see Subsection ``Datasets'') and we consider time increments between one and twenty years. For each time increment, we build a linear regression model of the economic growth between the start and end of the period as a function of the initial values of GDP, population size, life expectancy, and years of education. 

We focus on the expected economic growth of a median country. We set the initial values of GDP, population size, life expectancy, and years of education to the median values for the time increment and we compute the first differences between low (20\% quantile) and high (80\% quantile) ECI. Using Zelig~\cite{imai2009zelig}, we perform 1000 simulations to account for the fundamental and estimation uncertainty.

\begin{figure}[t!]
	\centering
	\includegraphics[width=0.85\columnwidth]{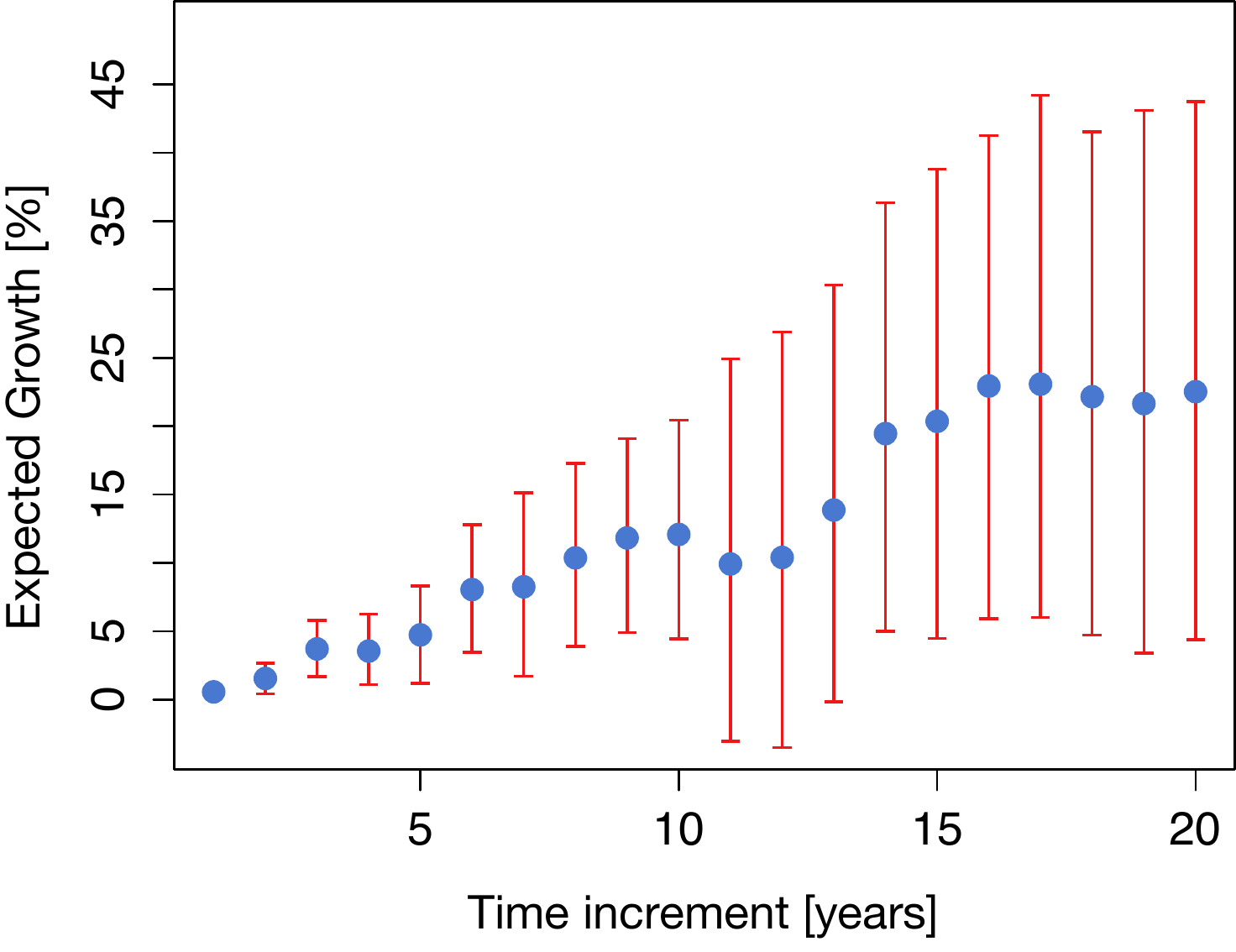}
	\caption{Estimate of the first differences in expected growth between high and low ECI for a median country. The bars correspond to 5\% confidence intervals obtained from 1000 simulations.}
	\label{fig:future_growth}
\end{figure}

Figure~\ref{fig:future_growth} shows the estimations of the expected growth and the 5\% confidence intervals. As one may expect, the expected economic growth in the next year is small (i.e., 1\%). An increase in the ECI results in expect growth of 5\% and 12\% in the next five and ten years, respectively. 
In the long term, an increase in the economic complexity is expected to generate a growth of 20\% and 23\% in the following 15 and 20 years, respectively.

Note that time increments bigger than ten years have larger confidence intervals. This is because we have less observations for these increments (e.g., there is one 15-year time increment between 1985 and 2005, but there are four 5-year increments).




\subsection{Summary and Discussion}
Many published work have discussed the relationship between a country's current production status and how this effects the long term growth possibilities. More recently, Hidalgo and Hausmann~\cite{hidalgo2009building} has introduced a very robust and useful tool to compute the \textit{economic complexity} of a country and the sophistication of a product. In this work, we investigate the main features of the $ECI$ and show that this single measure can capture many information regarding capabilities. We also find that the economic complexity measure significantly explain current per capita GDP. Through rigorous tests of the robustness of the measure, we find this significant statistical relationship to persists even when we include detailed controls for the country's population, life expectancy, years of education and the size of the economy. Finally, we find that the complexity measure highly predicts future long-term growth. This conclusion is supported by a wide body of literature in economic growth theory which states that the relationship between production and growth materializes over the long term. 

Finally, combining our results with the previous work in the literature can provide policy-makers with detailed insights to better design industrial policies that help economies diversify in a systematic way and accumulate important capabilities, which would generate sustained growth and prosperity.

%
%
%
%
%


\section*{Conclusion}
The this thesis highlighted the need for frameworks that take into account the complex structure of labor market interactions. We provide evidence to support the theory of idea flow, which predicts that the emergence of innovative ideas or the coordination of collective action depends on network structure. In particular, we explored the benefits of leveraging advancements and tools from computational social science, network science, and data-driven theories to  measure the opportunity/information flow in the context of the labor market. 

We started by investigating our key hypothesis, which is that opportunity/information flow through weak ties (at the individual level), and this flow is a key determinant of the length of unemployment. This implies that behaviors that promote social exploration and/or lower the cost of this exploration are key strategies for job-seekers to find employment. We provided evidence relating unemployment to individual behavior by examining the signatures of unemployment through mobile phone usage and constructed a simple model that produced accurate, easily interpretable reconstructions of district-level unemployment. We also investigated the role of social cohesion in increasing the effectiveness of incentives. Finally, we extended the idea of opportunity/information flow beyond job-seekers to clusters of other economic activities. Similarly, we expect the flow within clusters of related activities to be higher than within isolated activities. We quantified the opportunity/information flow using a ``complexity measure'' of two economic activities (i.e. jobs and exports). We found that this measure indeed predicts economic growth.

\chapter*{Appendix}

\section*{Chapter 1 Supporting Information}
\subsection*{Datasets}
For this study, we used anonymized mobile phone meta data known as Call Detail Records (CDRs), combined with records from an unemployment benefit program and census information.
\subsubsection{The CDRs dataset} Consists of one full month of records for the entire country, with 3 billion mobile activities to over $10,000$ unique cell towers, provided by a single telecommunication service provider~\cite{alhasouncity,aleissawired}. Each record contains: i) an anonymized user identifier; ii) the type of activity (i.e., call or data etc); iii) the identifier of the cell tower facilitating the service; iv) duration; and v) timestamp of the activity. Each cell tower is spatially mapped to its latitude and longitude and the reception area is approximated by Voronoi cell. The dataset studied records the identity of the closest tower at the time of activity; thus, we can not identify the position of a user within a Voronoi cell. For privacy considerations, user identification information has been anonymized by the telecommunication operator. Unlike standard CDRs, this dataset does not include the cell tower identity of the receiver end of the activity (i.e., only the location of the caller is approximated). The operator that provided the call data records had around 48\% market share at the time of data acquisition.

\subsubsection{The unemployment benefit program dataset} 
We derive the unemployment information from the data provided by the Ministry of Labor and the Human Resources Development Fund (HRDF) in Saudi Arabia regarding the new social security program, known as ``Hafiz'' (meaning incentive in English). The program provides job-seekers with financial help and to motivate them to try to get permanent jobs. Under this program each beneficiary receives SR 2,000 (\$533) per month~\cite{aldid,ottaway2012saudi}. More than 4M individuals applied and approximately 1.4M beneficiaries were approved, accounting for $\approx 7\%$ of the total national population.. The number varied from month to month as the beneficiaries entered and exited the program due to their gaining and losing the eligibility of receiving the benefit. Those accepted would have to prove they were really looking for employment and be ready to accept training and take offered jobs. If not, they would be dropped from the program after one year~\cite{ottaway2012saudi}. 

Each Hafiz application consists of: 1) personal information such as age, gender, social status and disability; 2) Spatial information including home address at the district level and the city of birth; and 3) educational \& work background. Therefore, from this dataset we derive spatial socio-economic status of the unemployed population at the regional level (i.e., 13 Administrative areas), city level (i.e., 61 cities), and down to the district level (i.e., 1277 districts). In the present work, we focus on the 148 districts within Riyadh, the capital city of Saudi Arabia (KSA). 

As the records contained anonymized applicant information including their home address (down to the district level), therefore, we are able to derive spatial socio-economic status of unemployed populations at the regional level (i.e., 13 Administrative areas), city level (i.e., 61 cities), and down to the district level (i.e., 1277 districts). In the present work, we focus on the 148 districts within Riyadh, the capital city of Saudi Arabia (KSA).

\subsubsection{Census information}
Census data at the levels of the administrative areas and cities were obtained from the Central Department of Statistics \& Information (CDSI)~\footnote{(http://www.cdsi.gov.sa/; accessed December 15th, 2014)}. Riyadh census data was obtained from the High Commission for Development of Arriyadh – or also known as the ArRiyadh Development Authority (ADA) at the Traffic Analysis Zones (TAZs) level. The administrative areas and city level census information were matched using their identifier codes. The district level information was obtained by mapping the TAZ information to the district boundaries. The average spatial resolution (i.e., square root of the land area divided by the number of land units) for the districts and TAZs in Riyadh is $2.6 km$ and $0.04 km$, respectively.

\subsection*{Prediction Directionality of ties}
\label{sec:Predicting Reciprocity}

Previous studies have investigated numerous factors that could have an influence on the reciprocity of friendships. This would include socio-economic status~\cite{ellenbogen1997peer}, gender differences~\cite{eder1978sex} and ethnic or racial origin~\cite{chen1992correlates}.

Here we are interested in the predictability of reciprocity (i.e., reciprocal versus unilateral) and directionality (i.e., incoming versus outgoing edge) of ties based on the topological structure of the underlying \textit{undirected} and \textit{unweighted} social network, independently of additional information such as gender, race, tie strength, etc.
Such additional information is often not available when analyzing communication networks~\cite{eagle2006reality,onnela2007structure}, trust networks~\cite{shmueli2014sensing}, or similar data.
Therefore, it is important to know the structural characteristics that allow effective intervention strategies.

Two possible social factors that can be used to predict the reciprocity and directionality of a friendship tie between two individuals are:
(i) Social Embeddedness: the extent to which their friendship circles overlap; and
(ii) Social Centrality: the difference in their social hierarchical organizational status.

The Social Embeddedness (SE) of a pair of individuals captures the idea that the behavior individuals choose are significantly constrained by the social relations within which they function~\cite{Granovetter85economicaction}. 
Inspired by this idea, we expect social embeddedness to explain reciprocity as an effect of the network transitivity property~\cite{uzzi1997social}.
Here, we use the number of common friends the two individuals share to capture their SE in the network. The finding that friends share more common neighbors has been extensively studied in the literature (e.g.,~\cite{granovetter1973strength,newman2001,onnela2007structure,gilbert2009predicting}).

In this section, we explore the effect of Social Embededdness (SE) and Social Centrality (SC) on the Probability of an ego to form a reciprocal tie or be perceived as a friend (i.e., incoming edge) in the Friends and Family dataset. Control variables include whether the dyad have the same country of origin, ethnicity, and gender. In both cases, we use a logit model for the friendship ties in the network as follows:
\begin{align*} 
\label{eq:probability}
Y_i  &= Bernoulli(\pi_i) \\
\pi_i &= \frac{1}{1+e^{-X_i\beta}}
\end{align*}
Where $X_i$ represents the vector of control variables including whether the dyad have the same country of origin (values 0 or 1), ethnicity (values 0 or 1), and gender (values 0 or 1).

Where for the SE (Recall that $\left\vert \Gamma(v_1) \cap \Gamma(v_2) \right\vert$ is the number of common friends):
\[
X_i\beta  = \beta_0 + X_{i,gender}\beta_1 + X_{i,ethnicity}\beta_2 + X_{i,origin}\beta_3 + X_{i,\left\vert \Gamma(v_1) \cap \Gamma(v_2) \right\vert}\beta_4\\
\]
While for the SC logit model, we incorporate the difference in degree centrality $\Delta C_{d}$:
\[
X_i\beta  = \beta_0 + X_{i,gender}\beta_1 + X_{i,ethnicity}\beta_2 + X_{i,origin}\beta_3 + X_{i,\Delta C_d}\beta_4\\
\]

Table~\ref{table:coefficients} shows the coefficients for both models. In order to compute the expected probabilities of forming reciprocal ties or incoming-ties for each number of common friends or difference in centrality value, we simulate $\tilde{\beta}$ from the sampling distribution of the estimated $\hat{\beta}$:
\[
\tilde{\beta} \sim MVN(\hat{\beta},\hat{V}(\hat{\beta}))
\]
Where $\hat{V}(\hat{\beta})$ is the variance covariance matrix. We then choose one value for the explanatory variable (i.e., specific number of common friends) and hold all other covariates at their median value and donate the vector of values $X_c$. Using $X_c$ and $\tilde{\beta}$, we calculate the systematic component $\tilde{\pi}$ as:
\[
\tilde{\pi} = \frac{1}{1+e^{-X_c\tilde{\beta}}}
\]
Then for each $\tilde{\pi}$ draw, we simulate from the stochastic component:
\[
\tilde{Y_c}  = Bernoulli(\tilde{\pi})
\]
Then we report the mean of these simulations for each $\tilde{\pi}$, $E[Y_C]$ and the 95\% confidence interval using the 2.5\% and 97.5\% quantiles from 10,000 simulations.

\begin{table}[h!]
	\small
	\begin{center}
		\caption{Logit coefficients for the reciprocal ties and incoming unilateral tie.}
		\begin{tabular}{l c c }
			\hline
			& Reciprocal ties & Incoming ties \\
			\hline
			Number of common friends          & $0.19^{***}$  &               \\
			& $(0.02)$      &               \\
			Same gender              & $0.10$        & $-0.03$       \\
			& $(0.19)$      & $(0.21)$      \\
			Same country of origin & $1.08^{***}$  & $0.37$        \\
			& $(0.26)$      & $(0.26)$      \\
			Same religion            & $0.39$        & $0.32$        \\
			& $(0.23)$      & $(0.25)$      \\
			Same ethnic group           & $-0.09$       & $-0.27$       \\
			& $(0.23)$      & $(0.25)$      \\
			$\Delta degree$                        &               & $-3.89^{***}$ \\
			&               & $(1.08)$      \\
			(Intercept)                   & $-2.61^{***}$ & $-0.28$       \\
			& $(0.23)$      & $(0.19)$      \\
			\hline
			BIC                           & 737.81        & 548.05        \\
			Log Likelihood                & -349.26       & -256.18       \\
			Deviance                      & 698.52        & 512.36        \\
			Num. obs.                     & 698           & 383           \\
			\hline
			\multicolumn{3}{l}{\scriptsize{$^{***}p<0.001$, $^{**}p<0.01$, $^*p<0.05$}}
		\end{tabular}
		\label{table:coefficients}
	\end{center}
\end{table}

\begin{figure}[h!]
	\centering
		\includegraphics[width=1\columnwidth]{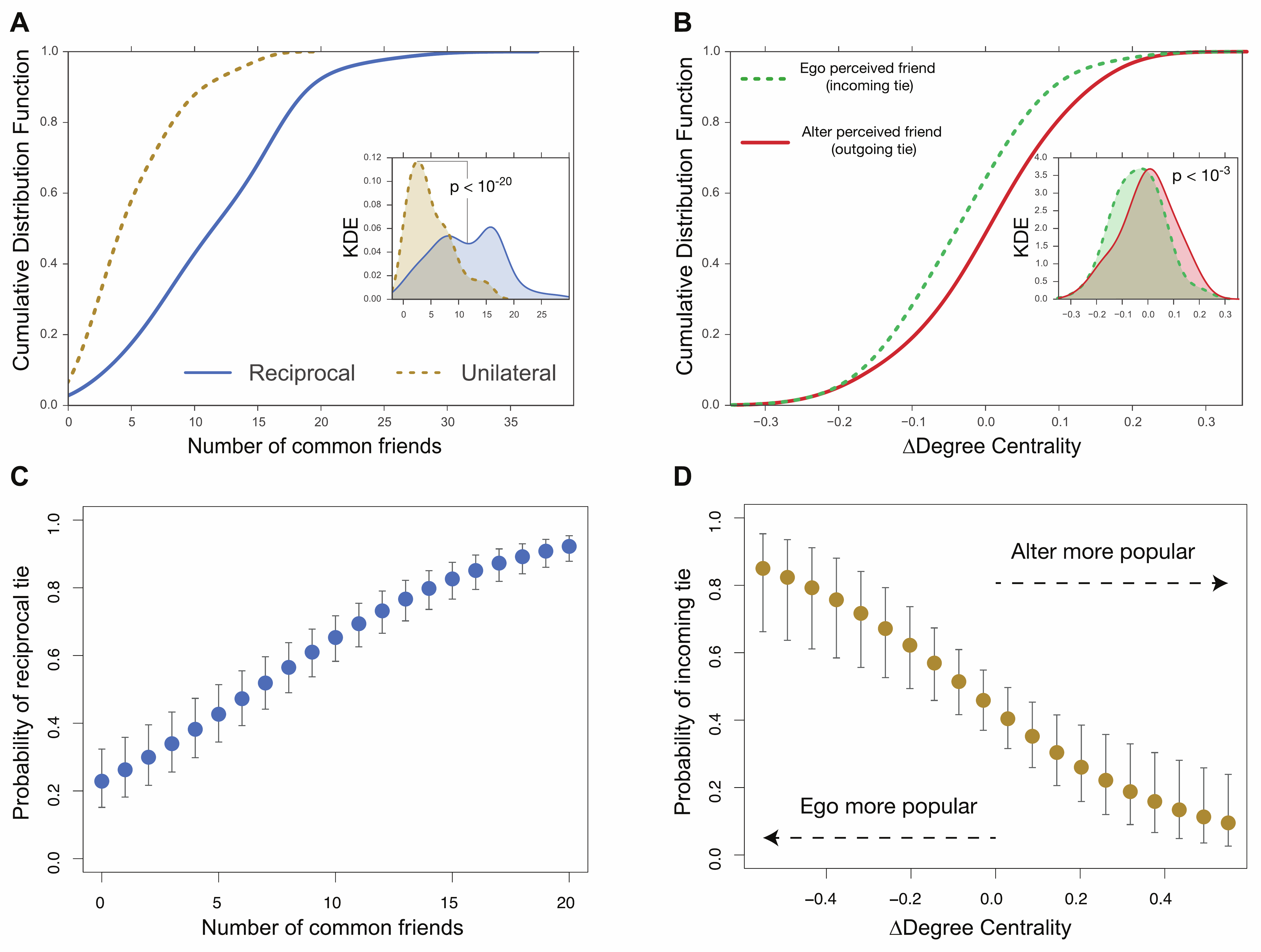}
	\caption{{The effect of Social Embeddendness (SE) and the difference in Social Centrality (SC) on determining the tie type in the Friends and Family dataset.}
		In Panels ({A}) and ({B}), we observe that SE (i.e., represented by the number of common friends) and SC (as $\Delta$ Degree centrality) are good discriminators between reciprocal and unilateral ties, as well as between the two directions of unilateral ties.
		Panels ({C}) and ({D}) show the effect of SE and SC on the Probability of an ego to form a reciprocal tie or be perceived as a friend, respectively.
		Control variables include the ethnic and racial origin as well as religious and gender differences.
		The vertical gray bars for both panels show 95\% confidence intervals based on $10,000$ drawn sets of estimates from the coefficient covariance matrix and with all other covariates held at their median.
	}
	\label{fig3}
\end{figure}

Fig.~\ref{fig3} highlights the effect of SE on the probability of an ego to form a reciprocal tie in the Friends and Family dataset while holding demographic covariates (e.g., gender differences, and ethnic and racial origin) at their median values. This result supports our hypothesis by showing that reciprocal ties exhibit higher average number of common friends.
We have also investigated five additional measures of SE and show that they perform similarly.

We also look at the difference in Social Centrality (SC) as a possible explanation of the directionality of unilateral friendship ties.
The Social Centrality (SC) of an individual captures the idea that hierarchical social organizations are highly characterized by ordered, linearly transitive social relationships~\cite{DavisMoore1945aa,ball2013friendship}.
Therefore, one would expect that the direction of unilateral ties would tend to flow from individuals with lower status to individuals with higher status~\cite{ball2013friendship}.
Here, we use the \textit{Degree Centrality} measure to capture the centrality of a node in the network and show that they generate similar results).
Fig.~\ref{fig3} supports our hypothesis by showing that unilateral ties tend to flow from the less central node to the more central node of the dyad. We find the effect to be persistent after controlling for the demographic attributes. 

\begin{figure}[h!]
	\centering	
	\includegraphics[width=1\columnwidth]{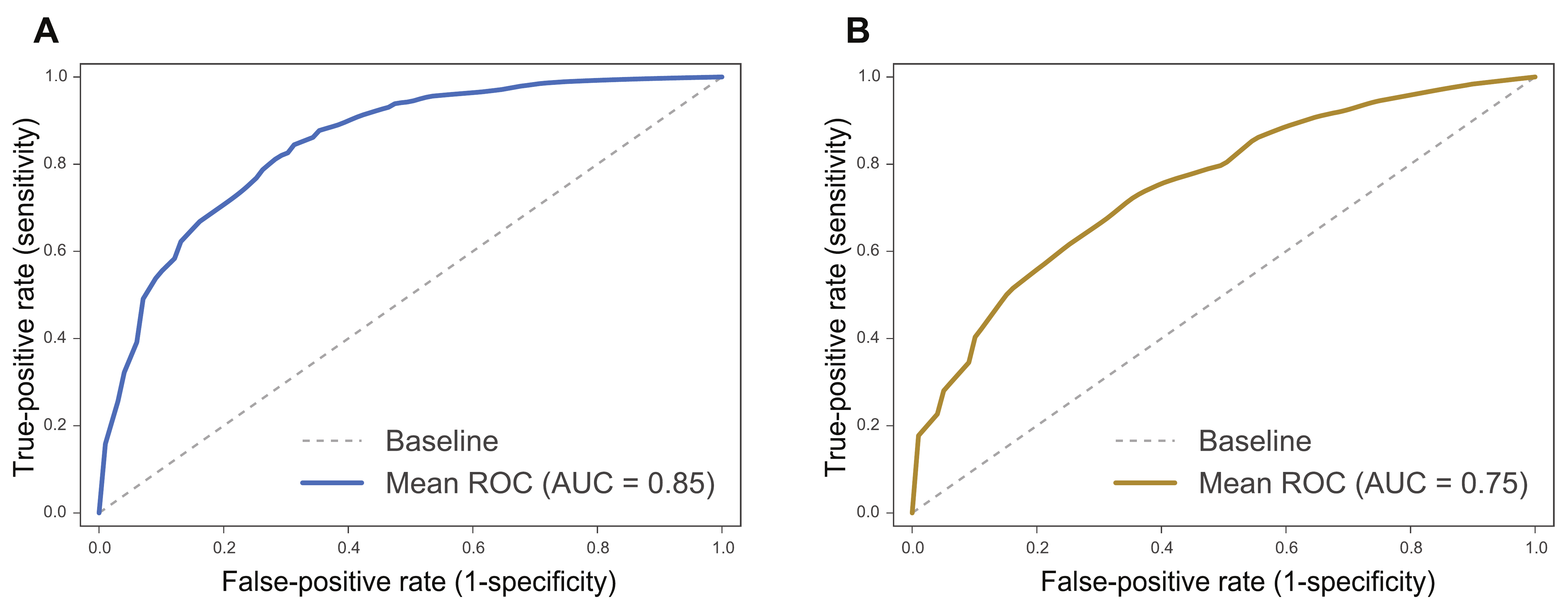}
	\caption{
		Mean ROC curves demonstrating the model performance in predicting ties type. Panel \textbf{(A)} shows the prediction performance for reciprocal ties (AUC = 0.85, 95\% CI: 0.82--0.87). Panel \textbf{(B)} shows the model performance in prediction incoming ties (AUC = 0.75, 95\% CI: 0.70--0.80).}
	\label{fig4}
\end{figure}

The conditional entropy for the two measures, H(reciprocity$\vert$number of common friends) = 1.8 and H(directionality$\vert \Delta$degree centrality) = 2.02568,  which implies good separation of the classes and that good classification is possible. To confirm this, we further analyzed the predictive power of these measures in the two classification tasks: (i) reciprocal vs unilateral ties; and (ii) the direction of unilateral ties. For each classification task, we train and test a classifier in K-fold-cross validation ($K=10$). First, we use a simple Logistic Regression classifier and a single feature for each classification task (i.e. number of common friends for the first task and difference in degree centrality for the second task).
When trying to identify reciprocal ties, the Logistic Regression classifier performed significantly better than the baseline, obtaining $0.81$ AUC (95\% CI: 0.77 -- 0.85).
As for identifying the direction of unilateral ties, the Logistic Regression classifier performed better than the baseline, although the results obtained were less compelling: an average AUC of $0.61$ (95\% CI: 0.58 -- 0.66). These results are consistent with the results of the conditional entropy, where the first task (i.e., predicting reciprocity) is more attainable than the second (i.e., predicting directionality).

Finally, we used all 10 identified features (6 SE features and 4 SC features as described in ~\cite{almaatouq2016b} to train a Random Forest classifier for each classification task, obtaining encouraging results for both classification tasks: an average AUC of 0.85 (95\% CI: 0.82--0.87) for the reciprocal classification task and an average AUC of 0.75 (95\% CI: 0.70--0.80) for the directionality task. The performances of the two classification tasks are reported in Fig.~\ref{fig4}. We further experimented with five additional friendship nomination networks, and both prediction tasks showed similar results .

\begin{table}[h!]
	\caption{Evaluation results of reciprocity and directionality classification on additional datasets.}
	\begin{tabular}{l | c c | c c}
		\hline
		\multirow{2}{*}{Dataset} & \multicolumn{2}{c|}{Reciprocal vs Unilateral}  & \multicolumn{2}{c}{Incoming vs Outgoing}\\
		\cline{2-5}
		& Average AUC & 95\% C.I. & Average AUC & 95\% C.I.\\
		\hline
		Friends and Family & 0.85 & [0.82--0.87] 		& 0.75  & [0.70--0.80]\\
		Strongest Ties		& 0.78		& [0.72--0.83]	& 0.80	&	[0.72--0.88]\\
		Reality Mining		& 0.57		&	[0.46--0.64]	& 0.71	& [0.47--0.86] \\
		Social Evolution 2008-09-09 & 0.77 &	[0.70--0.78] 	& 0.97 & [0.96--0.98]\\
		Social Evolution 2008-10-19	& 0.81 &	[0.78--0.84]	& 0.96 & [0.94--0.97]\\
		Social Evolution 2008-12-13	& 0.78 &	[0.76--0.80]	& 0.94 & [0.93--0.95]\\
		Social Evolution 2009-03-05	& 0.83 &	[0.80--0.84]	& 0.96 & [0.96--0.97]\\
		Social Evolution 2009-04-17 & 0.83 &	[0.79--0.85]	& 0.97 & [0.96--0.98]\\
		Social Evolution 2009-05-18 & 0.81 &	[0.79--0.83]	& 0.97 & [0.96--0.98]\\
		Personality Survey	& 0.90		& [0.89--0.91]	& 0.96	& [0.95--0.97]\\
		Reciprocity Survey	& 0.73	& [0.69--0.76]	& 0.69	& [0.63--0.76]\\
		\hline
	\end{tabular}
	\label{table:other_datasets}
\end{table}

\begin{figure}[h!]
	\centering
	\includegraphics[width=1\columnwidth]{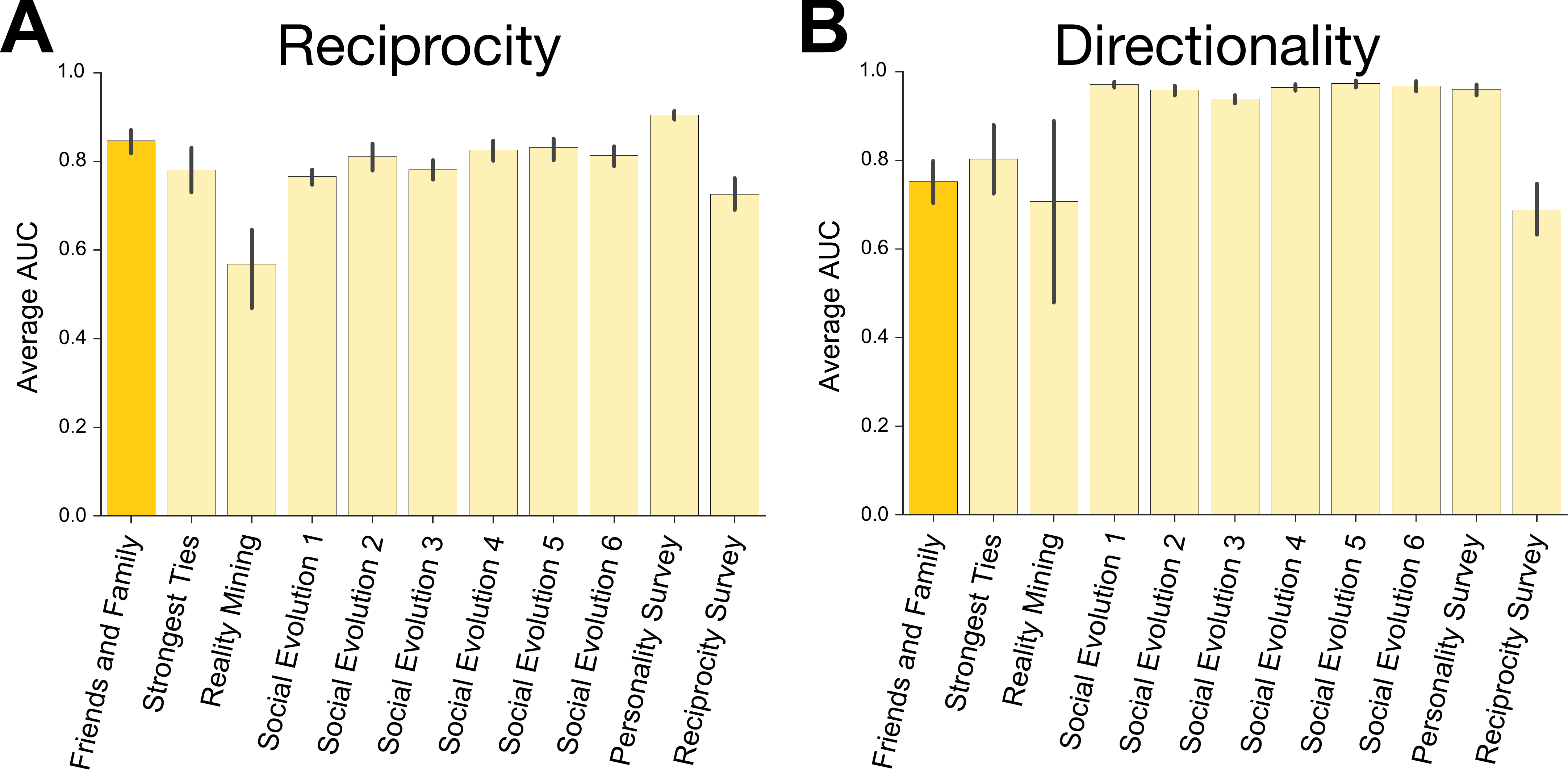}
	\caption{Comparison of classification results with additional datasets. We repeat the two classification tasks (reciprocal vs unilateral (\textbf{A}), incoming vs outgoing (\textbf{B}))  on additional datasets
		and report the results for the Friends and Family study (dark yellow) next to the additional datasets (light yellow): 
		\textit{Reality Mining}, \textit{Strongest Ties}, \textit{Social Evolution} study,
		a \textit{Personality Survey}, and our \textit{Reciprocity Survey}.
		The \textit{Social Evolution} study is split into six temporal slices, and we report each of them as a separate dataset.
		We report the average AUC of the results provided by a Random Forest evaluated with a 10-fold cross-validation method
		and the confidence intervals at 95\% computed via bootstrapping.}
	\label{fig:additional_ds}	
\end{figure}

We repeat the evaluation of both classification tasks on additional datasets (table~\ref{table:other_datasets}).
We obtain results very close to the ones with the Friends and Family dataset for all additional datasets, 
with a noticeable drop in performance only for the Reality Mining and Reciprocity Survey datasets.
In the first classification task (reciprocal vs unilateral, Fig.~\ref{fig:additional_ds}) the average AUC
for the additional datasets (0.78 on average) is generally slightly lower than in the Friends and Family dataset (0.85).
On the contrary, in the second classification task (incoming vs outgoing, Fig.~\ref{fig:additional_ds}) the average AUC for the additional datasets (0.89 on average) is generally higher than in the Friends and Family (0.75).

\begin{singlespace}
	\bibliography{main}
	\bibliographystyle{plain}
\end{singlespace}
\end{document}